\documentclass[preprint, review,3p,times, 10pt]{elsarticle}

\usepackage{natbib}
\usepackage[most]{tcolorbox}
\usepackage{float}
\usepackage{cite}
\usepackage{amsmath,amssymb,amsfonts}
\usepackage{algorithmic}
\usepackage{graphicx}
\usepackage{textcomp}
\usepackage{comment}
\usepackage{footnote}
\usepackage{comment}
\usepackage{threeparttable}
\usepackage{romannum}
\usepackage{longtable}
\usepackage{lscape}
\usepackage{soul}
\usepackage{array, makecell}
\usepackage{setspace}
\usepackage{tikz}
\usepackage{pgfplots}
\pgfplotsset{width=10cm,compat=1.9}
\usepackage{longtable}
\usepackage{threeparttablex} 
\usepackage{booktabs}

\usepackage[hyphens]{url}
\usepackage{hyperref}
\hypersetup{colorlinks=true,breaklinks=true,pdfauthor=author}

\usepackage{ragged2e}
\usepackage{blindtext}
\usepackage{verbatim}
\usepackage{sectsty}

\usepackage{subcaption}

\usepackage{etoolbox}
\usepackage{lmodern}

\usepackage{lipsum}
\usepackage{lineno}
\usepackage{enumitem}

\usepackage{pgf-umlsd}

\usepackage[rightcaption]{sidecap}

\usepackage[noabbrev]{cleveref}
\newcounter{rownumbers}
\newcommand\rownumber{\stepcounter{rownumbers}\arabic{rownumbers}}
\usepackage{svg}
\usepackage{multirow}
\usepackage{rotating}
\usepackage{paracol}
\usepackage{multicol}
\DeclareUnicodeCharacter{0307}{ }
\usepackage[utf8]{inputenc}
\usepackage{colortbl}

\usepackage{listings}
\usepackage{xcolor}

\definecolor{codegreen}{rgb}{0,0.6,0}
\definecolor{codegray}{rgb}{0.5,0.5,0.5}
\definecolor{codepurple}{rgb}{0.58,0,0.82}
\definecolor{backcolour}{rgb}{0.95,0.95,0.92}

\lstdefinestyle{java}{
commentstyle=\color{codegreen},
keywordstyle=\color{magenta},
numberstyle=\tiny\color{codegray},
stringstyle=\color{codepurple},
basicstyle=\linespread{0.8}\ttfamily\footnotesize,
breakatwhitespace=false, 
breaklines=true, 
captionpos=b,  
keepspaces=true, 
numbers=none,  
numbersep=0.5pt,  
showspaces=false,  
showstringspaces=false,
showtabs=false,  
tabsize=2
}
\definecolor{eclipseStrings}{RGB}{176,99,99}
\definecolor{eclipseKeywords}{RGB}{4,81,165}
\definecolor{eclipseBackground}{RGB}{191,191,191}
\colorlet{numb}{magenta!60!black}

\lstdefinelanguage{json}{
basicstyle=\linespread{0.9}\ttfamily\footnotesize,
commentstyle=\color{eclipseStrings}, % style of comment
stringstyle=\color{eclipseKeywords}, % style of strings
numbers=none,
numberstyle=\tiny\color{codegray},
numbers=none,
numberstyle=\scriptsize,
stepnumber=1,
numbersep=8pt,
showstringspaces=false,
breaklines=true,
string=[s]{"}{"},
comment=[l]{:\ "},
morecomment=[l]{:"},
literate=
*{0}{{{\color{numb}0}}}{1}
{1}{{{\color{numb}1}}}{1}
{2}{{{\color{numb}2}}}{1}
{3}{{{\color{numb}3}}}{1}
{4}{{{\color{numb}4}}}{1}
{5}{{{\color{numb}5}}}{1}
{6}{{{\color{numb}6}}}{1}
{7}{{{\color{numb}7}}}{1}
{8}{{{\color{numb}8}}}{1}
{9}{{{\color{numb}9}}}{1}
}
\usepackage{color}

\usepackage{tikz}
\usepackage{pgfplots}
\usepgfplotslibrary{statistics}
\pgfplotsset{
compat=1.18,
xlabel style={font=\scriptsize},
ylabel style={font=\scriptsize},
tick label style={font=\scriptsize},
}
\usepackage{pgfplotstable}
\usepackage{rotating}

\journal{Computers and Electronics in Agriculture}
\begin{document}

\begin{frontmatter}

%%%%%%%%%%%%%%%%%%%%%%%%%%%%%%%%%%%%%%%%%%%%%%%%%%%%%%%%
%%%%%                    Title                     %%%%%
%%%%%%%%%%%%%%%%%%%%%%%%%%%%%%%%%%%%%%%%%%%%%%%%%%%%%%%%
\title{\textit{LEISA}: A Scalable Microservice-based System for Efficient Livestock Data Sharing}

%%%%%%%%%%%%%%%%%%%%%%%%%%%%%%%%%%%%%%%%%%%%%%%%%%%%%%%%
%%%%%                   author                     %%%%%
%%%%%%%%%%%%%%%%%%%%%%%%%%%%%%%%%%%%%%%%%%%%%%%%%%%%%%%%

\author[1,2,5]{Mahir Habib\corref{correspondingauthor}}\ead{mhabib@csu.edu.au}
\author[1,2,5]{Muhammad Ashad Kabir}\ead{akabir@csu.edu.au}
\author[1,2,5]{Lihong Zheng}\ead{lzheng@csu.edu.au}

%%%%%%%%%%%%%%%%%%%%%%%%%%%%%%%%%%%%%%%%%%%%%%%%%%%%%%%%
%%%%%                   affiliation                %%%%%
%%%%%%%%%%%%%%%%%%%%%%%%%%%%%%%%%%%%%%%%%%%%%%%%%%%%%%%%
\affiliation[1]{organization={School of Computing, Mathematics and Engineering, Charles Sturt University}, city={Bathurst}, state={NSW}, postcode={2795}, country={Australia}}

\affiliation[2]{organization={Gulbali Institute for Agriculture, Water and Environment, Charles Sturt University}, city={Wagga Wagga}, state={NSW}, postcode={2678},  country={Australia}}

\affiliation[5]{organization={Food Agility CRC Ltd}, city={Sydney}, state={NSW}, postcode={2000},  country={Australia}}

%%%%%%%%%%%%%%%%%%%%%%%%%%%%%%%%%%%%%%%%%%%%%%%%%%%%%%%%
%%%%%              corresponding author           %%%%%
%%%%%%%%%%%%%%%%%%%%%%%%%%%%%%%%%%%%%%%%%%%%%%%%%%%%%%%%
\cortext[correspondingauthor]{Corresponding author: Charles Sturt University, Panorama Ave, Bathurst, NSW 2795. Ph.+61263386259}

%%%%%%%%%%%%%%%%%%%%%%%%%%%%%%%%%%%%%%%%%%%%%%%%%%%%%%%%
%%%%%                    abstract                  %%%%%
%%%%%%%%%%%%%%%%%%%%%%%%%%%%%%%%%%%%%%%%%%%%%%%%%%%%%%%%

\begin{abstract}

In the livestock sector, the fragmented data landscape across isolated systems presents a significant challenge, necessitating interoperability and integration. In this article, we introduce the Livestock Event Information Sharing Architecture (LEISA), a novel architecture designed to facilitate the sharing of livestock event information among various stakeholders, such as farmers and service providers. We comprehensively analysed both functional and non-functional requirements of such an architecture to ensure data integrity, standardisation, validation, ownership, and real-time data dissemination across various stakeholders while maintaining reliability, scalability, and interoperability. We designed and implemented LEISA using a cloud-based microservice architecture incorporating advanced technologies such as RESTful APIs and message brokers. We evaluated the efficiency and effectiveness of LEISA in terms of performance, scalability, and reliability, and examined its applicability through several real-world use-case scenarios. LEISA facilitates seamless data sharing, system interoperability, and enhanced decision-making capabilities, promoting synergy among stakeholders. 

\end{abstract}

%%%%%%%%%%%%%%%%%%%%%%%%%%%%%%%%%%%%%%%%%%%%%%%%%%%%%%%%
%%%%%                    keyword                   %%%%%
%%%%%%%%%%%%%%%%%%%%%%%%%%%%%%%%%%%%%%%%%%%%%%%%%%%%%%%%

\end{frontmatter}

%% \linenumbers
\pagenumbering{arabic}
%%%%%%%%%%%%%%%%%%%%%%%%%%%%%%%%%%%%%%%%%%%%%%%%%%%%%%%%
%%%%%                Introduction                  %%%%%
%%%%%%%%%%%%%%%%%%%%%%%%%%%%%%%%%%%%%%%%%%%%%%%%%%%%%%%%
\section{Introduction}\label{sec:intro}

Recent technological progress, which encompasses the adoption of sensors, automated monitoring systems, precision agriculture tools, and the Internet of Things (IoT) devices, has facilitated the collection, generation, and storage of data. These technologies play a crucial role in promoting data sharing and information exchange for management in various fields, including livestock management~\citep{Hanrahan2017}. In the livestock domain, these technologies generate large amounts of real-time data on animal health, behaviour, insemination, and movement patterns. This data includes livestock event information such as weight, location, vaccination treatments, milking, observation, and insemination events. This event information plays a central role in improving decision-making, traceability,\footnote{Traceability refers to the ability to track livestock from birth to slaughter or when moving from one yard to another, which is essential for managing disease outbreaks and ensuring food safety.} quality control,\footnote{Quality control in livestock involves using technologies and methods to continuously monitor and manage livestock health and production parameters. It includes real-time data collection and analysis to maintain optimal conditions, prevent issues, and ensure consistent, high-quality livestock products.} and overall operational efficiency in livestock management~\citep{Stygar2021}.

Sharing data provides significant business advantages such as unlocking significant value by enhancing operational efficiency~\citep{hosoda2008there}, driving innovation~\citep{eckartz2014decision}, opening new revenue streams~\citep{sugathapala2015quantifying}, and enabling data-driven decision-making~\citep{huttunen2019benefits}. However, information about livestock events is isolated by data producers,\footnote{A data producer is a farmer who raises sheep or cattle, or alternatively, systems and technologies that collect, analyse, and share information on livestock events.} such as farmers and Agtech providers, within various unconnected software systems that lack sufficient interoperability and integration. This data isolation in livestock management leads to operational inefficiencies and significantly restricts the comprehensive use of data by consumers,\footnote{A data consumer, also referred to as a data subscriber, represents the ultimate recipient of an information event.} such as researchers, regulators, sale yards and veterinarians, for informed decision-making, innovation, and in-depth analysis and research~\citep{Bahlo2021LivestockAustralia}.

To effectively share livestock event information and overcome the challenges of data isolation, a robust data-sharing system architecture is necessary. Addressing both functional and non-functional requirements is crucial for designing this architecture. The functional requirements include real-time data sharing to ensure that data adhere to predefined standards and formats, allowing data producers to control how their data is shared, maintaining a clear separation between data producers and consumers, and enabling consumers to operate without needing continuous online connectivity. From a non-functional perspective, the system must be reliable, scalable to handle increased workloads, and capable of integrating seamlessly with existing and future agricultural data systems.

In this paper, we have designed and implemented the Livestock Event Information Sharing Architecture (LEISA) to fulfil the aforementioned functional and non-functional requirements. LEISA employed the microservice architecture (MSA), which involves breaking down applications into smaller, independently deployable services that communicate through well-defined application programming interfaces (APIs). This modular approach allows for more efficient handling of diverse event types (e.g., animal health records, movement data, and genetic and breeding) and sources (e.g., sensors, manual input, and automated systems), which is crucial in smart farming environments~\citep{trilles2020iot}. LEISA facilitates the sharing of information on livestock events among stakeholders such as producers and consumers. It is designed to decouple these stakeholders and treat them as external services. By decoupling services, LEISA facilitates easier updates and maintenance, reducing downtime, and improving system resilience.

LEISA is constructed to ensure data standardisation, integrity, and efficient routing.
It includes 14 internal microservices, such as a validator for data standardisation, message\footnote{A message is a distinct unit of data that is sent between applications or services.} publishing, message consumption, and queue mapping to route messages. In addition, it features a message broker for real-time information sharing and scalability between producers and consumers, along with a database for their registration in LEISA. This design enables livestock producers to retain control over the routing of their data. All LEISA components are implemented within a cloud infrastructure, providing high availability, serverless computing, and automatic scaling, ensuring robust and cost-effective operations.

Several studies, such as those by~\citet{Spanaki2021},~\citet{shonhe2021sharing}, and~\citet{Wysel2021}, emphasise the importance of data sharing in agriculture but primarily focus on crop-related data rather than livestock data. A few studies, including those by~\citet{lupu2023attitudes},~\citet{zhu2021agricultural}, and~\citet{kumar2022deep}, explore data sharing specifically within the livestock domain but target only single livestock events and utilise technologies such as IoT and blockchain rather than focusing on the software architecture. Studies that addressed data sharing for livestock focused mainly on internal users (i.e., the producers themselves) rather than external consumers~\citep{zhang2021will,Taneja2019SmartHerdFarming,Aydin2022DesignData}. Many studies propose IoT architectures for specific data management tasks, such as health monitoring, GPS (Global Positioning System) tracking, and movement tracking, but do not emphasise the need for a comprehensive data sharing architecture~\citep{durrant2021might, kharel2022linking}.

It is important to note that, unlike many studies~\citep{abbasi2022towards,stacey2022fair,leme2020secure} that emphasise the importance of data security and privacy when discussing data sharing, our study is primarily concerned with the mechanisms and processes that facilitate data sharing. The contributions of this article are fourfold:

\begin{itemize}

\item We identified both functional and non-functional requirements crucial for effective data sharing in the livestock industry. These requirements have laid a solid foundation for the architecture, ensuring its long-term viability and effectiveness in practical applications.

\item We proposed a design of the Livestock Event Information Sharing Architecture (LEISA) integrating 14 microservices. These microservices manage operations ranging from producer and consumer registration to event information (message) validation and advanced routing of messages.

\item We implemented LEISA on Azure, leveraging several of its advanced services. We incorporated a message broker that facilitates real-time data exchanges between producers and consumers, ensuring efficient and seamless communication.

\item We evaluated LEISA's performance, scalability, and reliability, and investigated its applicability by examining various real-world use-case scenarios.
\end{itemize}

The remainder of this paper is structured as follows. \Cref{sec:requirements-analysis} identifies the architectural requirements essential for addressing these challenges effectively. \Cref{sec:related-work} offers a comprehensive review of pertinent literature, encompassing established frameworks and recent data-sharing advancements within livestock management and related domains. Our primary contribution, the Livestock Event Information Sharing Architecture (LEISA), is elucidated in~\Cref{sec:LEISA}, which delineates the architecture's layered design, focusing on its key design principles, such as microservices and message broker implementation. \Cref{sec:implementation} expounds upon the implementation strategies employed in the development of LEISA. \Cref{sec:leisa applications} outlines real-world applications for LEISA, accentuating highlighting its interoperability and exemplifying its capacity to improve livestock management processes, improve data standardisation, and streamline data sharing efficiency. \Cref{sec:evaluation} delves into the evaluation methodology, results, and findings derived from evaluating the five main core microservices of LEISA.~\Cref{sec:threats} addresses potential threats to the validity of our findings. Finally,~\Cref{sec:conclusion} concludes the paper and outlines future research directions.

%%%%%%%%%%%%%%%%%%%%%%%%%%%%%%%%%%%%%%%%%%%%%%%%%%%%%%%%
%%%%%         Requirements Analysis                %%%%%
%%%%%%%%%%%%%%%%%%%%%%%%%%%%%%%%%%%%%%%%%%%%%%%%%%%%%%%%
\section{Requirements Analysis}\label{sec:requirements-analysis}

In developing an effective data sharing architecture, it is essential to perform a thorough requirements analysis. This analysis involves identifying the functional and non-functional requirements the architecture must meet to support data sharing needs within the livestock sector~\citep{abbasi2022towards}. \Cref{sec:freq,sec:nonfreq} detail the functional and non-functional requirements that have been identified.

\subsection{Functional Requirements}\label{sec:freq}

The functional requirements specify what a system should do, i.e., the tasks it performs, the data it handles, and the direct outcomes expected from these processes.

\begin{enumerate}[label=\textbf{FR \#\arabic*.}, leftmargin=*, labelwidth=*, align=left]

\item \textbf{Real-Time Data Sharing}: The architecture should enable real-time sharing and dissemination of livestock event information. This includes the capability to validate, standardise, and route data efficiently, utilising a message broker to ensure timely access to information for decision-making processes~\citep{Amiri-Zarandi2022}. For instance, in the weight event scenario (i.e., measuring the livestock weight), the architecture must handle weight data manually inputted through traditional means in portable formats like comma-separated values (CSV) or automatically transmitted through IoT devices. Real-time sharing of this weight data allows for the efficient assessment of livestock health and growth performance, demonstrating the requirement's practical application in enhancing agricultural productivity and decision-making accuracy.

\item \textbf{Data Standardisation and Validation}: A data validator is a component that ensures the integrity and correctness of data by verifying that it conforms to predefined formats and standards. Implementing a data validator within the architecture is critical for enhancing interoperability across systems and platforms. Compliance and adherence to standards in software architecture involve conforming to relevant industry regulations and standards, especially those pertaining to data sharing and agricultural practices. This compliance is essential for the architecture's long-term viability and to gain acceptance from various stakeholders, including regulatory bodies and users in the agricultural sector~\citep{habib2023lei}. The vaccination event is an example of the recording of vaccine administration details. This event necessitates standardising the data format (e.g., vaccine type, dosage, administration date) and validating it against standard. By applying real-time data standardisation and validation, the architecture ensures that vaccination records are consistently formatted and accurate, facilitating seamless sharing and compliance with standard regulations.

\item \textbf{Producer Control over Message Routing}: To preserve ownership, livestock data producers need control over their message routing when sharing data with consumers.\footnote{Data ownership involves legal rights and ethical responsibilities related to creating, controlling, and using data. It determines access, sharing, and benefits, ensuring security, privacy, and distribution control to maintain integrity and confidentiality.} Ownership allows producers to dictate access and sharing conditions, protecting sensitive business information. The architecture should provide mechanisms for producers to control data routing. For instance, in a livestock movement\footnote{A livestock movement event involves transporting animals between locations such as farms, markets, or slaughterhouses, essential for breeding, fattening, and disease management.} event involving the transfer of animals between locations or ownership,\footnote{Livestock ownership involves legal and practical rights over domesticated animals used for food production, income, and agricultural labour, including decisions about breeding, sale, and care of the animals.} producers specify who receives data about the transaction, including health records, ownership details, and movement logs. This control is crucial for privacy, traceability, and compliance, enabling producers to manage the dissemination of sensitive information effectively.

\item \textbf{Decoupling Producers from Consumers}: Decoupling producers from consumers is essential to maintain a clear separation of concerns between entities that generate data (i.e., producers) and those that use these data (i.e., consumers). The primary objective is to ensure that producers and consumers can operate, evolve, and scale independently without interfering with each other. Consider the scenario of the Location Event,\footnote{ Location Event is for recording the geographical positions of livestock using modern technologies. This includes monitoring animal movements, grazing patterns, and managing livestock for welfare purposes.} where IoT tracking devices collect real-time location data from livestock. In this case, farmers (producers) collect and share location data, while veterinarians and regulatory bodies (consumers) use these data for monitoring and management purposes. By decoupling the producers and consumers, farmers can change or enhance their data collection methods without affecting the operations of veterinarians and regulatory bodies.

\item \textbf{Consumer autonomy}: Consumer independence is crucial in enabling consumers to operate without the need for continuous online connectivity, particularly through using of asynchronous messaging systems. These systems allow consumers to receive and process messages or data even if they were offline when the messages were sent, thereby supporting a seamless communication experience. Implementing mechanisms like message queues or event buses can store messages until the consumer is ready to retrieve and process them, thus ensuring that communication is not dependent on constant connectivity~\citep{simpson2017high}. For instance, the status observed event demonstrates the importance of asynchronous data access, particularly during increased activity, such as breeding or selling seasons. Autonomy allows the consumer to access and process information independently, regardless of varying connectivity and operational conditions. As a result, consumers can manage data on their own schedules, enhancing overall efficiency and accommodating the diverse needs of the system.

\end{enumerate}

\subsection{Non-Functional Requirements}\label{sec:nonfreq}

Non-functional requirements describe how the architecture operates, focusing on the system's behaviour.

\begin{enumerate}[label=\textbf{NFR \#\arabic*.}, leftmargin=*, labelwidth=*, align=left]

\item \textbf{Reliability}: Reliability in a software architecture refers to the system's consistent and accurate operation over time. This characteristic is crucial in data sharing architectures to minimise risks associated with errors, system downtime, and performance issues, ensuring dependable system operation~\citep{mohamed2010architectural}. Using the Weight Event as an example, the system's reliability is paramount to ensure that real-time weight data, whether entered manually or captured through IoT devices, is processed and disseminated accurately and without interruption. This reliability supports effective livestock health and growth monitoring, enabling timely decisions based on accurate data.

\item \textbf{Scalability}: Scalability denotes the architecture's capacity to handle increased workloads, such as more users, data, or transactions, without sacrificing performance or reliability. This aspect is vital in data sharing architecture to accommodate growing numbers of producers, consumers, and messages, ensuring operational continuity and adaptability to changing usage patterns~\citep{Gorton2022}. The Vaccination Event highlights scalability, as the system must accommodate data from potentially thousands of vaccination records during peak health management periods without performance degradation. Scalability ensures that as the number of records grows, the system can still process, standardise, and validate each entry efficiently, maintaining performance levels even under increased strain.

\item \textbf{Interoperability}: Interoperability is the architecture's ability to seamlessly integrate with both existing and future agricultural data systems. This involves creating a modular architecture, facilitating the addition or modification of components, and implementing comprehensive and robust APIs for effective data communication and utilisation~\citep{Jepsen2021}. In the context of the Livestock Movement Event, interoperability is key for exchanging data between various stakeholders involved in the buying and selling process. The system must integrate with different databases, regulatory reporting tools, and other farm management systems to ensure smooth transactions. This includes standardising of data formats for easy access and processing by all parties, facilitating a streamlined transfer of ownership and movement records across diverse platforms.

\end{enumerate}

%%%%%%%%%%%%%%%%%%%%%%%%%%%%%%%%%%%%%%%%%%%%%%%%%%%%%%%%
%%%%%                Related work                  %%%%%
%%%%%%%%%%%%%%%%%%%%%%%%%%%%%%%%%%%%%%%%%%%%%%%%%%%%%%%%
\section{Related Work}\label{sec:related-work}
The landscape of data sharing within the livestock sector has advanced significantly due to the increasing adoption of precision livestock farming (PLF) technologies. This section explores the current research on data sharing in the livestock sector, focusing on the functional and non-functional requirements essential for effective data sharing. To provide a structured comparative analysis,~\Cref{tab:state-of-the-art} below evaluates the extent to which various studies have addressed specific criteria related to data sharing, domain specificity towards livestock, the implementation of MSA, and adherence to functional and non-functional requirements. This comparison reveals a strong emphasis on data sharing across the board yet highlights a relative scarcity of research specifically focused on the livestock domain and the full utilisation of MSA. Moreover, the diversity in addressing functional and non-functional requirements reflects the varying approaches and complexity of data management needs in the sector.

\begin{table*}[!ht]
\caption{Comparative analysis of research on data sharing in agriculture: Focus on livestock domain, use of microservice architecture, and adherence to functional and non-functional requirements}
\label{tab:state-of-the-art}
\centering
\resizebox{\textwidth}{!}{%
\begin{tabular}{lllllllllllll}
\hline
\multirow{2}{*}{Author/initiative} & \multirow{2}{*}{\shortstack{Data\\Sharing}} & \multirow{2}{*}{\shortstack{Livestock\\domain}} & \multirow{2}{*}{\shortstack{Using\\MSA}} & \multicolumn{5}{c}{{\shortstack{Functional Requirements}}} & & \multicolumn{3}{c}{{\shortstack{Non-functional Requirements}}} \\
\cline{5-9} \cline{11-13}
& & & & FR \#1 & FR \#2 & FR \#3 & FR \#4 & FR \#5 & & NF \#1 & NF \#2 & NF \#3 \\
\hline
\citet{Shen2019} & $\checkmark$ & $\times$ & $\sim$ & $\sim$ & $\sim$ & $\sim$ & $\checkmark$ & $\checkmark$ & & $\checkmark$ & $\checkmark$ & $\sim$ \\
\citet{Qiu2020} & $\checkmark$ & $\times$ & $\times$ & $\times$ & $\times$ & $\times$ & $\times$ & $\times$ & & $\times$ & $\times$ & $\times$ \\
\citet{Zhang2022} & $\checkmark$ & $\times$ & $\sim$ & $\sim$ & $\sim$ & $\checkmark$ & $\checkmark$ & $\checkmark$ & & $\checkmark$ & $\checkmark$ & $\sim$ \\
\citet{Hund2021} & $\checkmark$ & $\times$ & $\sim$ & $\sim$ & $\checkmark$ & $\checkmark$ & $\checkmark$ & $\sim$ & & $\checkmark$ & $\checkmark$ & $\checkmark$ \\
\citet{Yang2018} & $\checkmark$ & $\times$ & $\checkmark$ & $\sim$ & $\checkmark$ & $\sim$ & $\checkmark$ & $\sim$ & & $\checkmark$ & $\checkmark$ & $\checkmark$ \\
\citet{Hasan2021} & $\checkmark$ & $\times$ & $\checkmark$ & $\sim$ & $\sim$ & $\sim$ & $\checkmark$ & $\sim$ & & $\checkmark$ & $\checkmark$ & $\checkmark$ \\
\citet{Maia2020} & $\checkmark$ & $\times$ & $\checkmark$ & $\checkmark$ & $\checkmark$ & $\sim$ & $\checkmark$ & $\sim$ & & $\checkmark$ & $\checkmark$ & $\checkmark$ \\
\citet{Aydin2022DesignData} & $\sim$ & $\checkmark$ & $\checkmark$ & $\times$ & $\times$ & $\times$ & $\checkmark$ & $\checkmark$ & & $\checkmark$ & $\checkmark$ & $\checkmark$ \\
\citet{Mateo-Fornes2021} & $\sim$ & \checkmark & \checkmark & $\times$ & $\times$ & $\times$ & $\checkmark$ & $\checkmark$ & & $\checkmark$ & $\checkmark$ & $\checkmark$ \\
\citet{Taneja2019SmartHerdFarming} & $\sim$ & \checkmark & \checkmark & $\times$ & $\times$ & $\times$ & $\checkmark$ & $\checkmark$ & & $\checkmark$ & $\checkmark$ & $\checkmark$ \\
\citet{Shabani2022} & $\sim$ & \checkmark & \checkmark & $\times$ & $\times$ & $\times$ & $\checkmark$ & $\checkmark$ & & $\checkmark$ & $\checkmark$ & $\checkmark$\\
\citet{Trzec2022} & $\sim$ & $\times$ & \checkmark & $\checkmark$ & $\times$ & $\times$ & $\checkmark$ & $\checkmark$ & & $\checkmark$ & $\checkmark$ & $\checkmark$\\
\citet{Nguyen2022a} & \checkmark & $\times$ & \checkmark & $\times$ & $\times$ & $\times$ & $\checkmark$ & $\checkmark$ & & $\checkmark$ & $\checkmark$ & $\checkmark$ \\
\citet{Moysiadis2022} & $\sim$ & $\times$ & \checkmark & $\times$ & $\times$ & $\times$ & $\checkmark$ & $\checkmark$ & & $\checkmark$ & $\checkmark$ & $\sim$\\
\citet{Valecce2019} & $\sim$ & $\times$ & $\checkmark$ & $\checkmark$ & $\times$ & $\times$ & $\checkmark$ & $\checkmark$ & & $\checkmark$ & $\checkmark$ & $\sim$\\
\citet{Codeluppi2020} & $\sim$ & $\times$ & $\checkmark$ & $\checkmark$ & $\times$ & $\times$ & $\checkmark$ & $\checkmark$ & & $\checkmark$ & $\checkmark$ & $\sim$\\
\citet{Gkoulis2021} & $\sim$ & $\times$ & $\checkmark$ & $\sim$ & $\times$ & $\times$ & $\checkmark$ & $\checkmark$ & & $\checkmark$ & $\checkmark$ & $\checkmark$\\
\citet{Antonopoulos2019} & $\sim$ & $\times$ & $\checkmark$ & $\sim$ & $\times$ & $\times$ & $\checkmark$ & $\checkmark$ & & $\checkmark$ & $\checkmark$ & $\sim$\\
\citet{Oliver2018} & $\sim$ & $\times$ & $\checkmark$ & $\checkmark$ & $\times$ & $\times$ & $\checkmark$ & $\checkmark$ & & $\checkmark$ & $\checkmark$ & $\sim$\\
\citet{Moysiadis2022a} & $\sim$ & $\times$ & \checkmark & $\times$ & $\times$ & $\times$ & $\checkmark$ & $\checkmark$ & & $\checkmark$ & $\checkmark$ & $\sim$\\
\citet{Oliveira2021} & $\sim$ & $\times$ & \checkmark & $\checkmark$ & $\times$ & $\times$ & $\checkmark$ & $\checkmark$ & & $\checkmark$ & $\checkmark$ & $\checkmark$\\
% \citet{Wysel2021} & \checkmark & $\times$ & $\times$ & $\times$ & $\times$ \\
\citet{Yan2016} & $\checkmark$ & $\times$ & $\times$ & $\times$ & $\times$ & $\times$ & $\checkmark$ & $\checkmark$ & & $\checkmark$ & $\checkmark$ & $\sim$\\
\citet{Kim2015} & $\checkmark$ & $\times$ & $\sim$ & $\times$ & $\times$ & $\times$ & $\checkmark$ & $\checkmark$ & & $\checkmark$ & $\checkmark$ & $\sim$\\
\citet{Roussaki2023} & $\checkmark$ & $\times$ & $\sim$ & $\times$ & $\times$ & $\times$ & $\checkmark$ & $\checkmark$ & & $\checkmark$ & $\checkmark$ & $\checkmark$\\
Our work & \checkmark & \checkmark & \checkmark & \checkmark & \checkmark & \checkmark & \checkmark & \checkmark & &\checkmark & \checkmark & \checkmark \\
\hline
\end{tabular}
}
\begin{tablenotes}
\footnotesize
\item Note: FR \#1= Real-Time Data Processing utilising message broker, FR \#2 = Data Standardisation and Validation, FR \#3 = Producer Control over Message Routing, FR \#4 = Decoupling Producers from Consumers, FR \#5 = Consumer Independence, NFR \#1 = Reliability, NFR \#2 = Scalability, NFR \#3 = Interoperability

\item \checkmark indicates a direct match or presence
\item $\times$ indicates a direct absence or mismatch
\item $\sim$ indicates the concept is indirectly mentioned or inferred
\end{tablenotes}
\end{table*}

This analysis reviews the current landscape of data sharing research within agriculture, specifically focusing on the livestock domain. It evaluates the use of MSA and adherence to functional and non-functional requirements across various studies conducted from 2015 to 2023. A total of 26 studies, including a recent contribution labelled ``Our work," were systematically analysed. The studies were assessed based on their adherence to data sharing practices, relevance to the livestock domain, implementation of MSA, and conformance to specified functional and non-functional requirements.

The analysis of the ``Data Sharing" criterion across various studies, as presented in the overview table, illustrates a primary focus on integrating data sharing mechanisms. Twenty out of 26 studies incorporate data sharing, marked with a checkmark (\checkmark), underscoring its critical role in agricultural research and signifying a robust commitment to facilitating data exchange practices~\citep{Shen2019, Qiu2020, Zhang2022}.

However, the ``Livestock Domain" criterion reveals a significant gap, with only a few studies directly engaging with this sector. Only eight studies focus explicitly on livestock, as indicated by the checkmarks (\checkmark), suggesting broader applications of data sharing research within agriculture. This discrepancy highlights a notable research void in integrating data sharing frameworks specifically tailored to livestock management, underscoring the need for targeted research in this essential area \citep{Aydin2022DesignData, Mateo-Fornes2021, Taneja2019SmartHerdFarming}.

The utilisation of MSA is another key aspect examined in the table. This distinction helps identify studies employing this crucial technology, which is pivotal role in enhancing scalability and flexibility within agricultural data management systems \citep{Trzec2022, Roussaki2023}. The substantial allocation of checkmarks (\checkmark) under the MSA column indicates a growing recognition within the agricultural research community of the benefits offered by MSA. While not all studies explicitly state their use of MSA, 10 studies employ microservice architecture, indicating a growing but not yet mainstream trend. The inference of its principles from their methodologies suggests an emerging trend towards adopting such frameworks \citep{Nguyen2022a, Moysiadis2022}.

The analysis of functional requirements reveals diverse levels of implementation across the studies. Real-time data processing utilising a message broker (FR \#1) is implemented in 12 studies, indicating a moderate adoption rate. Data standardisation and validation (FR \#2) are adhered to in 18 studies, underscoring its recognised importance in ensuring data integrity and usability across different systems and applications. Producer control over message routing (FR \#3) is addressed in only 8 studies, highlighting a significant gap and potential area for improvement. This requirement is critical for managing data flow and ensuring that data producers can influence how their data are disseminated. Decoupling producers from consumers (FR \#4) is implemented in 16 studies, reflecting a strong recognition of the need for flexible and scalable data architectures that allow the independent evolution of data producers and consumers. Consumer independence (FR \#5) is ensured in 21 studies, demonstrating its high significance in the design of data sharing systems, as it allows consumers to access and utilise data without being tightly coupled to specific data sources.

Examining non-functional requirements reveals a high priority of reliability (NFR \#1), with 24 studies addressing this aspect. This focus underscores the critical need for dependable and consistent data sharing systems that users can trust. Scalability (NFR \#2) is also highlighted, with 23 studies addressing this requirement. This reflects the need for systems that can efficiently handle increasing amounts of data and a growing number of users without compromising performance. Interoperability (NFR \#3) is addressed in 22 studies, highlighting its importance in allowing diverse systems and applications to work seamlessly. This facilitates broader and more effective data sharing across different platforms and organisations.

This discussion provides valuable insights into the current state of data sharing research in agriculture, highlighting key trends and identifying areas for future exploration. The findings underscore the need to enhance the focus on livestock-specific research, increase the adoption of microservice architecture (MSA), and improve the adherence to functional requirements such as producer control over message routing and real-time data processing. Moreover, the review also highlights significant advancements and ongoing challenges in the livestock sector. It confirms the widespread emphasis on data sharing and the increasing adoption of MSA but also reveals a notable lack of studies addressing the unique requirements of the livestock domain. Both functional and non-functional areas require continued focus.

Adopting advanced technologies and interoperable standards promises to unlock the full potential of data-driven approaches in livestock management. The distribution of responses to functional and non-functional criteria indicates selective fulfilment of these essential requirements, with our work notably achieving complete adherence to all listed criteria.

%%%%%%%%%%%%%%%%%%%%%%%%%%%%%%%%%%%%%%%%%%%%%%%%%%%%%%%%
%%%%%                     LEISA                    %%%%%
%%%%%%%%%%%%%%%%%%%%%%%%%%%%%%%%%%%%%%%%%%%%%%%%%%%%%%%%
\section{Livestock Event Information Sharing Architecture}\label{sec:LEISA}

\begin{figure*}[!htb]
\centering
\includegraphics[width=1\textwidth]{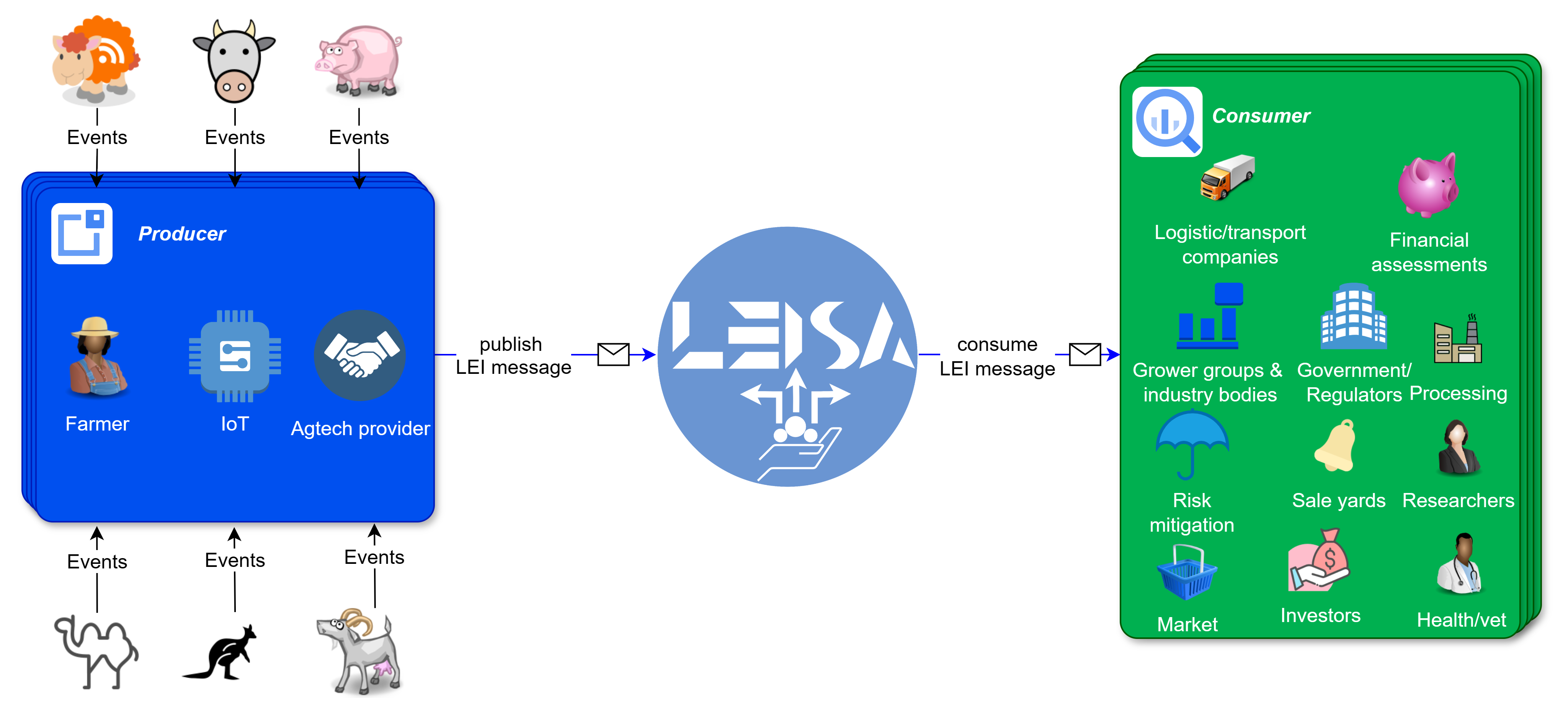}
\caption{The Livestock Event Information Sharing Architecture (LEISA) Overview}
\label{fig:datasharing}
\end{figure*}

The Livestock Event Information Sharing Architecture (LEISA) is a cloud-based solution designed for implementation in any cloud infrastructure, as illustrated in~\Cref{fig:datasharing}. It follows the principles of the microservice approach, addressing the need for sharing livestock event data among stakeholders in the red meat industry, including producers, consumers, service providers, and processors. By employing a hybrid design pattern that combines API\footnote{API is the standardised method to integrate software components.} patterns and message brokering,\footnote{Message broker is middleware software that handles communication between applications.} LEISA offers a robust solution to meet industry-wide data sharing needs.

The implementation of LEISA in a cloud infrastructure offers flexibility and adaptability across different cloud platforms and environments. By leveraging cloud services and technologies, LEISA efficiently handles, validates, and shares livestock event data. This cloud-based approach aligns with the growing trend of using internet and communication technologies to connect various entities (e.g., producers and consumers) and enable data-driven solutions.

Furthermore, microservices play a fundamental role in LEISA's design. This architecture promotes modularity and scalability by allowing the system to consist of loosely coupled and independently deployable services. Decoupling ensures that changes or updates in one service do not impact others, providing flexibility for stakeholders to evolve their systems independently while seamlessly exchanging livestock event data through LEISA.

Scalability is a critical requirement for LEISA due to the nature of the red meat industry. With the increasing number of data and stakeholders, LEISA must handle a significant amount of data and accommodate future growth. By utilising a microservices-based architecture, LEISA can scale horizontally by adding more instances of specific services as needed. This approach ensures that the system meets the data sharing requirements of the red meat industry efficiently, regardless of size or complexity~\citep{Dineva2021}.

The decoupling provided by the microservice architecture is crucial in the diverse environment of the red meat industry. With multiple stakeholders involved, each with its own systems and requirements, decoupling ensures that changes or updates in one service do not impact others. This flexibility allows stakeholders to evolve their systems independently while seamlessly exchanging livestock event data through LEISA. The decoupling also promotes modularity and reusability, facilitating easier integration with existing systems and future enhancements.

LEISA can be applied in various scenarios, such as tracking livestock health events, optimising supply chain processes, and ensuring traceability in the red meat industry. Future enhancements to LEISA could include integrating advanced analytics for predictive insights and expanding interoperability with additional industry platforms (see~\Cref{sec:leisa applications}).

%%%%%%%%%%%%%%%%%%%%%%%%%%%%%%%%%%%%%%%%%%%%%%%%%%%%%%%%
%%%%%                LEISA Layers                  %%%%%
%%%%%%%%%%%%%%%%%%%%%%%%%%%%%%%%%%%%%%%%%%%%%%%%%%%%%%%%
\subsection{LEISA Layers}\label{sec:layers}

In our architecture, producers and consumers are considered as distinct, self-contained entities, operating as external services. This design approach allows their independent development and management, minimising reliance on other architectural components. Producers and consumers engage with the core application architecture through well-defined interfaces and APIs. This strategy fosters modularity and the segregation of responsibilities, simplifying the autonomous development, maintenance, and scalability of various architectural components.

\begin{figure}[!htb]
\centering
\includegraphics[width=0.9\textwidth]{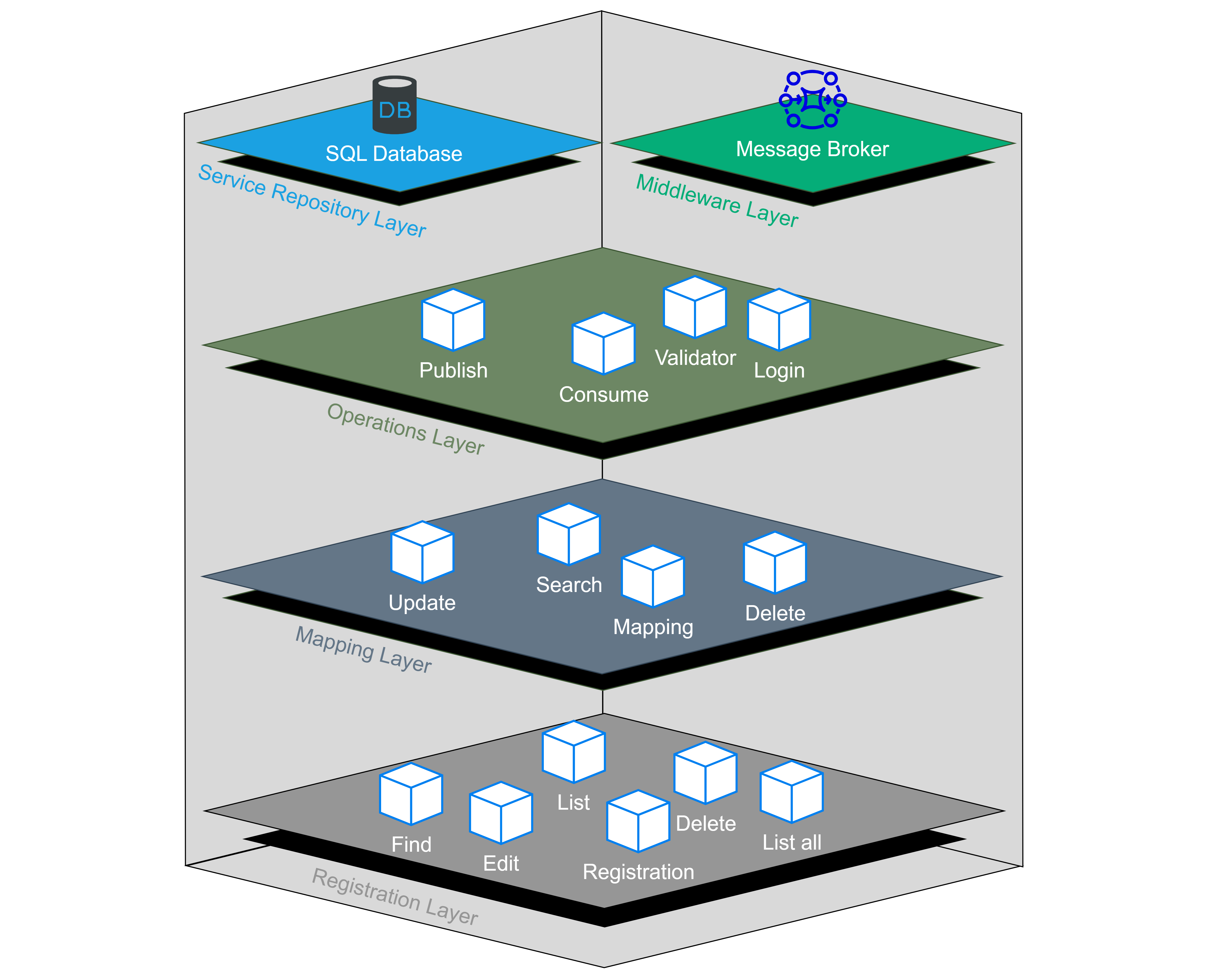}
\caption{LEISA layers overview}
\label{fig:leisalayers}
\end{figure}

LEISA is organised into multiple layers, each fulfilling specific roles and functions. \Cref{fig:leisalayers} breaks down these layers and their respective functions. The LEISA architecture comprises five distinct layers: the \emph{Registration Layer}, \emph{Mapping Layer}, \emph{Operations Layer}, \emph{Service Repository Layer}, and \emph{Middleware Layer}. Each layer plays a unique role in ensuring effective architecture operation.

The \textbf{Registration Layer} is a foundational component, providing a mechanism for external services to register and interact with LEISA. This Layer handles the initial onboarding process for external services, creating unique identifiers and credentials to ensure secure communication with LEISA. It manages the registration process, including verifying the identity and authenticity of external entities. In addition to registration, this layer provides Create, Read, Update, and Delete (CRUD) operations for registered services, allowing them to interact with and manipulate their data within LEISA.

The \textbf{Mapping Layer} connects producers and consumers through livestock events. It plays a critical role in data routing by mapping events to the correct consumer queues, ensuring that messages are directed to the appropriate consumers based on specific livestock events. It supports CRUD operations to maintain queue mappings, ensuring adaptability and scalability as the system evolves. Utilising message brokers and SQL databases, the Mapping Layer facilitates an event-driven architecture that processes messages triggered by specific events, enabling consumers to dynamically adapt to changing conditions.

The \textbf{Operations Layer} is the hub for the core functionalities that drive the system's business logic and data processing. This layer hosts microservices dedicated to publishing and consuming messages, validating messages against the LEI schema, and managing service login. It ensures the accurate flow and processing of information throughout the architecture. Key functions include managing the publication and retrieval of messages, validating data consistency and integrity, and securing access through robust authentication mechanisms.

The \textbf{Service Repository Layer} is the backbone for storing and managing information related to external services and mapping events. It provides a stable and secure environment for maintaining service-related data. The SQL database component offers efficient querying capabilities, essential for ensuring smooth operations and rapid response times. This layer stores credentials and essential information for external services, ensuring accurate authentication and authorisation, and manages data related to mapping events.

The \textbf{Middleware Layer} acts as a crucial communication bridge, facilitating the bidirectional flow of data between producers and consumers through asynchronous processing. Asynchronous message handling allows producers to send messages without requiring immediate processing by consumers, enhancing system responsiveness and preventing bottlenecks. This layer includes a message broker, which handles the routing of messages to the appropriate queues, ensuring scalable and reliable data transfer. Security within this layer is paramount, with implemented access controls to ensure that only authorised producers and consumers can interact with the message broker.

%%%%%%%%%%%%%%%%%%%%%%%%%%%%%%%%%%%%%%%%%%%%%%%%%%%%%%%%
%%%%%                LEISA Design                  %%%%%
%%%%%%%%%%%%%%%%%%%%%%%%%%%%%%%%%%%%%%%%%%%%%%%%%%%%%%%%

\subsection{Design}\label{sec:design}

LEISA's architecture is a testament to modern design, integrating a hybrid pattern that combines the immediacy of APIs with the robustness of message brokering. This innovative approach allows for synchronous interactions and asynchronous data flows, ensuring direct communication within LEISA and facilitating seamless external services to data sharing.

At the core of LEISA's communication strategy are APIs, which enable direct exchanges between its internal microservices and allow external services, such as third-party applications, to engage with LEISA. The data sharing is complemented by a message broker, which efficiently manages the data flow between data-producing external services (producers) and data-utilising external services (consumers). The message broker's role is crucial in enabling asynchronous communication, allowing external services to continue their operations without delay, and providing a scalable solution that supports adding or removing services without affecting LEISA's core functions.

The sophistication of LEISA's design is further highlighted by its microservices architecture, which consists of 14 distinct services, each with a specific role. These microservices are categorised as either Functions or RESTful APIs. Functions are standalone computational units that are event-driven and managed within a serverless environment. In contrast, RESTful APIs are based on the REST principles, using standardised HTTP methods for structured communication across different systems.

Certain microservices within LEISA require authentication to perform their tasks. This authentication process ensures only verified external services can interact with the internal microservices. Communication within or with LEISA primarily occurs through HTTP requests, which enable the transfer of data and execution results between the microservices of LEISA or between LEISA and the external service, thus simplifying the interaction process.

Parameters are essential for the operation of LEISA's microservices, providing the necessary data and context for their tasks. These parameters range from service identifiers to complex LEI JSON messages tailored to the specific requirements of each microservice. The outputs generated by the microservices are critical for the communication cycle, offering structured data, status updates, or error details to the producers or consumers, thus enabling effective communication across the architecture.

To provide a clear understanding of each microservice's role within LEISA,~\Cref{{tab:services}} summarises their names, types, authentication requirements, access methods, parameters, and the returns they provide. This table serves as a valuable reference for comprehending the functionalities and contributions of individual services to the overall architecture.

\setcounter{rownumbers}{0}
\begin{table*}[!ht]
\centering
\caption{Overview of the microservices within LEISA}
\label{tab:services}
\resizebox{\textwidth}{!}{%
\begin{tabular}{lllp{2cm}lp{3cm}p{3cm}}
\hline
No. & Microservice Name & Service Type & Authentication required & Access method & Parameters & Returns\\
\hline
\rownumber & ServiceRegistrationMicroservice & Function & $\times$ & HTTP request & N/A & HTTP response + service details or error details\\

\rownumber & ServiceLoginMicroservice & Function & $\times$ & HTTP request & N/A & HTTP request + service details\\

\rownumber & DeleteServiceMicroservice & Function & \checkmark & HTTP request & N/A & HTTP request or error details\\

\rownumber & FindServiceMicroservice & Function & \checkmark & HTTP request & Service ID & HTTP request + service details\\

\rownumber & ListAllServicesMicroservice & Function & \checkmark & HTTP request & N/A & HTTP response + services details \\

\rownumber & ListServiceMicroservice & Function & \checkmark & HTTP request & N/A & HTTP response + services details \\

\rownumber & MessageValidatorMicroservice & Function & $\times$ & HTTP request & LEI JSON message & HTTP response or error details in case message invalid\\

\rownumber & PublishMessageMicroservice & Function & \checkmark & HTTP request & Event name, LEI JSON message & HTTP response\\

\rownumber & UpdateServiceMicroservice & Function & \checkmark & HTTP request & Service ID & HTTP response\\

\rownumber & SetQueueMappingMicroservice & Function & \checkmark & HTTP request & Event name , service (consumer) ID & HTTP response\\

\rownumber & GetQueueMappingMicroservice & Function & \checkmark & HTTP request & N/A & HTTP response + queue mapping details\\

\rownumber & UpdateQueueMappingMicroservice & Function & \checkmark & HTTP request & Event name , service (consumer) ID & HTTP response\\

\rownumber & DeleteQueueMappingMicroservice & Function & \checkmark & HTTP request & N/A & HTTP response\\

\rownumber & ConsumeMessageAPI & Restful API & \checkmark & HTTP request & N/A & HTTP response\\
\hline

\end{tabular}
}
\end{table*}

The deployment strategy of the LEISA architecture on a cloud computing platform capitalises on the cloud's scalability, reliability, and distributed computing features. This strategic decision ensures the architecture can adapt to fluctuating loads and be dynamically scaled to meet evolving system demands.

In addition to cloud deployment, the LEISA architecture incorporates a specialised database serving two primary functions. Firstly, it facilitates the registration of producers and consumers, enabling dynamic service discovery and interaction. Secondly, it allows producers to designate message destinations, enhancing the system's message targeting and routing capabilities.

\Cref{tab:services functionalities} provides a detailed breakdown of the functionalities associated with each microservice. From creating and managing services to validating messages and handling queue mappings, each microservice contributes to the overall functionality and efficiency of the architecture.

\setcounter{rownumbers}{0}
\begin{table*}[!ht]
\centering
\caption{Overview of microservice functionalities in LEISA architecture}
\label{tab:services functionalities}
\resizebox{\textwidth}{!}{%
\begin{tabular}{llp{15cm}}
\hline
No. & Microservice Name & Functionality \\
\hline
\rownumber & ServiceRegistrationMicroservice & Responsible for creating new services within the LEISA architecture. It accepts HTTP requests and returns service details upon successful creation or error details if unsuccessful. \\
\rownumber & ServiceLoginMicroservice & Facilitates the authentication and login process for accessing services. It requires authentication via HTTP request and returns service details upon successful login. \\
\rownumber & DeleteServiceMicroservice & Handles the deletion of services. Authentication is required, and it responds to HTTP requests with success messages or error details. \\
\rownumber & FindServiceMicroservice & Enables the retrieval of service details based on a provided Service ID. Authentication is necessary, and it responds to HTTP requests with the requested service details. \\
\rownumber & ListAllServicesMicroservice & Lists all available services. Authentication is required, and it responds to HTTP requests with a list of service details. \\
\rownumber & ListServiceMicroservice & Lists specific services. Authentication is necessary, and it responds to HTTP requests with the details of the specified services. \\
\rownumber & MessageValidatorMicroservice & Validates LEI JSON messages received. It does not require authentication and responds to HTTP requests with validation results or error details if the message is invalid. \\
\rownumber & PublishMessageMicroservice & Publishes messages to the message broker after successful validation against a schema, receives event names and LEI JSON messages. Authentication is required, and it responds to HTTP requests confirming successful publication. \\
\rownumber & UpdateServiceMicroservice & Updates existing services based on provided Service ID. Authentication is necessary, and it responds to HTTP requests with update confirmation. \\
\rownumber & SetQueueMappingMicroservice & Establishes queue mappings for events and service consumers. Authentication is required, and it responds to HTTP requests confirming successful mapping. \\
\rownumber & GetQueueMappingMicroservice & Retrieves queue mapping details. Authentication is necessary, and it responds to HTTP requests with the details of queue mappings. \\
\rownumber & UpdateQueueMappingMicroservice & Updates queue mappings for events and service consumers. Authentication is required, and it responds to HTTP requests confirming successful update. \\
\rownumber & DeleteQueueMappingMicroservice & Deletes queue mappings. Authentication is necessary, and it responds to HTTP requests confirming successful deletion. \\
\rownumber & ConsumeMessageAPI & Facilitates the consumption of messages through a RESTful API. Authentication is required, and it responds to HTTP requests with the consumed message data. \\
\hline
\end{tabular}
}
\end{table*}

%%%%%%%%%%%%%%%%%%%%%%%%%%%%%%%%%%%%%%%%%%%%%%%%%%%%%%%%
%%%%%           Microservice in design             %%%%%
%%%%%%%%%%%%%%%%%%%%%%%%%%%%%%%%%%%%%%%%%%%%%%%%%%%%%%%%
\subsubsection{Microservice in design}
The development of microservices is supported by various methodologies that offer distinct advantages in system design, scalability, and maintenance. These methodologies include command-query-response segregation (CQRS), API-first design, event-driven architecture (EDA), model-driven design (MDD) and domain-driven design (DDD). Each methodology contributes to building robust, scalable, and maintainable microservice architectures. They emphasise the importance of aligning technical approaches with business needs and domain logic to create more effective and business-orientated solutions.

\textbf{CQRS} facilitates the development of scalable applications and has been shown to improve scalability through horizontal scaling, especially in cloud environments. It involves separating command operations (creating, updating, deleting data) from query operations (reading data). This approach allows for specialised optimisation of each type of operation and can be paired with event sourcing for effective communication and data management between services~\citep{lima2021improving}. 

\textbf{API-First Design} emphasises defining APIs at the beginning of the development process, ensuring that microservices have well-documented interfaces. This strategy creates loosely coupled systems that are easier to develop, test, and maintain. It fosters improved integration and collaboration across microservices~\citep{singjai2021practitioner}. 

\textbf{EDA} focuses on generating, detecting, and responding to system events. Encouraging asynchronous communication helps decouple services, improving flexibility and scalability. EDA has been shown to outperform synchronous communication models in terms of response time and error rates~\citep{shadija2017towards}. 

\textbf{MDD} leverages abstract models to guide the development process, potentially improving the conceptual clarity and maintainability of the system. This approach promotes efficient development by using model-driven methods and the Unified Modelling Language (UML) for design and integration~\citep{rademacher2019model}.

\textbf{DDD} focuses on structuring software design according to the domain it addresses. It has been increasingly applied in microservice architectures to improve the development and maintenance of complex systems. DDD helps identify the functional boundaries of microservices, promoting scalability, availability, reliability, and modifiability. Key concepts like bounded contexts, aggregates, and domain events help design microservices that are cohesive and deployable independently~\citep{singjai2022conformance}. 

\begin{figure}[!htb]
\centering
\includegraphics[width=0.8\textwidth]{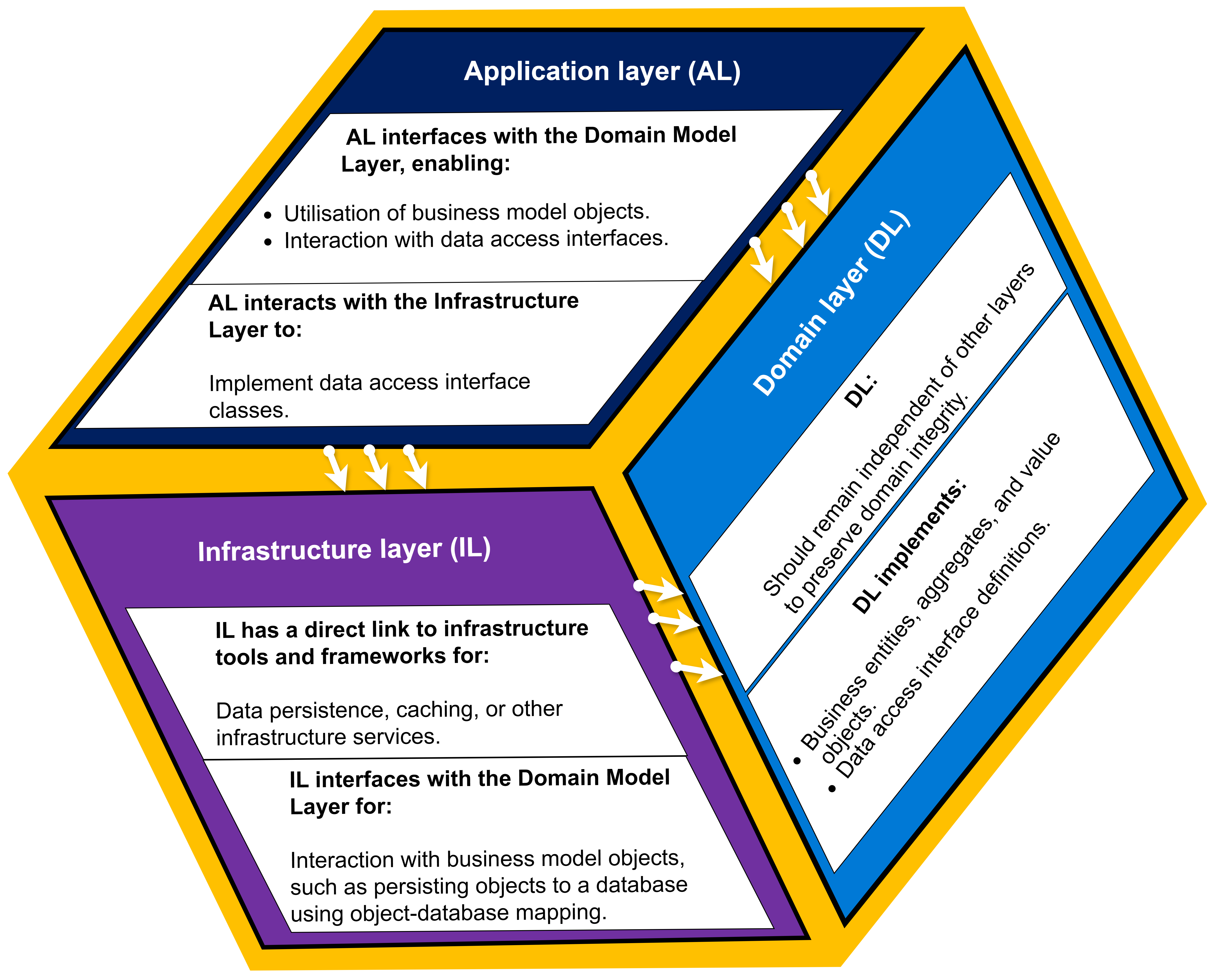}
\caption{Domain-Driven Design layers for microservice-based application (adopted from~\citep{torre2023designing})}
\label{fig:DDD}
\end{figure}

The design principles of LEISA microservices are based on DDD. In DDD, most applications with significant complexity are defined by multiple layers. These layers help manage the complexity in the code and are logical artefacts independent of the service deployment. The three layers commonly used in a DDD are the \textit{Application}, \textit{Domain}, and \textit{Infrastructure} layer. Each layer serves a specific purpose; the \textit{Application} layer handles external interactions, such as exposing APIs and handling user interface concerns. The \textit{Domain} layer contains the core business logic and entities, implementing the business rules and invariants. The \textit{Infrastructure} layer provides technical infrastructure components, such as data access and external service integrations. As shown in~\Cref{fig:DDD}, when designing a microservice using DDD, the Application layer relies on both the Domain and Infrastructure layers, while the Infrastructure layer relies on the Domain layer. However, the Domain layer is independent and does not rely on any layer. It is important to keep this layer design separate for each microservice.

\begin{figure}[!htb]
\centering
\includegraphics[width=1\textwidth]{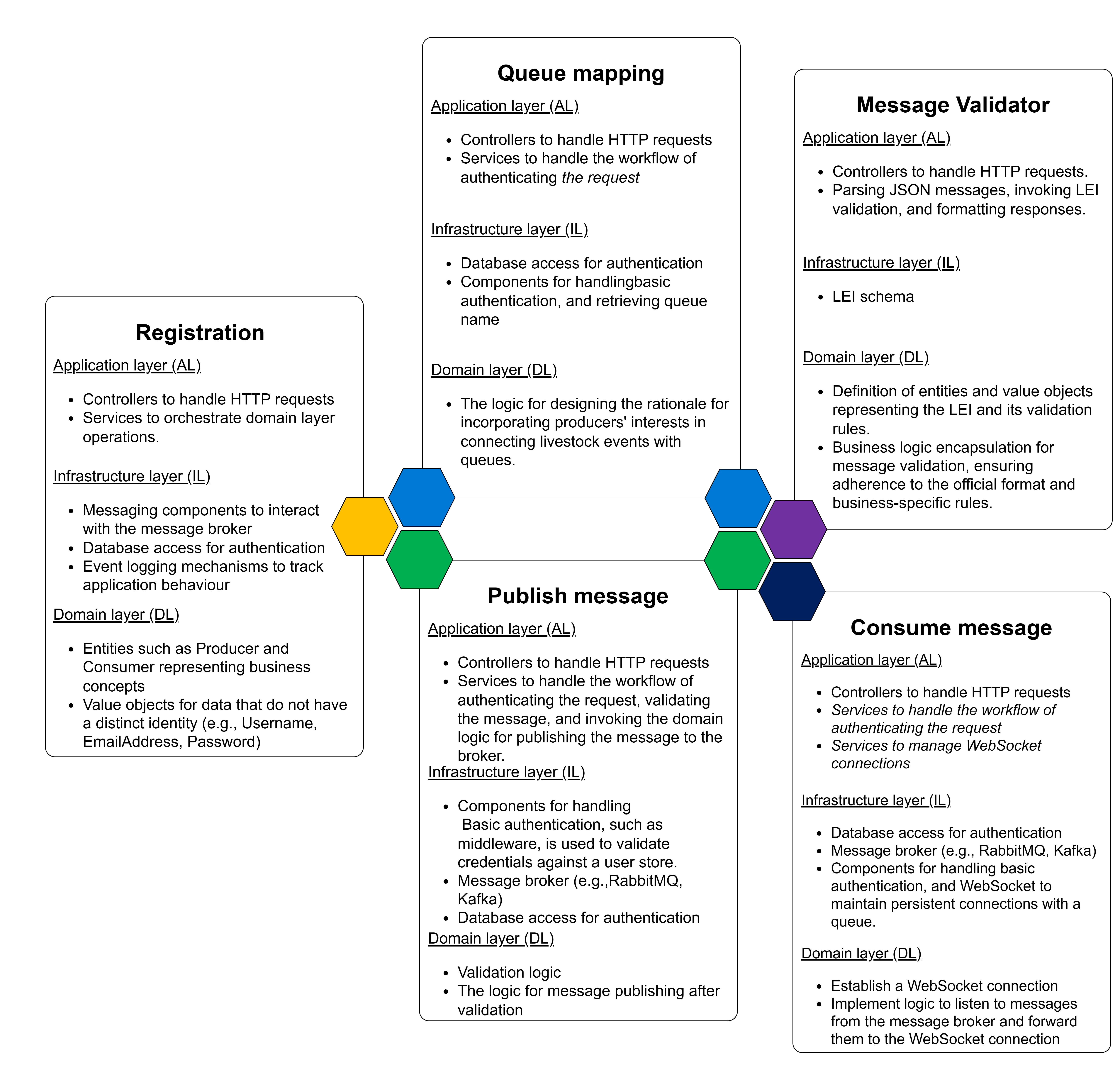}
\caption{DDD structure for five core microservices—\textit{MessageValidator}, \textit{ServiceRegistration}, \textit{SetQueueMapping}, \textit{PublishMessage}, and \textit{consumeMessage}—demonstrating the distinct responsibilities of each layer}
\label{fig:leisa ddd}
\end{figure}

~\Cref{fig:leisa ddd} delineates a layered architectural approach within a DDD for five distinct microservices: \textit{MessageValidator}, \textit{ServiceRegistration}, \textit{SetQueueMapping}, \textit{PublishMessage}, and \textit{consumeMessage}. Each microservice is structured into three strata: the Application Layer (AL), Domain Layer (DL), and Infrastructure Layer (IL), which are foundational to the DDD methodology.

In the \textit{Registration} microservice, the Application Layer is responsible for the interface handling user interactions, such as HTTP request processing and session state management. The Domain Layer encompasses the core business logic, including validation of registration data and enforcement of business invariants. The Infrastructure Layer provides foundational support, including database interactions for persistent storage and mechanisms for user authentication.

Regarding the \textit{Message Validator} microservice, the application layer receives and handles incoming data, such as messages that require validation. The communication of results back to the requester also occurs at this layer. The domain layer is where the core validation logic resides, encapsulating the rules and procedures necessary to determine the validity of messages according to the established schema. In the infrastructure layer, the mechanisms to carry out the validation are housed, including libraries that parse and check messages against the LEI schemas. Additionally, this layer may handle the log of validation actions.

The \textit{Queue Mapping} microservice is pivotal in LEISA, facilitating the dynamic routing and processing of messages, which, in turn, supports the system's scalability and adaptability. The application layer is tasked with providing a secure interface for client interactions, specifically to create and manage event-to-queue mappings. Handling client requests, authenticating users, orchestrating between client requests and domain operations, and delivering feedback regarding the status of the mapping operations are part of its responsibilities. Within the domain layer, the business logic is encapsulated to define the rules and constraints for how events correlate with specific queues. The infrastructure layer interacts with the database system to implement the queue mappings defined by the domain logic.

The \textit{Publish Message} microservice is similarly partitioned. The Application Layer manages the API endpoints for message publication and orchestrates the message validation and forwarding process. The Domain Layer validates the integrity and format of the messages against the business rules and LEI schema, acting as a gatekeeper to ensure only valid messages are published. The Infrastructure Layer interfaces with the message broker system to facilitate the actual message publication.

For the \textit{ConsumeMessage} microservice, the Application Layer oversees the message consumption endpoints and manages protocol transitions, such as upgrading HTTP connections to WebSocket channels for real-time communication. The Domain Layer processes consumed messages and determines how they should be handled or transformed. Finally, the Infrastructure Layer handles the low-level details of message consumption from the message broker and the distribution of messages to clients over WebSockets.

This DDD approach ensures a clean separation of concerns, each layer having a distinct set of responsibilities. This separation allows for easier maintenance and scalability of each microservice, as changes to one layer can often be made independently of the others. The microservices communicate with the message broker to publish and consume messages, facilitating a distributed system architecture that can handle complex workflows and data processing decoupled.

%%%%%%%%%%%%%%%%%%%%%%%%%%%%%%%%%%%%%%%%%%%%%%%%%%%%%%%%
%%%%%          Message broker in design            %%%%%
%%%%%%%%%%%%%%%%%%%%%%%%%%%%%%%%%%%%%%%%%%%%%%%%%%%%%%%%
\subsubsection{Message broker in design}
The LEISA design highlights the separation of producers and consumers by defining them as distinct microservices within their ecosystem. Encapsulating the message broker ensures that producers and consumers interact with it indirectly through task-specific microservices, facilitating scalability. Producers publish messages to specific queues based on predetermined criteria using a publish-subscribe pattern, ensuring efficient routing and delivery within the architecture.

\begin{figure*}[!htb]
\centering
\begin{subfigure}{\textwidth}
\centering
\includegraphics[width=0.9\textwidth,height=\textheight,keepaspectratio]{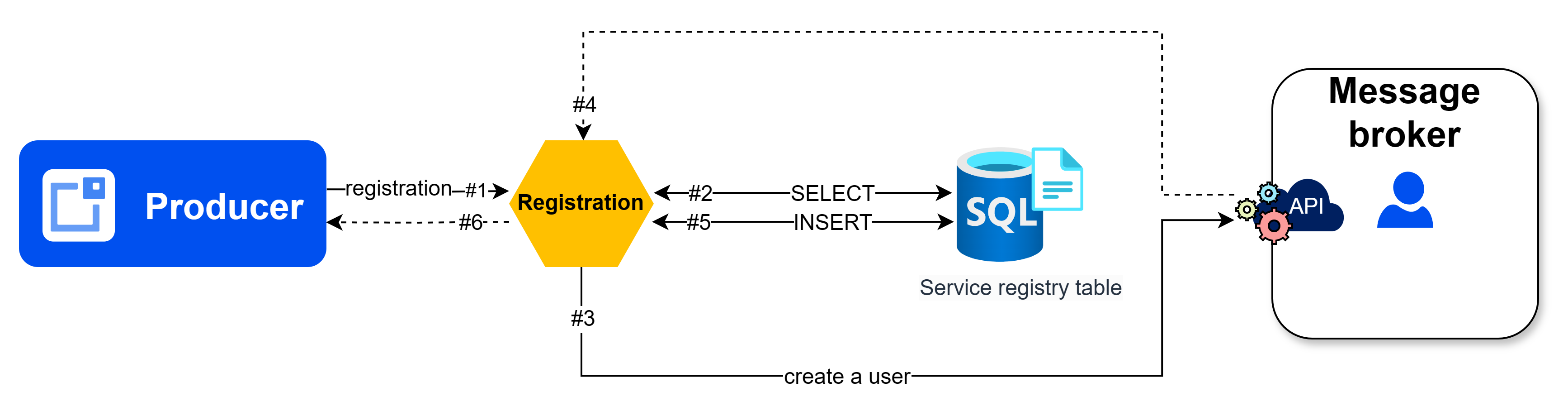}
\caption{Steps involved in registering a producer within the LEISA architecture}
\label{fig:producer registration}
\end{subfigure}

\begin{subfigure}{\textwidth}
\centering
\includegraphics[width=\textwidth,height=\textheight,keepaspectratio]{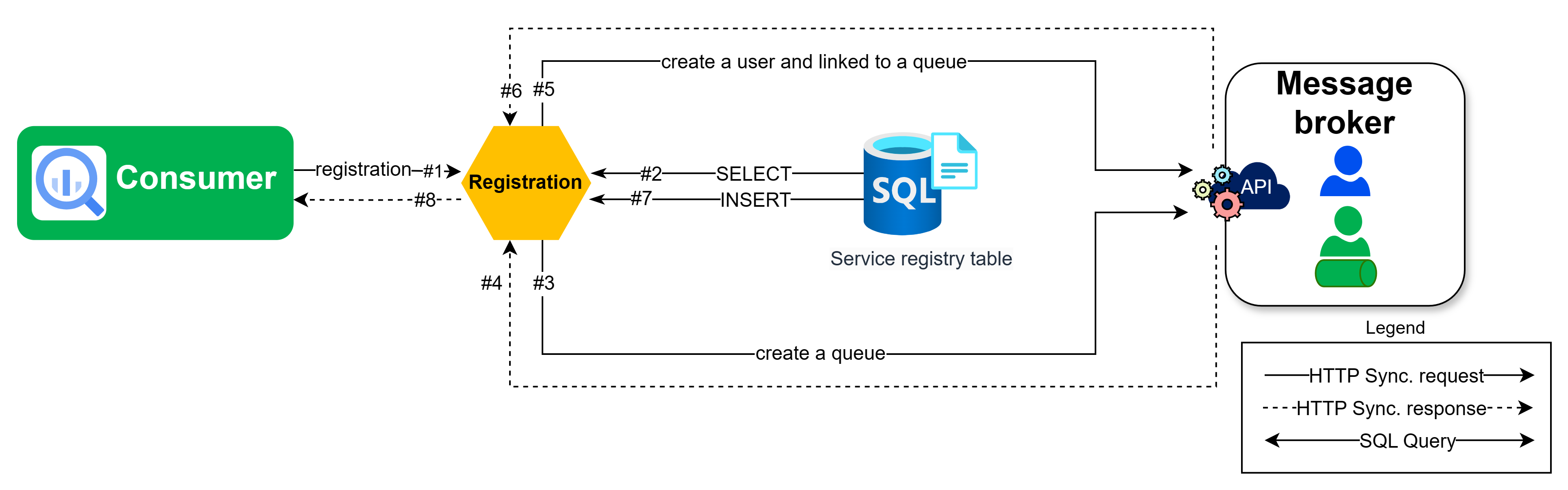}
\caption{Steps involved in registering a consumer within the LEISA architecture}
\label{fig:consumer registration}
\end{subfigure} 

\caption{Registration process for producers and consumers within the LEISA architecture, showing interaction with the \textit{ServiceRegistration} microservice to establish service credentials and queue associations}
\label{fig:Registrar of producer and consumer in LEISA}
\end{figure*}

To address the mechanism of message publication into queues within the architecture, a systematic approach is implemented during the registration of both producers and consumers, as depicted in~\Cref{fig:Registrar of producer and consumer in LEISA}. 

Initially, the producer/consumer initiates the registration process by sending a synchronous HTTP request to the \textit{ServiceRegistration} microservice. The microservice interacts with an SQL database to verify the presence of the producer or consumer.

For consumer registration, the \textit{ServiceRegistration} microservice sends a synchronous HTTP request to the Message Broker to set up a new message queue. This task is handled by the Message Broker's management plugin API, which configures the queues. After the queue is established, the Message Broker communicates back to the \textit{ServiceRegistration} microservice to confirm the setup. 

Next, the \textit{ServiceRegistration} microservice creates a user within the broker's system, granting them specific privileges. For consumers, this includes linking the newly created queue to their profile, while producers are not assigned any queue, allowing them to send messages to any queue without the ability to read.

An insert operation on the Service Registry table follows, registering the new service and, if relevant, the associated message queue. This step ensures effective management and coordination of users and their respective queues within the architecture.

The \textit{ServiceRegistration} microservice completes the registration process by sending a synchronous HTTP response to the producer/consumer, confirming successful registration and readiness of the message queue for use. In case of errors, the producer/consumer is notified of an unsuccessful registration.

\begin{figure*}[!htb]
\centering
\includegraphics[width=\textwidth]{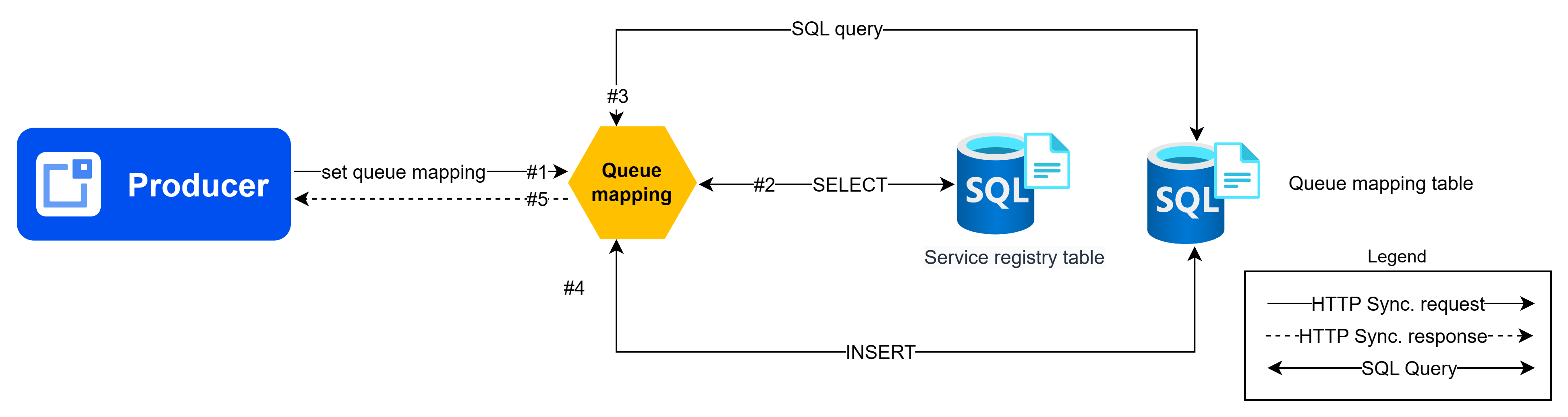}
\caption{Setup process for queue mapping within the LEISA architecture, allocating producers to designated consumer queues based on livestock events} 
\label{fig:Queue mapping sequence diagram}
\end{figure*}

LEISA incorporates a specialised database table known as the queue mapping table, which is crucial for managing the flow of messages between a producer and a consumer queue related to a specific livestock event. This table orchestrates message distribution throughout the architecture, as illustrated in~\Cref{fig:Queue mapping sequence diagram}. 

The process begins when the Producer sends a synchronous HTTP request to the Queue Mapping microservice, which performs an authentication query against the SQL database to verify credentials. Upon successful authentication, the Queue Mapping microservice extracts essential data from the request, including the producer's identity, a list of consumer queues, and the associated livestock event.

If new queue mapping is required, the Queue Mapping microservice inserts the details into the Queue Mapping table. The process concludes when the Queue Mapping microservice returns a synchronous HTTP response to the Producer, confirming the successful association with specific consumer queues and the categorisation of the livestock event.

\begin{figure*}[!htb]
\centering
\begin{subfigure}{\textwidth}
\centering
\includegraphics[width=\textwidth,height=\textheight,keepaspectratio]{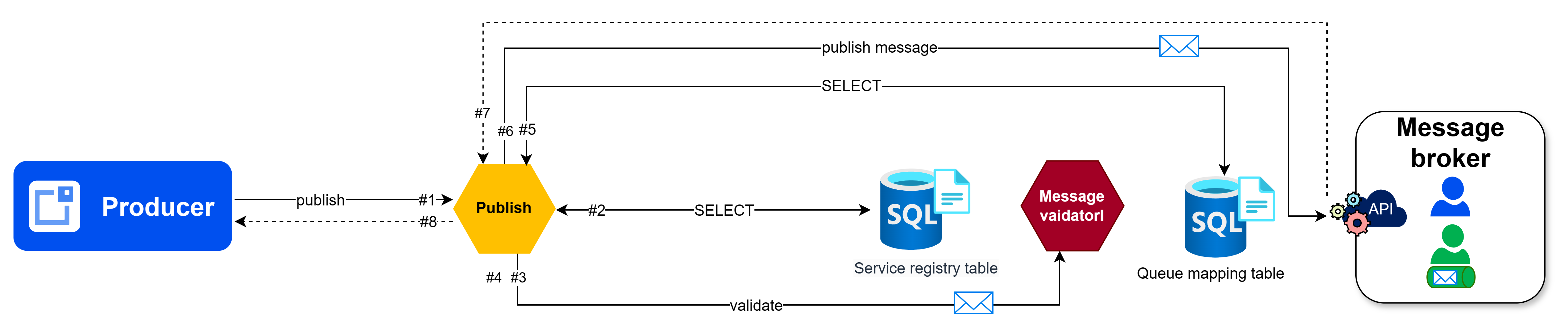}
\caption{Message publishing workflow in LEISA}
\label{fig:publish}
\end{subfigure}

\begin{subfigure}{\textwidth}
\centering
\includegraphics[width=0.8\textwidth,height=\textheight,keepaspectratio]{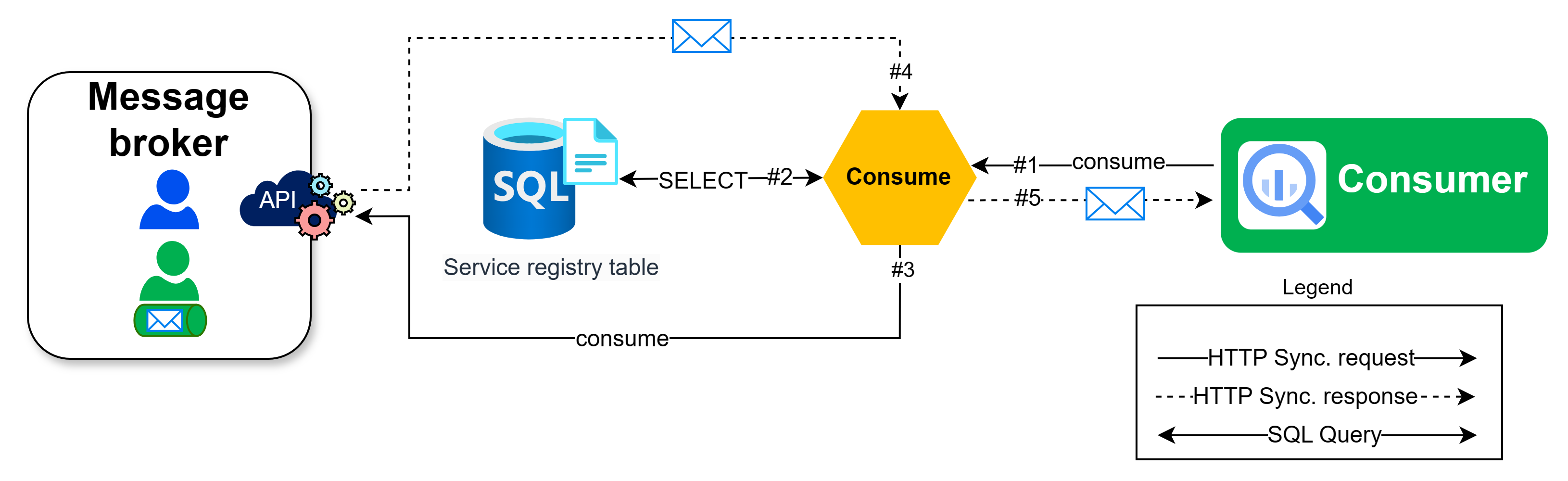}
\caption{Message consuming workflow in LEISA}
\label{fig:consume}
\end{subfigure} 

\caption{Message flow between producers and consumers}
\label{fig:Publish and consume message sequence diagram}
\end{figure*}

\Cref{fig:Publish and consume message sequence diagram} delineates the process flow for message publishing and consuming.~\Cref{fig:publish} illustrates the systematic procedure for message publication within the architecture. 

For message publication, the Producer sends a synchronous HTTP request to the Publish microservice, including the message, livestock event type, and credentials. Upon receipt, the Publish microservice conducts an authentication check and validates the message against a predefined schema using the \textit{MessageValidator} microservice. Once validated, the Publish microservice identifies the appropriate consumer queue and publishes the message using the Message Broker's management plugin API. The Producer receives an HTTP response acknowledging the successful publication.

For message consumption, the Consumer sends a synchronous HTTP request to the Consume microservice, which verifies credentials and fetches messages from the specified queue using the Message Broker. The Consume microservice then forwards the messages to the Consumer.

This detailed design ensures efficient, scalable, and secure communication between producers and consumers within the LEISA architecture.

%%%%%%%%%%%%%%%%%%%%%%%%%%%%%%%%%%%%%%%%%%%%%%%%%%%%%%%%
%%%%%                Implementation                %%%%%
%%%%%%%%%%%%%%%%%%%%%%%%%%%%%%%%%%%%%%%%%%%%%%%%%%%%%%%%
\section{Implementation}\label{sec:implementation}

In this section, we delve into the practical aspects of our architecture, breaking down the implementation into discrete subsections. We explore the cloud infrastructure used then examine the selection and configuration of messaging middleware. Next, we discuss the database setup and finally address the deployment of the five core microservices within the framework of serverless computing.

\subsection{Azure cloud utilisation}
Our architecture was implemented in Microsoft's Azure cloud computing service, using the Azure for Students offering, which provides access to Azure's services without cost, subject to specific limits.\footnote{\url{https://azure.microsoft.com/en-us/free/students/}} This allowed us to construct and deploy our architecture without a financial outlay. We follow the guidelines of the Azure Well-Architected Framework to ensure the durability and performance of the architecture, which offer recommendations for building secure, efficient, high performance, and resilient architectures.\footnote{\url{https://docs.microsoft.com/en-us/azure/architecture/framework/}}

\begin{figure*}[!htb]
\centering
\includegraphics[width=\textwidth,height=0.85\textheight,keepaspectratio]{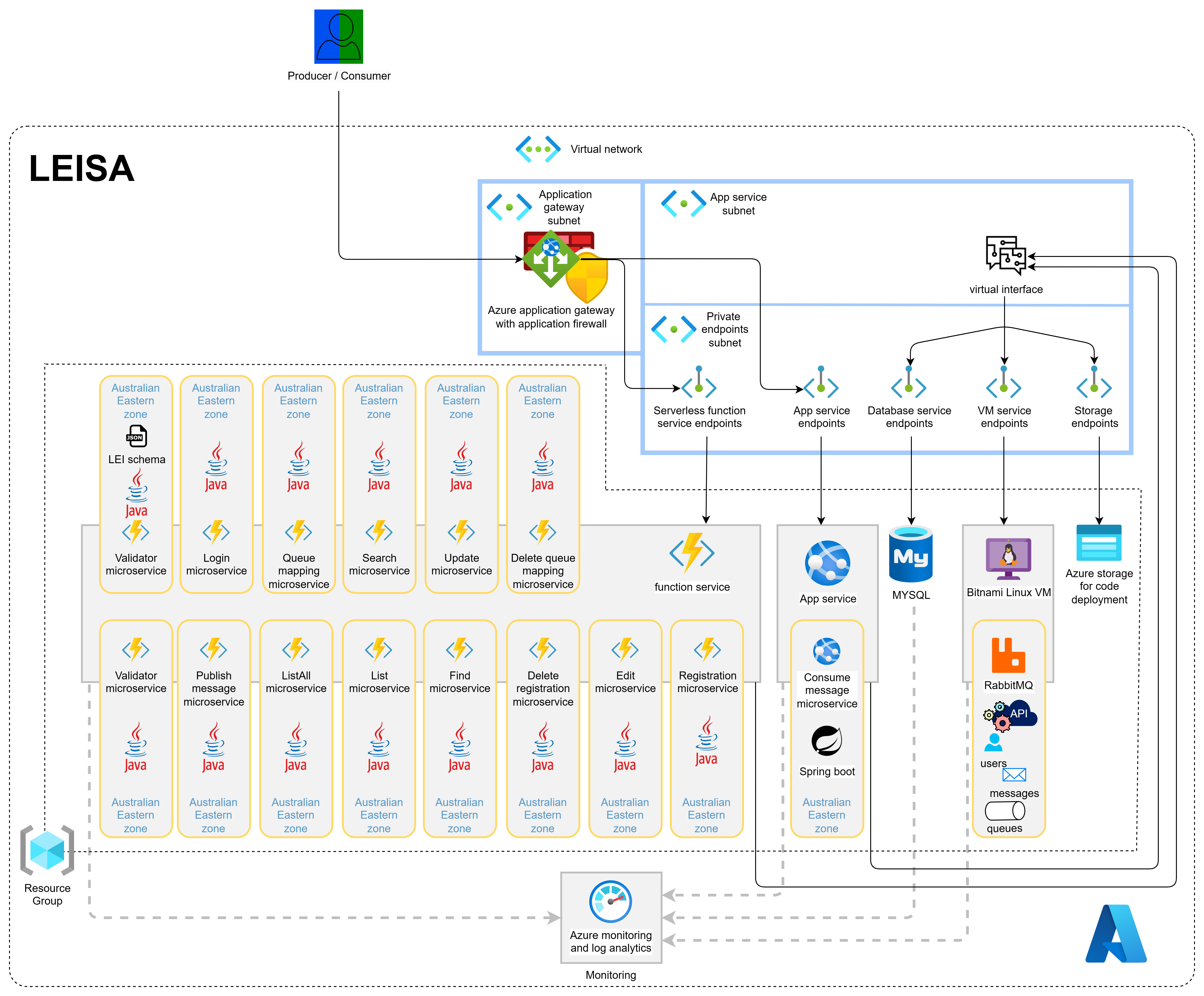}
\caption{LEISA implementation in Azure cloud}
\label{fig:cloud structure}
\end{figure*}

\Cref{fig:cloud structure} represents a sophisticated diagram detailing a LEISA deployment in Microsoft Azure, starting with a Resource Group, which serves as a container for logically organising the resources required for cloud solutions. The architecture spans the Eastern Australian Zone, suggesting a regional focus for deployment to ensure compliance with data residency and latency optimisation.

At the core of the architecture are a series of Java-based microservices, each responsible for a distinct aspect of the application's functionality. These include a message validator microservice and various operational services such as login, queue mapping, searching, updating, and deleting functions, indicative of a modular and scalable system.

The Azure Application Gateway, with an integrated application firewall, serves as a critical entry point for web traffic. It uses load balancing and traffic management while offloading encryption tasks to enhance performance. The firewall component offers protection against web vulnerabilities, underscoring the importance of security in this implementation.

Adjacent to this gateway are subnets that separate cloud services, including serverless function endpoints, app service endpoints, database service endpoints, virtual machine service endpoints, and storage endpoints. These endpoints allow secure communication and the operation of services within a virtual network, ensuring that all communication occurs within a secure and isolated environment.

Data storage and management components, such as MySQL for relational databases and a Bitnami Linux Virtual Machine for RabbitMQ, are depicted on the right side of the figure. RabbitMQ, featured alongside its API, highlights the asynchronous data sharing mechanism between producers and consumers.

Azure monitoring and log analytics emphasise the oversight and analysis of the system's health and performance, ensuring that operational metrics are tracked and the system's behaviour is within expected parameters.

\subsection{Messaging Middleware Configuration}
We use asynchronous messaging, allowing the producer to share data with the consumer without requiring simultaneous availability. RabbitMQ was chosen for its maturity and feature-rich capabilities, which support multiple protocols such as AMQP, MQTT, and STOMP. It handles high throughput and is commonly used for tasks such as handling background jobs, long-running tasks, or communication between microservices to avoid bottlenecks~\citep{ionescu2015analysis}.

To use RabbitMQ within our architecture, we configured it to optimise message delivery, durability, and system resilience. Durable queues ensure message persistence during network failures or service restarts, enhancing reliability. Messages are configured as persistent to prevent loss even if the broker restarts.

Since Microsoft Azure does not offer RabbitMQ as a standalone managed service, we utilised a Bitnami package for RabbitMQ on Azure. This package simplifies deployment, providing a pre-configured environment that reduces setup time and technical overhead, allowing developers to focus more on application development.

\subsection{Database Management}
Azure services, including Azure Database for MySQL, were utilised for database management. Azure Database for MySQL is a managed database service that offers scalability, flexibility, and cost-effectiveness while ensuring security and performance. It automates many administrative tasks, allowing developers to focus on application development and optimisation.

\begin{figure}[!htb]
\centering
\includegraphics[width=0.5\textwidth,keepaspectratio]{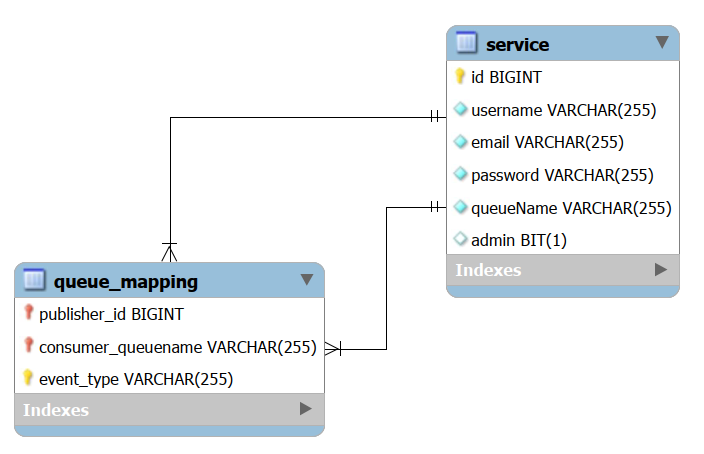}
\caption{MySQL Database Schema for LEISA}
\label{fig:mysql}
\end{figure}

The MySQL database schema for LEISA, depicted in~\Cref{fig:mysql}, enables efficient information sharing between producers and consumers by associating producer services with consumer services. The schema includes two tables: \textit{service} and \textit{queue\_mapping}, each optimising data flow and service coordination.

The \textit{service} table stores credentials and roles for external services, including fields for unique identifiers, user authentication details, a designated queue name for message routing, and administrative privileges. This table is crucial for managing external service interactions within the LEISA framework.

The \textit{queue\_mapping} table links services to appropriate consumer queues based on event types, categorising and directing the flow of messages effectively and efficiently.

\subsection{Microservices Deployment}
We utilised Azure Functions to create a serverless computing environment, eliminating the need for infrastructure management when executing code in response to triggers. This promotes optimal resource utilisation, as functions are only triggered by specific events such as HTTP requests.

Azure Web Apps provided a scalable managed platform for deploying our APIs. This service offers minimal infrastructure management and supports various programming languages, frameworks, and technologies. It enables continuous integration and delivery pipelines, supporting a DevOps approach to software development.

To organise and manage the Azure services employed in the LEISA architecture, we utilised Azure Resource Groups. This strategy facilitated the streamlined management, deployment and monitoring of resources that are part of the same application lifecycle.

\Cref{tab:servicesimpelemnation} encapsulates a comprehensive delineation of various microservices orchestrated within LEISA, deployed on the Microsoft Azure cloud platform. The table lists microservices, their operational status, deployment region, pricing tier, service type, and GitHub links.

\setcounter{rownumbers}{0}
\begin{table*}[!htbp]
\centering
\caption{Microservice Information}
\label{tab:servicesimpelemnation}
\resizebox{\textwidth}{!}{%
\begin{tabular}{llllllll}
\hline
No. & Microservice Name & Status & Region & Pricing Tier & Service Type & GitHub Link \\
\hline
\rownumber & ServiceRegistrationMicroservice & Running & Australia East & Dynamic & Function App & \url{https://github.com/mahirgamal/ServiceRegistrationMicroservice} \\

\rownumber & DeleteServiceMicroservice & Running & Australia East & Dynamic & Function App & \url{https://github.com/mahirgamal/ServiceDeleteMicroservice} \\

\rownumber & FindServiceMicroservice & Running & Australia East & Dynamic & Function App & \url{https://github.com/mahirgamal/ServiceSearchMicroservice} \\

\rownumber & ListAllServicesMicroservice & Running & Australia East & Dynamic & Function App & \url{https://github.com/mahirgamal/ServiceListAllMicroservice} \\

\rownumber & ListServiceMicroservice & Running & Australia East & Dynamic & Function App & \url{https://github.com/mahirgamal/ServiceListMicroservice} \\

\rownumber & MessageValidatorMicroservice & Running & Australia East & Dynamic & Function App & \url{https://github.com/mahirgamal/MessageValidatorMicroservice} \\

\rownumber & PublishMessageMicroservice & Running & Australia East & Dynamic & Function App & \url{https://github.com/mahirgamal/PublishMessageMicroservice} \\

\rownumber & ServiceLoginMicroservice & Running & Australia East & Dynamic & Function App & \url{https://github.com/mahirgamal/ServiceLoginMicroservice} \\

\rownumber & UpdateServiceMicroservice & Running & Australia East & Dynamic & Function App & \url{https://github.com/mahirgamal/ServiceUpdateMicroservice} \\

\rownumber & SetQueueMappingMicroservice & Running & Australia East & Dynamic & Function App & \url{https://github.com/mahirgamal/SetQueueMappingMicroservice} \\

\rownumber & GetQueueMappingMicroservice & Running & Australia East & Dynamic & Function App & \url{https://github.com/mahirgamal/GetQueueMappingMicroservice} \\

\rownumber & UpdateQueueMappingMicroservice & Running & Australia East & Dynamic & Function App & \url{https://github.com/mahirgamal/UpdateQueueMappingMicroservice} \\

\rownumber & DeleteQueueMappingMicroservice & Running & Australia East & Dynamic & Function App & \url{https://github.com/mahirgamal/DeleteQueueMappingMicroservice} \\

\rownumber & ConsumeMessageAPI & Running & Australia East & Free & Web App & \url{https://github.com/mahirgamal/consume_message_API} \\
\hline

\end{tabular}
}
\end{table*}

The designation ``Microservice Name" in the table corresponds to individual microservices, each tailored to execute a distinct function within the broader architecture. These microservices are designed to be small, autonomously deployable units that collaboratively contribute to the functionality of the entire application, catering to diverse operational needs such as service creation, deletion, discovery, listing, and message validation.

The ``Status" column reflects the operational state of each microservice, with ``Running" indicating that the service is active, operational, and poised to receive and process requests. This operational readiness ensures uninterrupted service delivery and responsiveness to user requests.

Spatial deployment considerations are addressed under the ``Region" column, specifying the geographical location of the Azure data centre hosting each microservice. Optimal placement of services in proximity to the target user base is crucial for minimising latency and enhancing the overall performance of the application. The uniform deployment in the ``Australia East" region suggests a strategic focus on optimising service delivery for users in this geographical area.

``Pricing Tier" delineates the cost structure associated with hosting each microservice on Azure, reflecting a balance between cost, performance, and available features. The prevalence of the ``Dynamic" tier for Function Apps highlights a consumption-based pricing model, wherein costs are predicated on the actual usage and resource consumption of the functions. Conversely, the ``Free" tier for the Web App indicates a no-cost hosting solution, albeit with certain limitations on resources and functionalities.

The ``Service Type" categories the hosting service utilised for each microservice, differentiating between ``Function App" and ``Web App". ``Function App", pertaining to Azure Functions, represents a serverless computing service that enables the execution of code in response to events, obviating the need for explicit infrastructure management. ``Web App", indicative of Azure Web Apps, exemplifies a platform as a service (PaaS) offering that facilitates the hosting and management of web applications, abstracting away the underlying infrastructure concerns. Finally, the ``GitHub Link" provides a conduit to the GitHub repositories where the source code for each microservice is maintained.

\begin{figure*}[!htbp]
\centering
\begin{subfigure}{0.32\textwidth}
\centering
\includegraphics[width=\textwidth,height=\textheight,keepaspectratio]{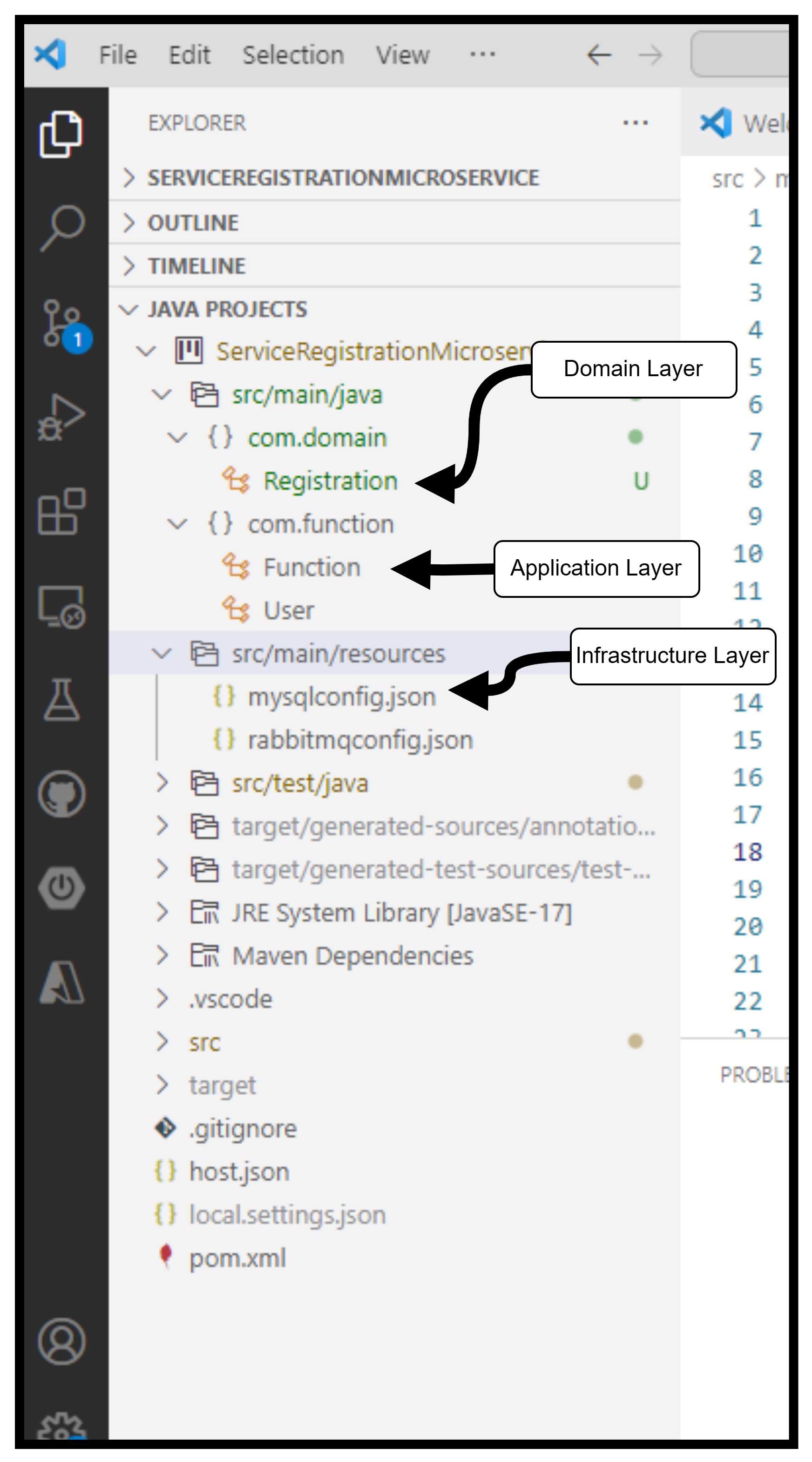}
\caption{\textit{ServiceRegistration} microservice}
\label{fig:registration ms}
\end{subfigure}
\begin{subfigure}{0.32\textwidth}
\centering
\includegraphics[width=\textwidth,height=\textheight,keepaspectratio]{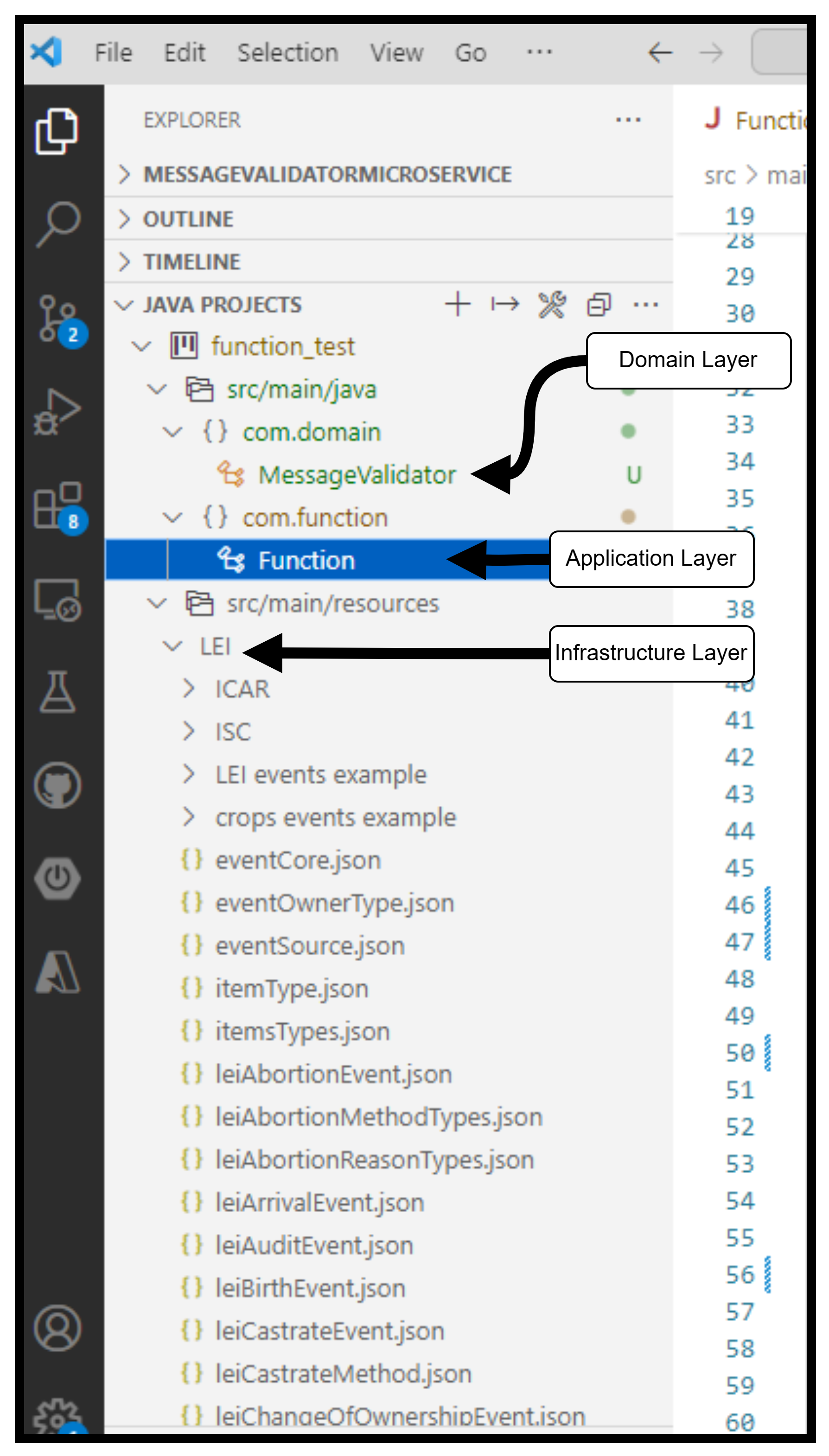}
\caption{\textit{MessageValidator} microservice}
\label{fig:message validator ms}
\end{subfigure}
\begin{subfigure}{0.32\textwidth}
\centering
\includegraphics[width=\textwidth,height=\textheight,keepaspectratio]{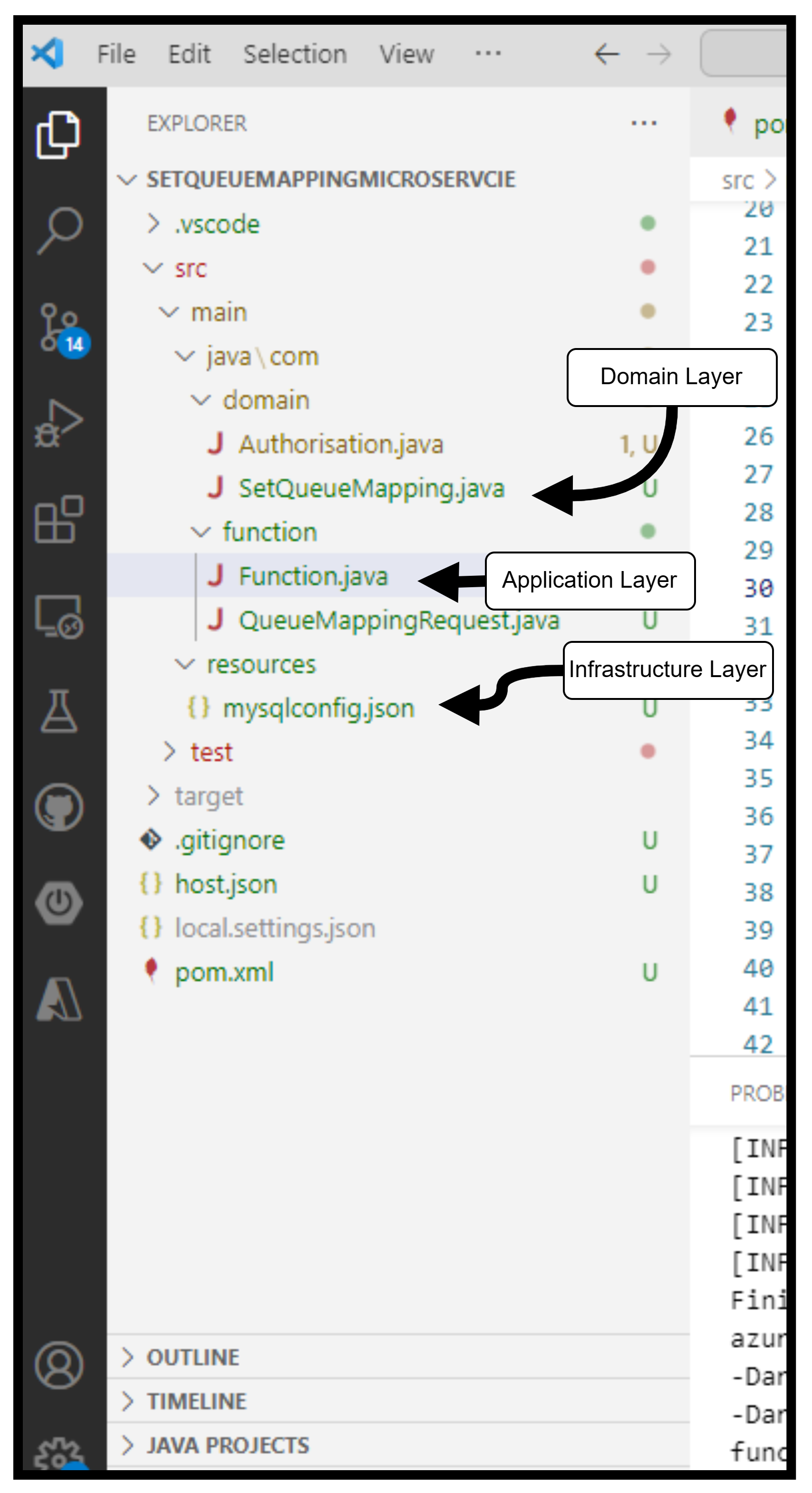}
\caption{\textit{SetQueueMapping} microservice}
\label{fig:set queue mapping ms.png}
\end{subfigure}

\begin{subfigure}{0.32\textwidth}
\centering
\includegraphics[width=\textwidth,height=\textheight,keepaspectratio]{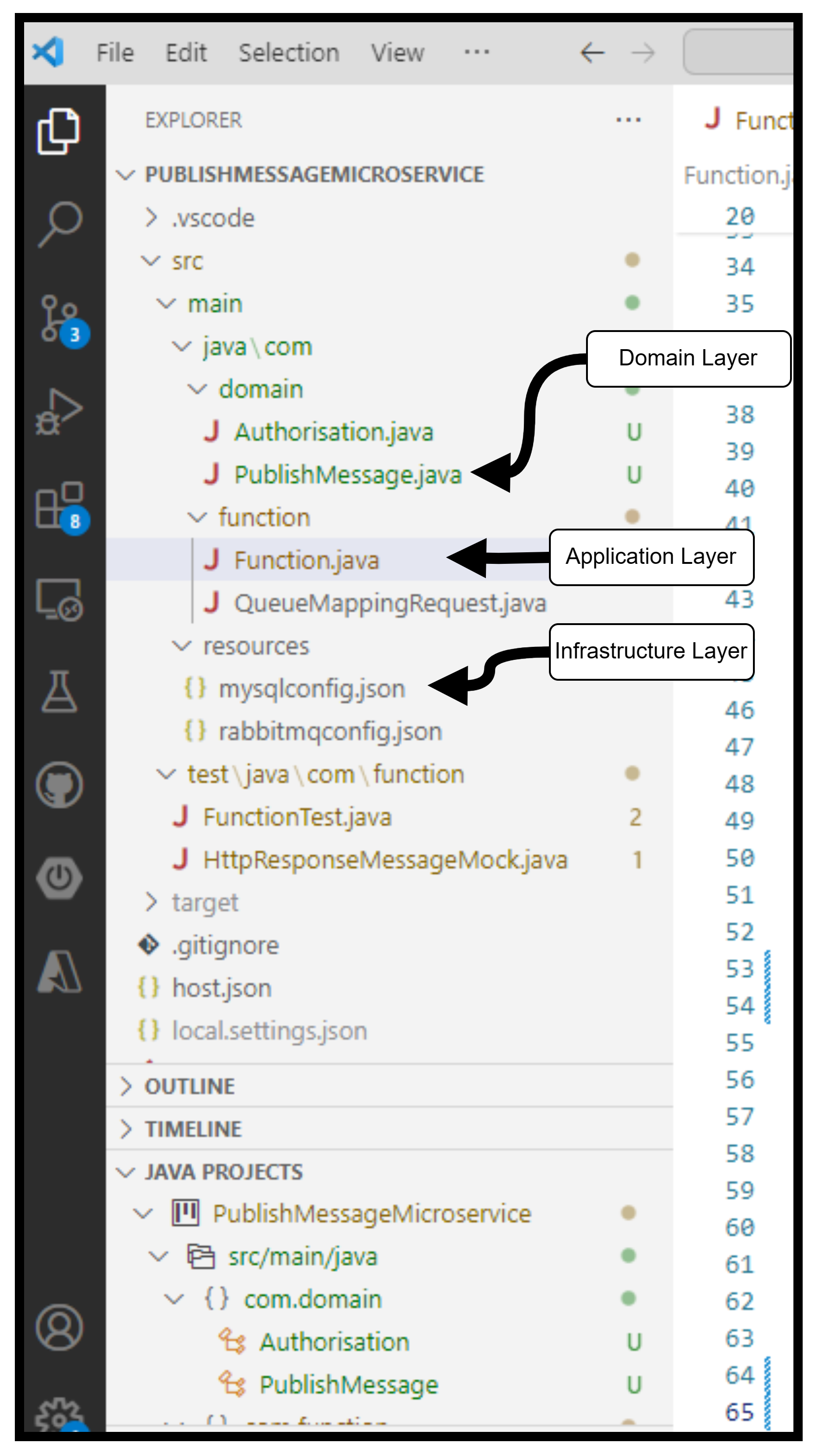}
\caption{\textit{PublishMessage} microservice}
\label{fig:publish message ms}
\end{subfigure}
\begin{subfigure}{0.32\textwidth}
\centering
\includegraphics[width=\textwidth,height=\textheight,keepaspectratio]{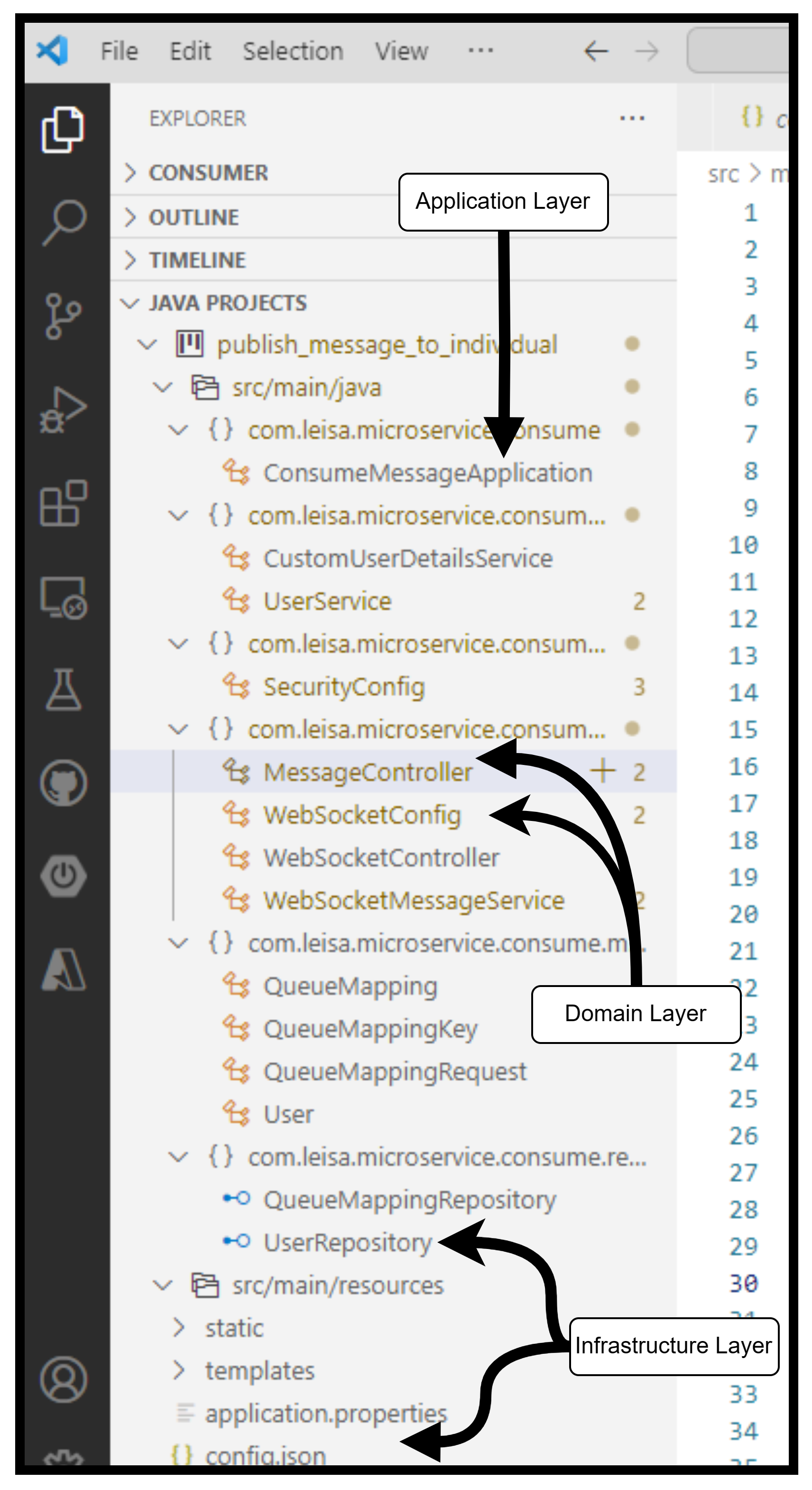}
\caption{\textit{ConsumeMessage} microservice}
\label{fig:consume message ms}
\end{subfigure}
\caption{Microservices DDD layers}
\label{fig:Microservices DDD layers}
\end{figure*}

\Cref{fig:Microservices DDD layers} comprises several subfigures, each depicting the structure and hierarchy of the five core microservices.

\Cref{fig:registration ms} showcases the \textit{ServiceRegistration} microservice, structured into various layers, including controllers, services, and repositories, indicating a clean separation of concerns that is essential in DDD. Similarly,~\Cref{fig:message validator ms} presents the \textit{MessageValidator} microservice, which includes components to validate the format and content of incoming messages, reflecting the autonomous nature of the microservice architecture by isolating the validation logic into a separate service. \Cref{fig:set queue mapping ms.png} delineates the \textit{SetQueueMapping} microservice, which maps messages to the appropriate queues and ensures they are directed to the correct consumers. The \textit{PublishMessage} microservice, as seen in~\Cref{fig:publish message ms}, manages the publication of messages on RabbitMQ, ensuring that producer services send messages to be consumed asynchronously by other parts of the system. Lastly,~\Cref{fig:consume message ms} illustrates the \textit{ConsumeMessage} microservice, which retrieves and processes messages from the queues, embodying the consumer aspect of the messaging pattern. Each microservice maintains its domain logic, adhering to the DDD principles, where each bounded context (each microservice) is a self-contained functional boundary.

%%%%%%%%%%%%%%%%%%%%%%%%%%%%%%%%%%%%%%%%%%%%%%%%%%%%%%%%
%%%%%                    LEI2JSON                  %%%%%
%%%%%%%%%%%%%%%%%%%%%%%%%%%%%%%%%%%%%%%%%%%%%%%%%%%%%%%%
\subsection{LEI2JSON}

LEI2JSON is designed to allow producers to convert their data from CSV format into JSON. This conversion is particularly valuable for those utilising Google Sheets, facilitating an efficient transition to JSON, a format more suitable for data sharing and processing~\citep{habib2023lei2json}. Using Google Sheets as an interface simplifies the adoption process for users, making it easier for producers to input and manage their data. This approach, demonstrated in the LEI2JSON application, offers a user-friendly way to enhance data sharing and interoperability within the livestock sector.

The latest extension, LEI2JSON version 2.0,\footnote{\url{https://github.com/mahirgamal/LEI2JSON}} marks a significant improvement in streamlining data sharing within the LEISA architecture. This update introduces advanced features and functionalities, including a new submenu labelled ``LEISA" under the main ``LEI2JSON" menu. This facilitates LEISA operations such as registration/login and comprehensive queue mapping operations, including setting, deleting, and updating queues. Additionally, the ``Publish to LEISA" button within the ``Validation Result" tab simplifies the process of distributing generated JSON data directly to the desired consumers' queue with just a single click.

\begin{figure*}[!htbp]
\centering
\includegraphics[width=\textwidth,height=\textheight,keepaspectratio]{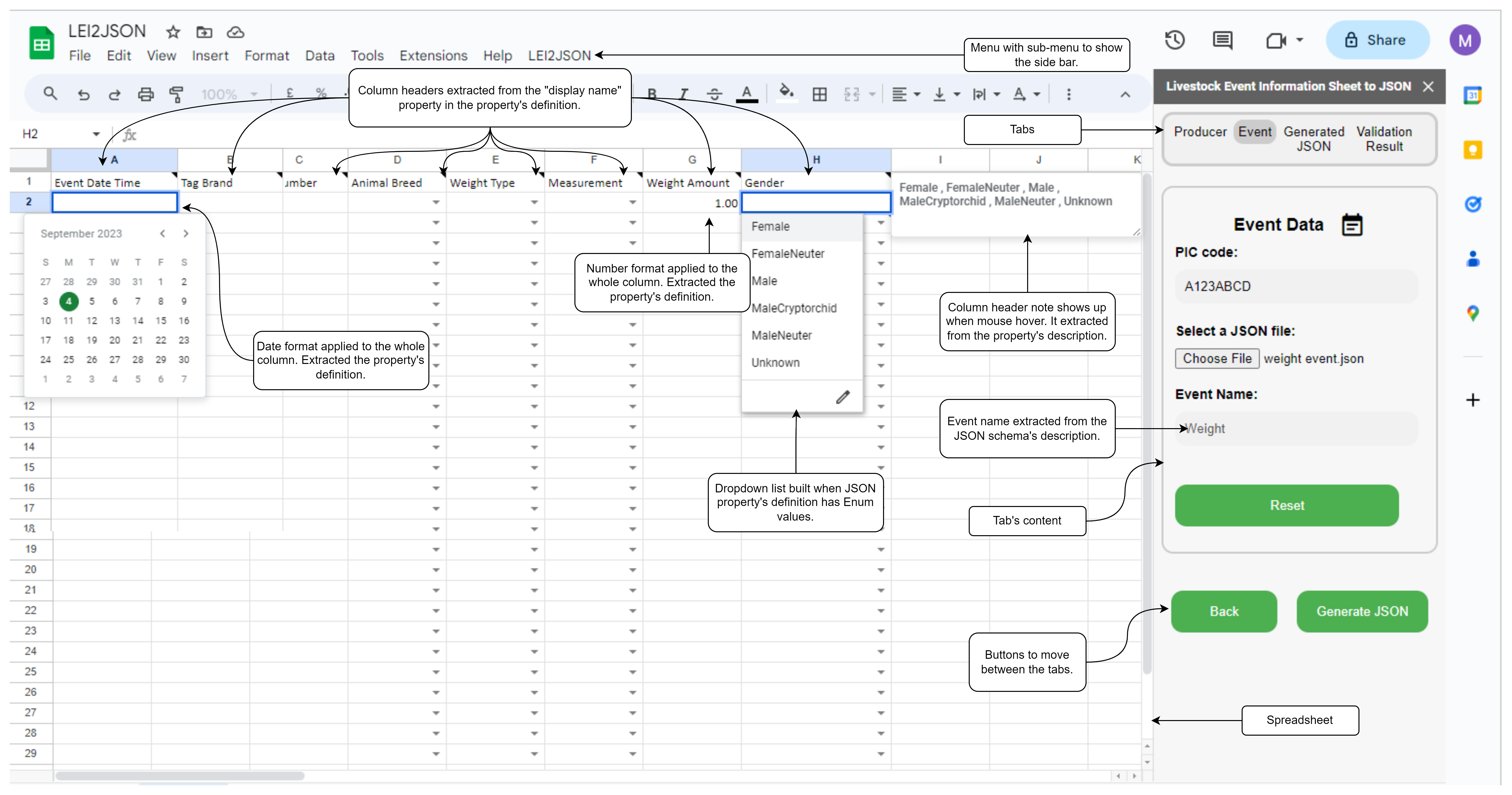}
\caption{Spreadsheet template featuring column headers, notes, validation rules, and formatting derived from JSON schema}
\label{fig:LEI2JSON}
\end{figure*}

Building upon its predecessor, LEI2JSON extends its functionalities by incorporating data sharing capabilities. This enhancement marks a significant step in developing tools designed to support efficient data management and sharing within the agricultural sector. By facilitating a more streamlined process for data standardisation and sharing, LEI2JSON plays a crucial role in advancing the adoption of standardised data practices, fostering improved data management, sharing, and ultimately more informed decision-making within the livestock sector.

%%%%%%%%%%%%%%%%%%%%%%%%%%%%%%%%%%%%%%%%%%%%%%%%%%%%%%%%
%%%%%           Applications of LEISA              %%%%%
%%%%%%%%%%%%%%%%%%%%%%%%%%%%%%%%%%%%%%%%%%%%%%%%%%%%%%%%
\section{Applications of LEISA}\label{sec:leisa applications}

In this section, we explore the transformative impacts of LEISA through detailed use-case scenarios. LEISA's structured schema significantly enhances data sharing across different software systems and platforms, ensuring consistent data formatting that is crucial for accurate interpretation and analysis.

We present three use-case scenarios to illustrate the broader implications of adopting the LEISA architecture. These scenarios exemplify how effectively managed and utilised data can serve as a foundation for developing solutions that tackle complex challenges in livestock management. Such innovations underscore the potential of digital technologies and the LEISA architecture to drive sustainability and advancement in agriculture, contributing significantly to the sector's overall resilience and progress.

%%%%%%%%%%%%%%%%%%%%%%%%%%%%%%%%%%%%%%%%%%%%%%%%%%%%%%%%
%%%%%            cattle Location Monitor           %%%%%
%%%%%%%%%%%%%%%%%%%%%%%%%%%%%%%%%%%%%%%%%%%%%%%%%%%%%%%%
\subsection{Cattle Location Monitor}

The practical application of LEISA in real-world scenarios is exemplified by the development of consumer applications that serve as a bridge between data collection and informed decision-making, beneficial both to livestock and their caregivers. The \textit{Cattle Location Monitor}\footnote{\url{https://github.com/mahirgamal/Cattle-Location-Monitor.git}} is a pioneering tool designed to visualise geographical locations related to cattle within a paddock, enabled by the shared \textit{LEI Location Event}.\footnote{\url{https://github.com/mahirgamal/LEI-schema/blob/main/leiLocationEvent.json}} Developed in Python, the \textit{Cattle Location Monitor} uses data visualisation techniques to precisely map cattle positions, offering valuable information about their locations.

\begin{figure}[!htb]
\centering
\begin{subfigure}{0.47\textwidth}
\centering
\includegraphics[width=\textwidth,height=\textheight,keepaspectratio]{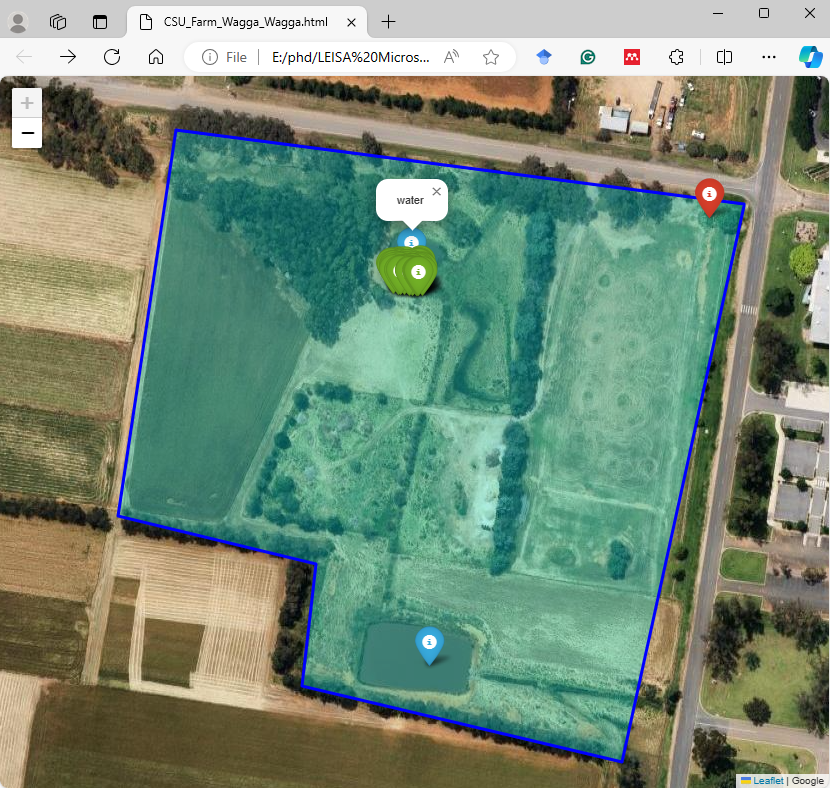}
\caption{Early morning gathering: cattle near water sources}
\label{fig:location morning}
\end{subfigure}
\hfill
\begin{subfigure}{0.47\textwidth}
\centering
\includegraphics[width=\textwidth,height=\textheight,keepaspectratio]{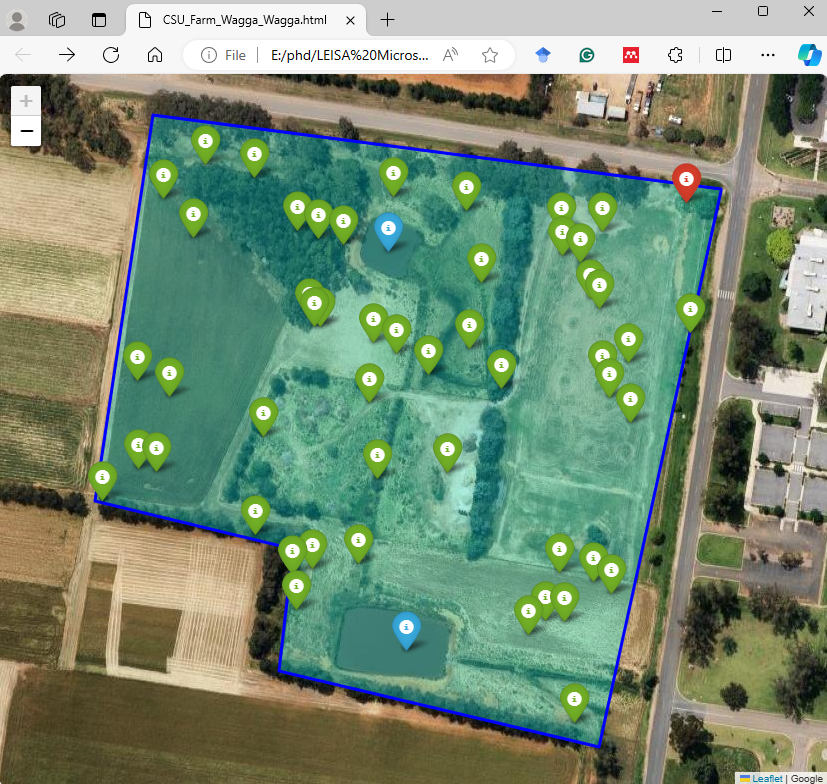}
\caption{Midday distribution: cattle grazing across the farm}
\label{fig:location mid day}
\end{subfigure}
\caption{Dynamic monitoring of cattle locations throughout the day, with markers indicating specific points: red markers denote administrative zones and operational centres of the farm, blue markers highlight water sources, and green markers represent cattle locations. }
\label{fig:location monitoring}
\end{figure}

\Cref{fig:location monitoring} illustrates the application's ability to display maps, locate cattle, and identify areas of interest, such as water sources, within specific boundaries on a farm. During the cooler hours of the morning, it is common for cattle to gather near water sources to hydrate after a night of sleep, as shown in~\Cref{fig:location morning}. In contrast,~\Cref{fig:location mid day} showcases the distribution of cattle throughout the farm during midday, highlighting their grazing habits.

By integrating real-time data visualisation with sophisticated tracking algorithms, the \textit{Cattle Location Monitor} not only improves the efficiency of farm operations but also contributes significantly to the well-being of cattle, which proves to be an indispensable tool in modern sustainable agriculture.

%%%%%%%%%%%%%%%%%%%%%%%%%%%%%%%%%%%%%%%%%%%%%%%%%%%%%%%%
%%%%%          AgriVet Treatment Grapher           %%%%%
%%%%%%%%%%%%%%%%%%%%%%%%%%%%%%%%%%%%%%%%%%%%%%%%%%%%%%%%
\subsection{AgriVet Treatment Grapher}
\begin{figure*}[!htb]
\centering
\includegraphics[width=\textwidth,height=\textheight,keepaspectratio]{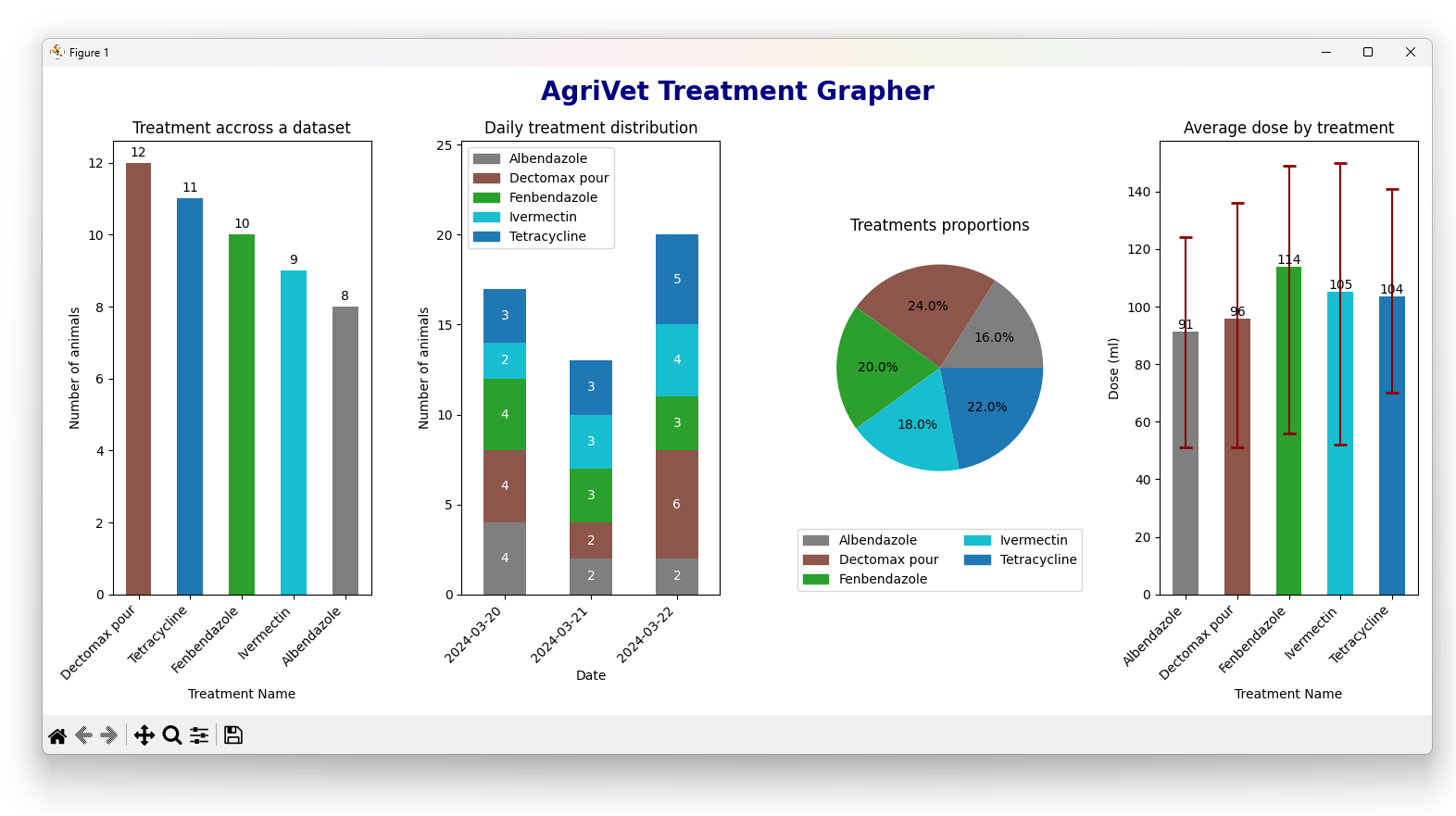}
\caption{Visualisations generated by the AgriVet Treatment Grapher to aid in treatment analysis and decision-making.}
\label{fig:AgriVet Treatment Grapher}
\end{figure*}

Another significant Python application of LEISA in real-world scenarios, as depicted in~\Cref{fig:AgriVet Treatment Grapher}, is the \textit{AgriVet Treatment Grapher}.\footnote{\url{https://github.com/mahirgamal/AgriVet-Treatment-Grapher.git}} This tool serves as an invaluable resource for veterinarians, farm managers, and researchers by allowing them to visualise treatment patterns, assess the frequency and distribution of different treatments, and analyse dosage information. The \textit{AgriVet Treatment Grapher} utilises messages in the \textit{LEI Treatment Event}\footnote{\url{https://github.com/mahirgamal/LEI-schema/blob/main/leiTreatmentEvent.json}} format to aid in decision-making processes, enhance treatment protocols, and contribute to overall animal health management. Its application in agricultural and veterinary contexts emphasises the importance of data sharing approaches in optimising treatment strategies and ensuring the well-being of animals under care.

The \textit{AgriVet Treatment Grapher} generates a series of plots that provide insights into shared messages. These include a bar graph showing the frequency of different treatments, a stacked bar chart illustrating the distribution of treatments on different dates, a pie chart representing the proportion of different treatments within the dataset, and another bar chart detailing average doses by treatment, including error bars for minimum and maximum doses.

%%%%%%%%%%%%%%%%%%%%%%%%%%%%%%%%%%%%%%%%%%%%%%%%%%%%%%%%
%%%%%            Experimental Evaluation           %%%%%
%%%%%%%%%%%%%%%%%%%%%%%%%%%%%%%%%%%%%%%%%%%%%%%%%%%%%%%%
\section{Experimental Evaluation}\label{sec:evaluation}
This section outlines the comprehensive methodology and framework employed to assess the effectiveness and efficiency of LEISA. This evaluation includes a detailed description of the objectives in~\Cref{sec:objective}, experimental settings in~\Cref{sec:experimental settings}, evaluation metrics in~\Cref{sec:evaluation metrics}, and the results in~\Cref{sec:results} obtained from the evaluation. Through a structured approach, this section aims to validate the performance, scalability, and reliability of LEISA, demonstrating its potential to revolutionise livestock event information sharing by enhancing decision-making processes, improving data standardisation, and facilitating more efficient data sharing among stakeholders in the agriculture sector.
%%%%%%%%%%%%%%%%%%%%%%%%%%%%%%%%%%%%%%%%%%%%%%%%%%%%%%%%
%%%%%                     Objectives              %%%%%
%%%%%%%%%%%%%%%%%%%%%%%%%%%%%%%%%%%%%%%%%%%%%%%%%%%%%%%%
\subsection{Objectives}\label{sec:objective}

In evaluating LEISA, the primary objective is to rigorously assess its effectiveness and efficiency in the context of data sharing. This involves an in-depth examination of the system's architecture and its components, focusing on key areas such as performance, scalability, and reliability.

\textbf{Performance} evaluation is paramount for determining the system's efficiency in processing and managing livestock event data. This metric is crucial in the agricultural domain, where timely and accurate data processing is directly correlated with the effectiveness of decision-making processes affecting livestock management. A high-performance system ensures stakeholders have access to real-time or near-real-time information, facilitating operational efficiency and supporting critical management decisions.

\textbf{Scalability}, as a key metric, addresses the system’s capability to adapt to varying loads, such as increased data volume, user base expansion, or heightened transaction rates, without detrimental effects on performance. Given the dynamic nature of agricultural operations, the ability to scale efficiently is vital, ensuring the system's long-term viability and its ability to accommodate growth and change.

\textbf{Reliability}, another cornerstone of the evaluation, involves examining the dependability of the system, focusing on uptime, error rates, and the accuracy of data handling processes. Ensuring the system is robust against failures and capable of graceful recovery from unexpected issues underpins the trust stakeholders place in it, which in turn, influences its adoption and utilisation.

Moreover, prioritising these metrics over others in the evaluation of LEISA is not arbitrary but is a strategic decision rooted in the understanding of the system's goals and the operational realities of its intended domain. Although metrics such as usability, security, and maintainability are undoubtedly significant, the foundational nature of performance, scalability, and reliability directly influences the system's ability to meet its core objectives of enhancing decision-making, improving data standardisation, and facilitating efficient data sharing among agricultural stakeholders. This focused approach ensures that the architecture meets the immediate needs of its users and is positioned for long-term success and adaptability within the ever-evolving landscape of agricultural technology.

Through this nuanced and exhaustive evaluation, the objective is to substantiate LEISA's revolutionary potential in the domain of livestock event information sharing, aiming to significantly enhance decision-making processes, elevate data standardisation, and streamline data sharing among a diverse array of stakeholders in the agriculture sector. This assessment is pivotal in demonstrating LEISA's capability to contribute meaningfully to the evolution of livestock management practices through advanced data sharing architectures.
%%%%%%%%%%%%%%%%%%%%%%%%%%%%%%%%%%%%%%%%%%%%%%%%%%%%%%%%
%%%%%            Experimental settings           %%%%%
%%%%%%%%%%%%%%%%%%%%%%%%%%%%%%%%%%%%%%%%%%%%%%%%%%%%%%%%
\subsection{Experimental settings}\label{sec:experimental settings}
In the evaluation conducted, the performance, scalability, and reliability of five core microservices within our architecture were rigorously assessed. These were message validator, registration to LEISA, set queue mapping, publish a message, and consume a message. The chosen microservices play an essential role in the data sharing and flow of the architecture, governing the accuracy of the validation, registration, and the transfer of messages between producers and consumers.

\begin{figure*}[!ht]
\centering
\begin{subfigure}{0.31\textwidth}
\centering
\includegraphics[width=\textwidth,height=\textheight,keepaspectratio]{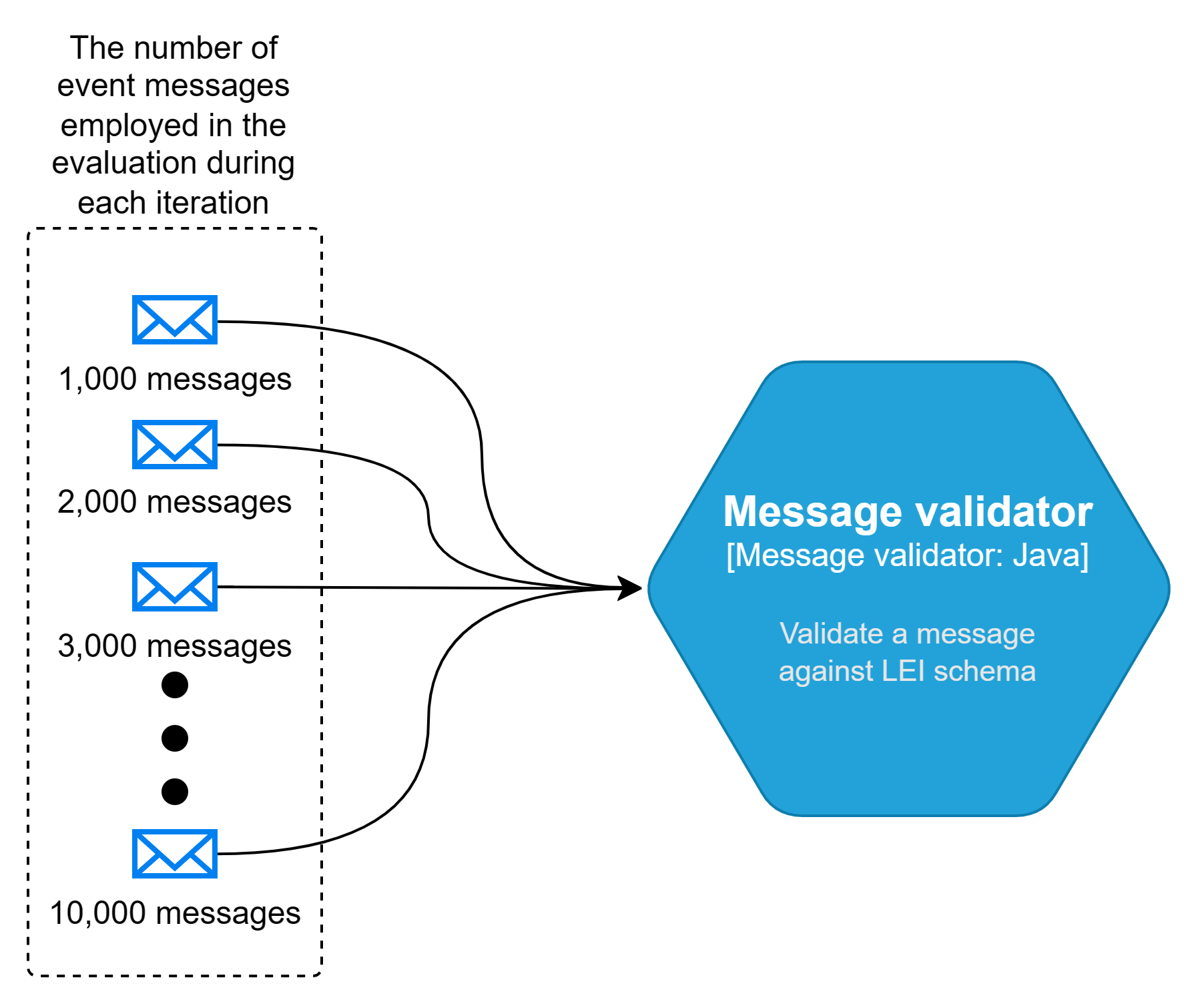}
\caption{Validator evaluation}
\label{fig:validator evaluation}
\end{subfigure}
\hspace{0.13\textwidth}
\begin{subfigure}{0.33\textwidth}
\centering
\includegraphics[width=\textwidth,height=\textheight,keepaspectratio]{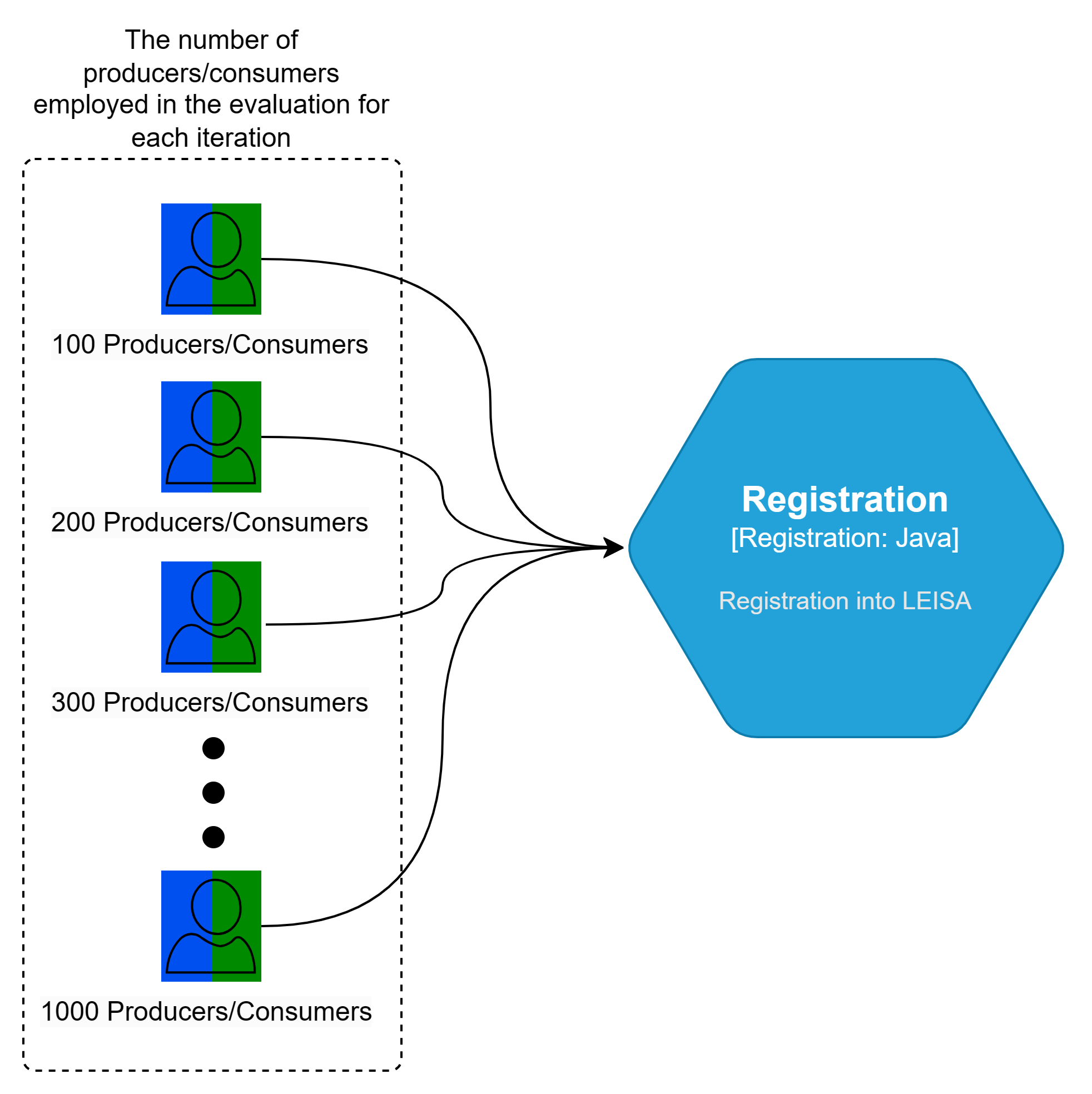}
\caption{Registration evaluation}
\label{fig:registration evaluation}
\end{subfigure}

\begin{subfigure}{0.44\textwidth}
\centering
\includegraphics[width=\textwidth,height=\textheight,keepaspectratio]{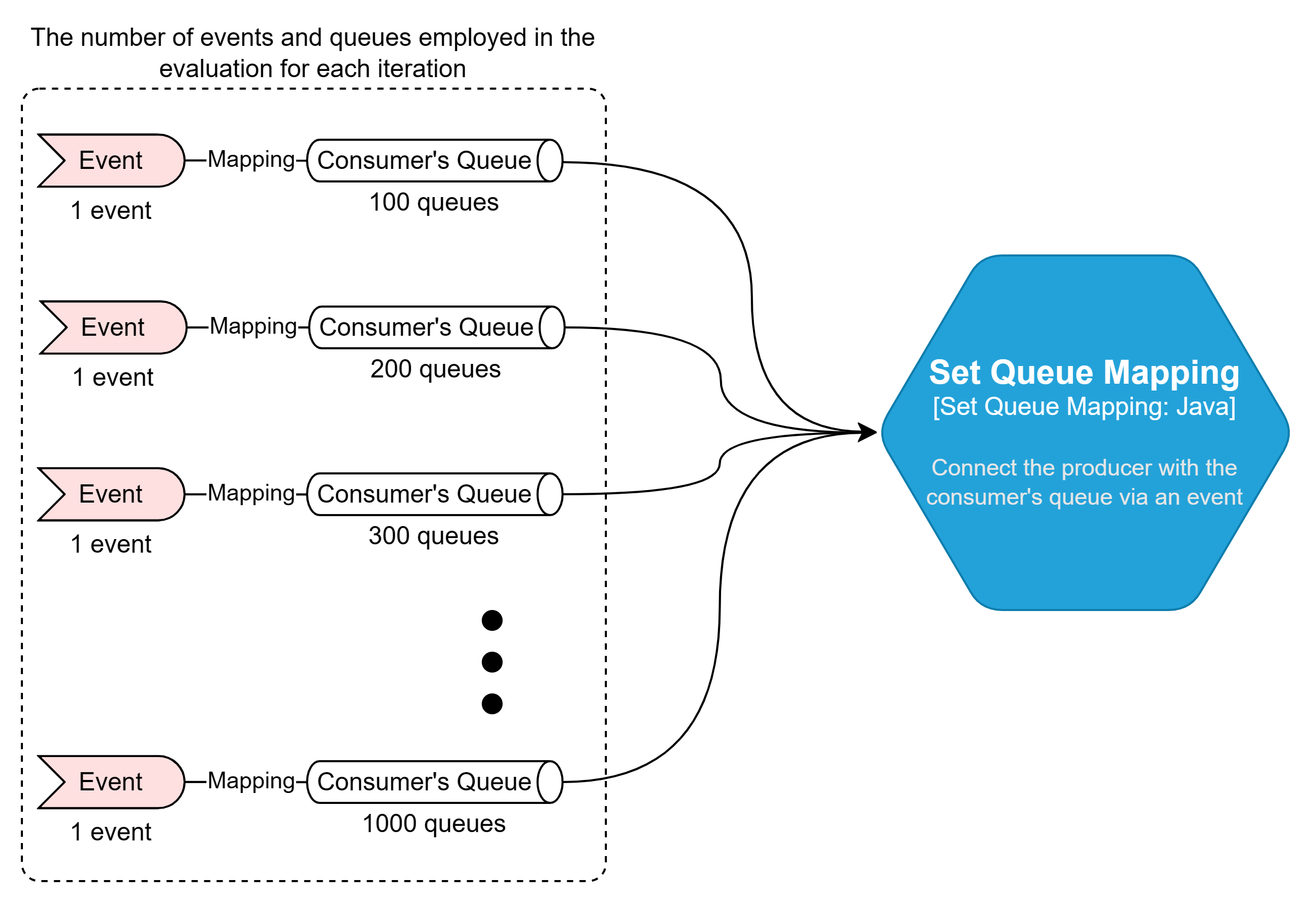}
\caption{Set queue mapping evaluation}
\label{fig:set queue mapping evaluation}
\end{subfigure}
\hspace{0.02\textwidth}
\begin{subfigure}{0.31\textwidth}
\centering
\includegraphics[width=\textwidth,height=\textheight,keepaspectratio]{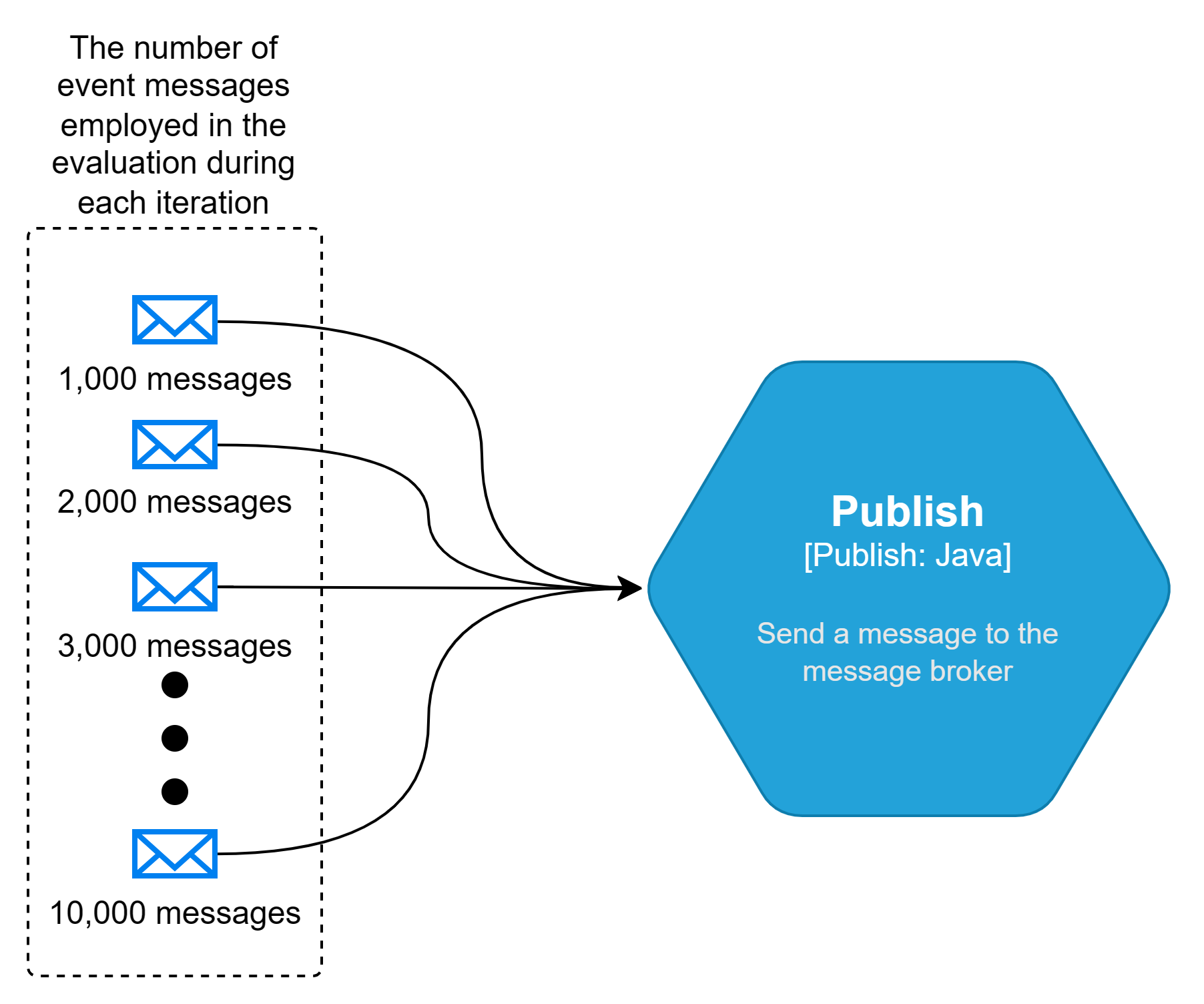}
\caption{Publish evaluation}
\label{fig:publish evaluation}
\end{subfigure}

\begin{subfigure}{0.31\textwidth}
\centering
\includegraphics[width=\textwidth,height=\textheight,keepaspectratio]{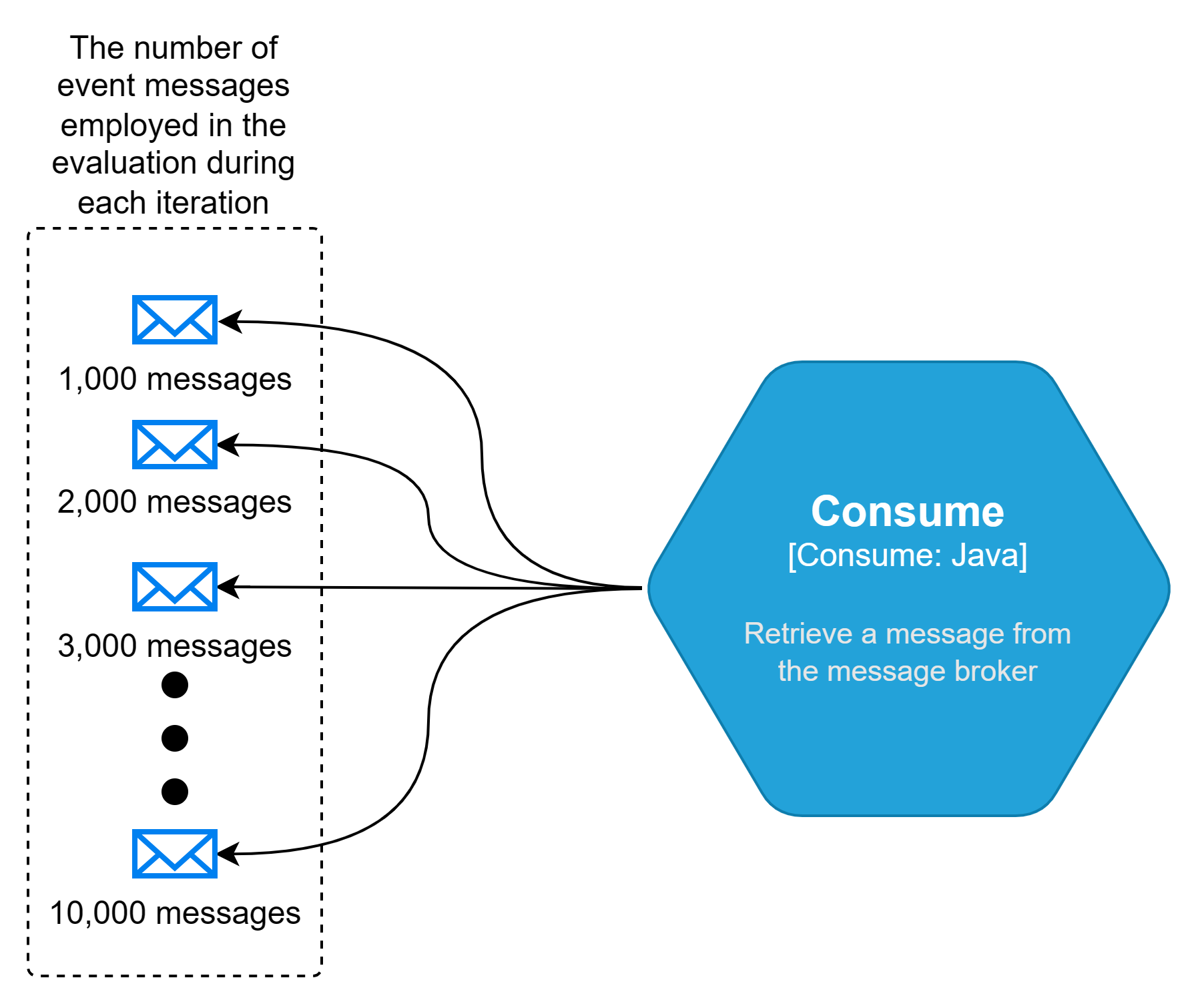}
\caption{Consume evaluation}
\label{fig:consume evaluation}
\end{subfigure}

\caption{Evaluation settings}
\label{fig:evaluation settings}
\end{figure*}

The evaluation is depicted in~\Cref{fig:evaluation settings}. For each microservice, a series of iterative tests were performed. Each iteration increased the load systematically to observe the microservice's behaviour under progressively heavier demands. Each iterative was repeated 10 times. This approach helped and allowed for a comprehensive understanding of the microservices' performance under a range of conditions.

~\Cref{fig:validator evaluation} illustrates the approach taken to assess the message validator, which entailed validating a spectrum of message loads to ensure compliance with the LEI schema. The \textit{MessageValidator} microservice was tested by subjecting it to increasing loads of messages, ranging from 1,000 to 10,000, to ensure data integrity and adherence to the LEI schema. The \textit{ServiceRegistration} microservice's evaluation capacity, shown in~\Cref{fig:registration evaluation}, was also evaluated by incrementing the number of producers and consumers from 100 to 1,000, examining the service's performance, scalability, and reliability.

\Cref{fig:set queue mapping evaluation} showcases the evaluation of the set queue mapping, which scrutinised the association of an event with a series of consumer queues, gauging the efficiency of the distribution system. The assessment for queue mapping examined the relationship of a single event to numerous consumer queues, varying from 100 to 1,000.

The evaluation of the publish microservice, depicted in~\Cref{fig:publish evaluation}, was conducted by progressively increasing the number of messages sent to the broker and monitoring how the system handled the communication burden. The capability of publishing was evaluated by sending different quantities of messages (ranging from 1,000 to 10,000) to the message broker to evaluate the data distribution. Finally, the proficiency of the consume microservice was evaluated, as detailed in~\Cref{fig:consume evaluation}, again with varying message volumes to simulate different load scenarios.

%%%%%%%%%%%%%%%%%%%%%%%%%%%%%%%%%%%%%%%%%%%%%%%%%%%%%%%%
%%%%%               valuation metrics              %%%%%
%%%%%%%%%%%%%%%%%%%%%%%%%%%%%%%%%%%%%%%%%%%%%%%%%%%%%%%%
\subsection{Evaluation metrics}\label{sec:evaluation metrics}
In examining of architecture akin to LEISA, applying of advanced metrics offers a granular insight into performance, scalability, and reliability, essential for understanding and optimising system behaviour under varied conditions.

When assessing performance, two essential statistical measures provide valuable insights: the mean and the standard deviation. In the context of microservices, understanding how well a microservice responds under various conditions is crucial.

The mean $\bar{x}$ provides a central tendency of response times, offering an average value, as demonstrated in~\Cref{eq:Mean Execution Time}, while the standard deviation ($\sigma$) indicates the dispersion or variability of the data from this mean, as illustrated in~\Cref{eq:standard deviation}. Understanding both is crucial to accurately describe the performance characteristics of microservices, as $\sigma$ reveals how consistent the performance is around the mean response time~\citep{lee2015standard}. For instance, a microservice might exhibit an acceptable average response time (mean), but a high standard deviation could suggest highly inconsistent response times, impacting the user experience.

\begin{equation}\label{eq:Mean Execution Time}
\bar{x} = \frac{1}{n}\sum_{i=1}^{n}t_i
\end{equation}

Where $\bar{x}$ is the mean execution time, $n$ is the number of requests, and $t_i$ is the execution time of the $i^{th}$ request.

\begin{equation}\label{eq:standard deviation}
\sigma = \sqrt{\frac{1}{n-1}\sum_{i=1}^{n}(t_i - \bar{x})^2}
\end{equation}

Where $\sigma$ is the standard deviation, $t_i$ is the execution time of the $i^{th}$ request, $\bar{x}$ is the mean execution time, and n is the total number of requests.

Measuring scalability is critical to understanding how well a service can handle increasing loads and adapt to growing demands. Scalability is a key attribute of microservice architecture, promoting the ability to dynamically increase or decrease resources and services based on current needs. Throughput, as illustrated in~\Cref{eq:Throughput}, becomes a critical measure, reflecting the system’s ability to handle a growing number of requests within a specific timeframe. This metric is instrumental in determining the system's capability to scale resources effectively in response to increased demand. Lower throughput under increased workload suggests that the system is struggling to cope with the demand, which can indicate scalability issues. In other words, if the throughput decreases as workload increases, it may signify that the system is not scaling effectively to meet the growing demands.~\citep{blinowski2022monolithic}.

\begin{equation}\label{eq:Throughput}
\Theta = \frac{n}{T}
\end{equation}

Where $\Theta$ represents throughput, $n$ is the number of requests and $T$ is the total time it takes to process these requests.

Measuring the reliability of a microservice using the coefficient of variation (CV) involves assessing the consistency and precision of the service's performance over time. The CV, a standardised measure of dispersion, is particularly useful in contexts where the mean alone cannot describe data variability. A lower CV indicates a system that performs consistently, underscoring its reliability and the predictability of its service quality, regardless of the fluctuating demands~\citep{wang2020workflow}.

\begin{equation}\label{eq:cv}
CV = \frac{\sigma}{\bar{x}}
\end{equation}

Where CV represents the coefficient of variation,$ \sigma$ is the standard deviation of execution times, and $\bar{x}$ is the mean execution time.

These metrics, woven together in a cohesive narrative, construct a comprehensive picture of the system's operational dynamics. This scientific approach uncovers the intricacies of system behaviour in the face of diverse challenges and guides strategic enhancements to ensure that the system meets its intended objectives effectively, thereby affirming its viability and efficacy in the designated application domain.

%%%%%%%%%%%%%%%%%%%%%%%%%%%%%%%%%%%%%%%%%%%%%%%%%%%%%%%%
%%%%%                       Results                %%%%%
%%%%%%%%%%%%%%%%%%%%%%%%%%%%%%%%%%%%%%%%%%%%%%%%%%%%%%%%
\subsection{Results}\label{sec:results}
LEISA's implementation and subsequent evaluation revealed several key findings with implications for both the software architecture and agricultural sectors. LEISA’s design incorporated a range of microservices, each crucial for the architecture's overall data processing and message flow, and was deployed using Microsoft Azure services.

\begin{figure*}[htbp]
\centering
\begin{subfigure}{0.45\textwidth}
\centering
\begin{tikzpicture}
\begin{axis}[
xlabel={Number of the events},
ylabel={Time in sec},
width=\textwidth,
height=5cm,
ymajorgrids=true,
yminorgrids=true,
grid=major,
legend pos=south east,
xtick=data,
xticklabels from table={\validate}{events},
scaled y ticks=false,
scaled x ticks=false,
xticklabel style={rotate=90, anchor=east}, 
]
\addplot+[
mark=none,
orange,
error bars/.cd,
y explicit,
y dir=both,
]
table [
x=events,
y expr=\thisrow{validate_mean}/1000,
y error expr=\thisrow{validate_stddev}/1000
] {\validate};
\end{axis}
\end{tikzpicture}
\caption{\textit{MessageValidator} microservice execution time}
\label{fig:validate_performance}
\vspace{1em}
\end{subfigure}
%%%%%%%%%%%%%%%%%%%%%%%%%%%%%%%%%%%%%%
%%%%%%%%%%%%%%%%%%%%%%%%%%%%%%%%%%%%%%
\begin{subfigure}{0.45\textwidth}
\centering
\begin{tikzpicture}
\begin{axis}[
xlabel={Number of registration requests},
ylabel={Time in sec},
width=\textwidth,
height=5cm,
ymajorgrids=true,
yminorgrids=true,
grid=major,
legend pos=south east,
xtick=data,
xticklabels from table={\registration}{events},
scaled y ticks=false,
scaled x ticks=false,
xticklabel style={rotate=90, anchor=east},
 ytick={5, 10, 15, 20, 25, 30},
]
\addplot+[
mark=none,
cyan,
error bars/.cd,
y explicit,
y dir=both,
]
table [
x=events,
y expr=\thisrow{registration_mean}/1000,
y error expr=\thisrow{registration_stddev}/1000
] {\registration};
\end{axis}
\end{tikzpicture}
\caption{\textit{ServiceRegistration} microservice execution time}
\label{fig:registration_performance}
\vspace{1em}
\end{subfigure}
%%%%%%%%%%%%%%%%%%%%%%%%%%%%%%%%%%%%%%
%%%%%%%%%%%%%%%%%%%%%%%%%%%%%%%%%%%%%%

\begin{subfigure}{0.45\textwidth}
\centering
\begin{tikzpicture}
\begin{axis}[
xlabel={Number of setQueueMapping requests},
ylabel={Time in sec},
width=\textwidth,
height=5cm,
ymajorgrids=true,
yminorgrids=true,
grid=major,
legend pos=south east,
xtick=data,
xticklabels from table={\setQueueMapping}{events},
scaled y ticks=false,
scaled x ticks=false,
xticklabel style={rotate=90, anchor=east},
]
\addplot+[
mark=none,
teal,
error bars/.cd,
y explicit,
y dir=both,
]
table [
x=events,
y expr=\thisrow{setQueueMapping_mean}/1000,
y error expr=\thisrow{setQueueMapping_stddev}/1000
] {\setQueueMapping};
\end{axis}
\end{tikzpicture}
\caption{\textit{SetQueueMapping} microservice execution time}
\label{fig:setQueueMapping_performance}
\vspace{1em}
\end{subfigure}
%%%%%%%%%%%%%%%%%%%%%%%%%%%%%%%%%%%%%%
%%%%%%%%%%%%%%%%%%%%%%%%%%%%%%%%%%%%%%
\begin{subfigure}{0.45\textwidth}
\centering
\begin{tikzpicture}
\begin{axis}[
xlabel={Number of events},
ylabel={Time in sec},
width=\textwidth,
height=5cm,
ymajorgrids=true,
yminorgrids=true,
grid=major,
legend pos=south east,
xtick=data,
xticklabels from table={\publishMessage}{events},
scaled y ticks=false,
scaled x ticks=false,
xticklabel style={rotate=90, anchor=east},
]
\addplot+[
mark=none,
magenta,
error bars/.cd,
y explicit,
y dir=both,
]
table [
x=events,
y expr=\thisrow{publishMessage_mean}/1000,
y error expr=\thisrow{publishMessage_stddev}/1000
] {\publishMessage};
\end{axis}
\end{tikzpicture}
\caption{\textit{PublishMessage} microservice execution time}
\label{fig:publishMessage_performance}
\vspace{1em}
\end{subfigure}
%%%%%%%%%%%%%%%%%%%%%%%%%%%%%%%%%%%%%%
%%%%%%%%%%%%%%%%%%%%%%%%%%%%%%%%%%%%%%

\begin{subfigure}{0.45\textwidth}
\centering
\begin{tikzpicture}
\begin{axis}[
xlabel={Number of events},
ylabel={Time in sec},
width=\textwidth,
height=5cm,
ymajorgrids=true,
yminorgrids=true,
grid=major,
legend pos=south east,
xtick=data,
xticklabels from table={\consumeMessage}{events},
scaled y ticks=false,
scaled x ticks=false,
xticklabel style={rotate=90, anchor=east},
]
\addplot+[
mark=none,
blue,
error bars/.cd,
y explicit,
y dir=both,
]
table [
x=events,
y expr=\thisrow{consumeMessage_mean}/1000,
y error expr=\thisrow{consumeMessage_stddev}/1000
] {\consumeMessage};
\end{axis}
\end{tikzpicture}
\caption{\textit{ConsumeMessage} microservice execution time}
\label{fig:consumeMessage_performance}
\vspace{1em}
\end{subfigure}
\caption{The execution time for five core microservices: \textit{MessageValidator}, \textit{ServiceRegistration}, \textit{SetQueueMapping}, \textit{PublishMessage}, and \textit{consumeMessage}. Time measurements are in seconds (sec) and represent the mean execution time. The error bars indicate the standard deviation}
\label{fig:execution_time}
\end{figure*}
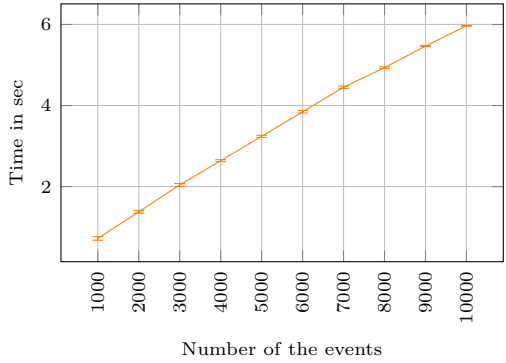
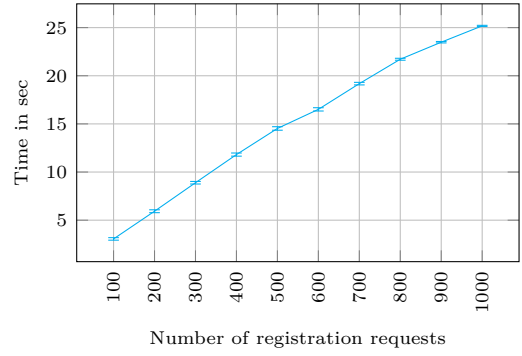
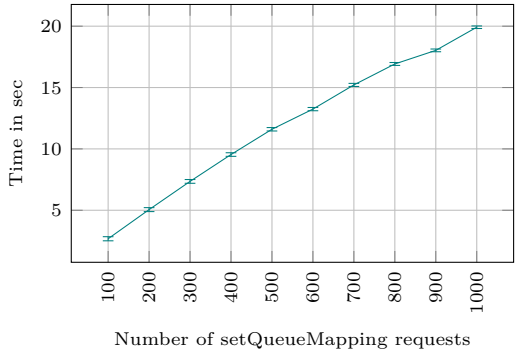
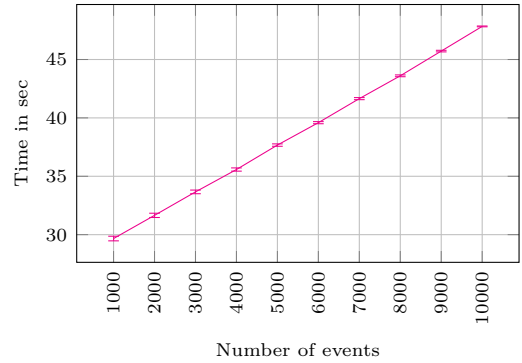
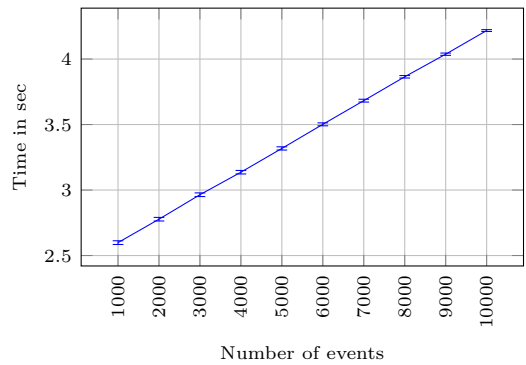

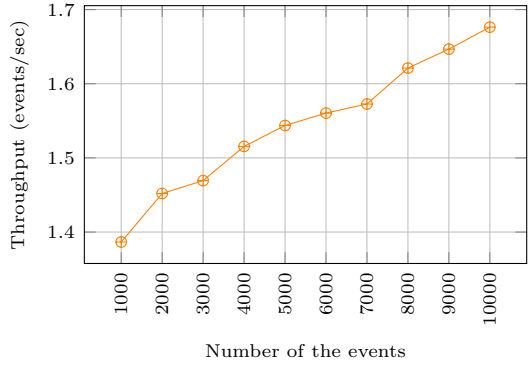
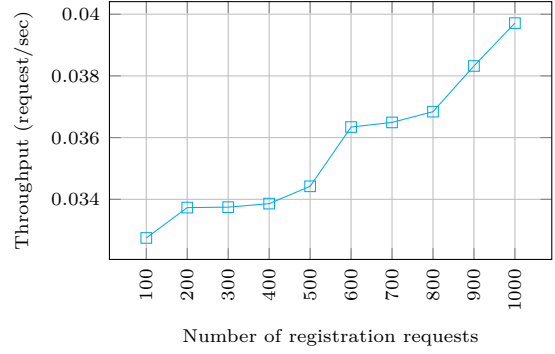
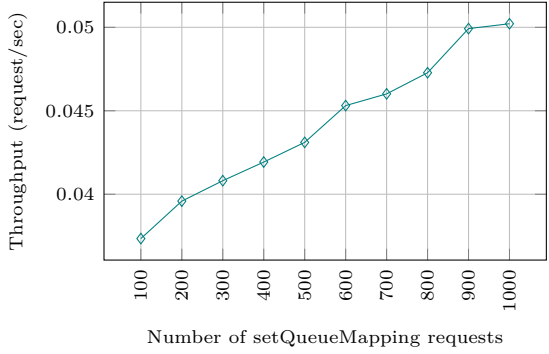
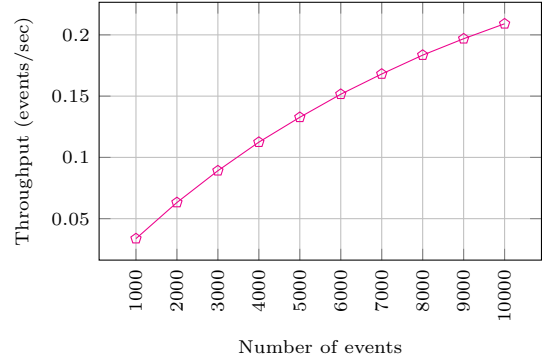
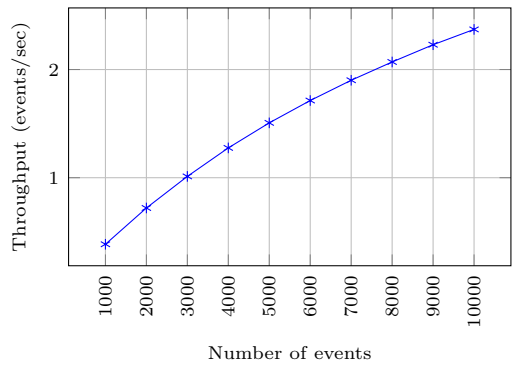
\begin{figure*}[htbp]
\centering

\begin{subfigure}{0.45\textwidth}
\centering
\begin{tikzpicture}
\begin{axis}[
xlabel={Number of the events},
ylabel={Throughput (events/sec)},
width=\textwidth,
height=5cm,
ymajorgrids=true,
yminorgrids=true,
grid=major,
legend pos=south east,
xtick=data,
xticklabels from table={\validate}{events},
scaled y ticks=false,
scaled x ticks=false,
xticklabel style={rotate=90, anchor=east},
yticklabel style={
        /pgf/number format/fixed,
        /pgf/number format/precision=3,
    },
]
\addplot+[
mark=oplus,
orange,
error bars/.cd,
y explicit,
y dir=both,
]
table [
x=events,
y expr=\thisrow{validate_throughput}/1000
] {\validate};
\end{axis}
\end{tikzpicture}
\caption{\textit{MessageValidator} microservice throughput}
\label{fig:validate_throughput}
\vspace{1em}
\end{subfigure}
%%%%%%%%%%%%%%%%%%%%%%%%%%%%%%%%%%%%%%
%%%%%%%%%%%%%%%%%%%%%%%%%%%%%%%%%%%%%%
\begin{subfigure}{0.45\textwidth}
\centering
\begin{tikzpicture}
\begin{axis}[
xlabel={Number of registration requests},
ylabel={Throughput (request/sec)},
width=\textwidth,
height=5cm,
ymajorgrids=true,
yminorgrids=true,
grid=major,
legend pos=south east,
xtick=data,
xticklabels from table={\registration}{events},
scaled y ticks=false,
scaled x ticks=false,
xticklabel style={rotate=90, anchor=east},
yticklabel style={
        /pgf/number format/fixed,
        /pgf/number format/precision=3,
    },
]
\addplot+[
mark=square,
cyan,
error bars/.cd,
y explicit,
y dir=both,
]
table [
x=events,
y expr=\thisrow{registration_throughput}/1000
] {\registration};
\end{axis}
\end{tikzpicture}
\caption{\textit{ServiceRegistration} microservice throughput}
\label{fig:registratioin_throughput}
\vspace{1em}
\end{subfigure}
%%%%%%%%%%%%%%%%%%%%%%%%%%%%%%%%%%%%%%
%%%%%%%%%%%%%%%%%%%%%%%%%%%%%%%%%%%%%%

\begin{subfigure}{0.45\textwidth}
\centering
\begin{tikzpicture}
\begin{axis}[
xlabel={Number of setQueueMapping requests},
ylabel={Throughput (request/sec)},
width=\textwidth,
height=5cm,
ymajorgrids=true,
yminorgrids=true,
grid=major,
legend pos=south east,
xtick=data,
xticklabels from table={\setQueueMapping}{events},
scaled y ticks=false,
scaled x ticks=false,
xticklabel style={rotate=90, anchor=east}, 
 yticklabel style={
        /pgf/number format/fixed,
        /pgf/number format/precision=3,
    },
]
\addplot+[
mark=diamond,
teal,
error bars/.cd,
y explicit,
y dir=both,
]
table [
x=events,
y expr=\thisrow{setQueueMapping_throughput}/1000
] {\setQueueMapping};
\end{axis}
\end{tikzpicture}
\caption{\textit{SetQueueMapping} microservice throughput}
\label{fig:setQueueMapping_throughput}
\vspace{1em}
\end{subfigure}
%%%%%%%%%%%%%%%%%%%%%%%%%%%%%%%%%%%%%%
%%%%%%%%%%%%%%%%%%%%%%%%%%%%%%%%%%%%%%
\begin{subfigure}{0.45\textwidth}
\centering
\begin{tikzpicture}
\begin{axis}[
xlabel={Number of events},
ylabel={Throughput (events/sec)},
width=\textwidth,
height=5cm,
ymajorgrids=true,
yminorgrids=true,
grid=major,
legend pos=south east,
xtick=data,
xticklabels from table={\publishMessage}{events},
scaled y ticks=false,
scaled x ticks=false,
xticklabel style={rotate=90, anchor=east}, 
 yticklabel style={
        /pgf/number format/fixed,
        /pgf/number format/precision=3,
    },
]
\addplot+[
mark=pentagon,
magenta,
error bars/.cd,
y explicit,
y dir=both,
]
table [
x=events,
y expr=\thisrow{publishMessage_throughput}/1000
] {\publishMessage};
\end{axis}
\end{tikzpicture}
\caption{\textit{PublishMessage} microservice throughput}
\label{fig:publishMessage_throughput}
\vspace{1em}
\end{subfigure}
%%%%%%%%%%%%%%%%%%%%%%%%%%%%%%%%%%%%%%
%%%%%%%%%%%%%%%%%%%%%%%%%%%%%%%%%%%%%%

\begin{subfigure}{0.45\textwidth}
\centering
\begin{tikzpicture}
\begin{axis}[
xlabel={Number of events},
ylabel={Throughput (events/sec)},
width=\textwidth,
height=5cm,
ymajorgrids=true,
yminorgrids=true,
grid=major,
legend pos=south east,
xtick=data,
xticklabels from table={\consumeMessage}{events},
scaled y ticks=false,
scaled x ticks=false,
xticklabel style={rotate=90, anchor=east},
 yticklabel style={
        /pgf/number format/fixed,
        /pgf/number format/precision=3,
    },
]
\addplot+[
mark=asterisk,
blue,
error bars/.cd,
y explicit,
y dir=both,
]
table [
x=events,
y expr=\thisrow{consumeMessage_throughput}/1000
] {\consumeMessage};
\end{axis}
\end{tikzpicture}
\caption{\textit{ConsumeMessage} microservice throughput}
\label{fig:consumeMessage_throughput}
\vspace{1em}
\end{subfigure}
\caption{The throughput for five core microservices: \textit{MessageValidator}, \textit{ServiceRegistration}, \textit{SetQueueMapping}, \textit{PublishMessage}, and \textit{consumeMessage}}
\label{fig:throughput}
\end{figure*}

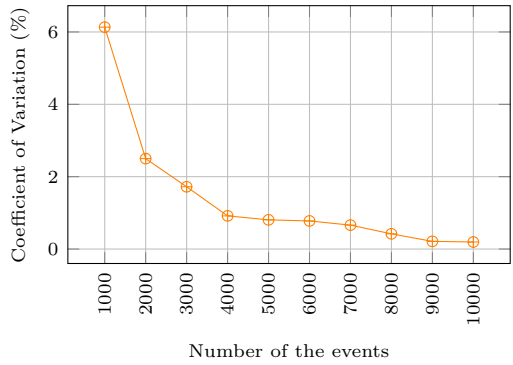
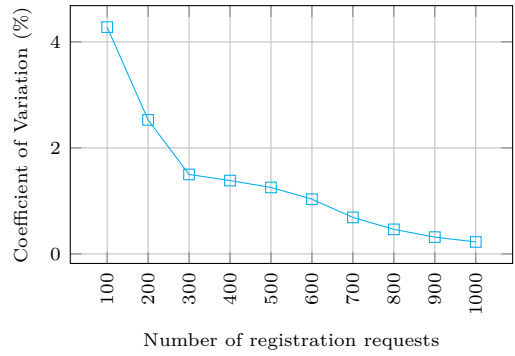
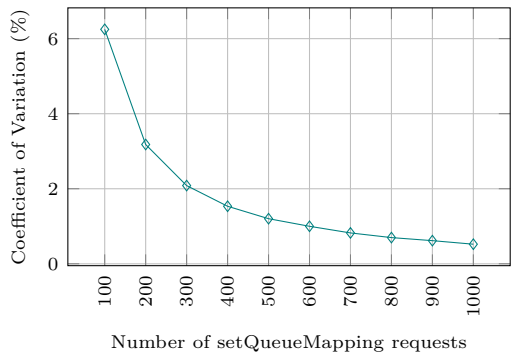
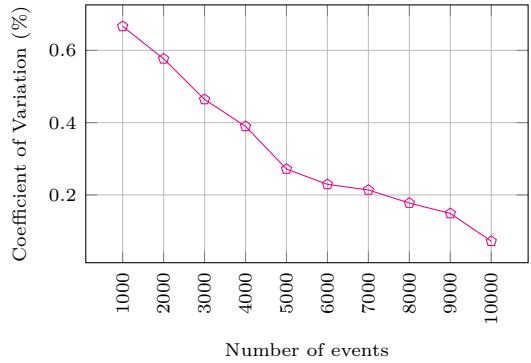
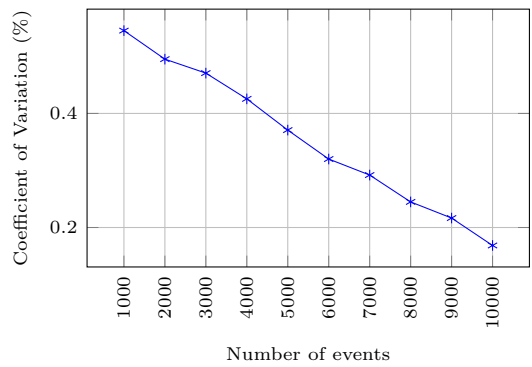
\begin{figure*}[htbp]
\centering

\begin{subfigure}{0.45\textwidth}
\centering
\begin{tikzpicture}
\begin{axis}[
xlabel={Number of the events},
ylabel={Coefficient of Variation (\%)},
width=\textwidth,
height=5cm,
ymajorgrids=true,
yminorgrids=true,
grid=major,
legend pos=south east,
xtick=data,
xticklabels from table={\validate}{events},
scaled y ticks=false,
scaled x ticks=false,
xticklabel style={rotate=90, anchor=east}, 
]
\addplot+[
mark=oplus,
orange,
error bars/.cd,
y explicit,
y dir=both,
]
table [
x=events,
y=validate_cv
] {\validate};
\end{axis}
\end{tikzpicture}
\caption{\textit{MessageValidator} microservice CV}
\label{fig:validate_cv}
\vspace{1em}
\end{subfigure}
%%%%%%%%%%%%%%%%%%%%%%%%%%%%%%%%%%%%%%
%%%%%%%%%%%%%%%%%%%%%%%%%%%%%%%%%%%%%%
\begin{subfigure}{0.45\textwidth}
\centering
\begin{tikzpicture}
\begin{axis}[
xlabel={Number of registration requests},
ylabel={Coefficient of Variation (\%)},
width=\textwidth,
height=5cm,
ymajorgrids=true,
yminorgrids=true,
grid=major,
legend pos=south east,
xtick=data,
xticklabels from table={\registration}{events},
scaled y ticks=false,
scaled x ticks=false,
xticklabel style={rotate=90, anchor=east}, 
]
\addplot+[
mark=square,
cyan,
error bars/.cd,
y explicit,
y dir=both,
]
table [
x=events,
y=registration_cv
] {\registration};
\end{axis}
\end{tikzpicture}
\caption{\textit{ServiceRegistration} microservice CV}
\label{fig:registration_cv}
\vspace{1em}
\end{subfigure}
%%%%%%%%%%%%%%%%%%%%%%%%%%%%%%%%%%%%%%
%%%%%%%%%%%%%%%%%%%%%%%%%%%%%%%%%%%%%%

\begin{subfigure}{0.45\textwidth}
\centering
\begin{tikzpicture}
\begin{axis}[
xlabel={Number of setQueueMapping requests},
ylabel={Coefficient of Variation (\%)},
width=\textwidth,
height=5cm,
ymajorgrids=true,
yminorgrids=true,
grid=major,
legend pos=south east,
xtick=data,
xticklabels from table={\setQueueMapping}{events},
scaled y ticks=false,
scaled x ticks=false,
xticklabel style={rotate=90, anchor=east}, 
]
\addplot+[
mark=diamond,
teal,
error bars/.cd,
y explicit,
y dir=both,
]
table [
x=events,
y=setQueueMapping_cv
] {\setQueueMapping};
\end{axis}
\end{tikzpicture}
\caption{\textit{SetQueueMapping} microservice CV}
\label{fig:rsetQueueMapping_cv}
\vspace{1em}
\end{subfigure}
%%%%%%%%%%%%%%%%%%%%%%%%%%%%%%%%%%%%%%
%%%%%%%%%%%%%%%%%%%%%%%%%%%%%%%%%%%%%%
\begin{subfigure}{0.45\textwidth}
\centering
\begin{tikzpicture}
\begin{axis}[
xlabel={Number of events},
ylabel={Coefficient of Variation (\%)},
width=\textwidth,
height=5cm,
ymajorgrids=true,
yminorgrids=true,
grid=major,
legend pos=south east,
xtick=data,
xticklabels from table={\publishMessage}{events},
scaled y ticks=false,
scaled x ticks=false,
xticklabel style={rotate=90, anchor=east}, 
]
\addplot+[
mark=pentagon,
magenta,
error bars/.cd,
y explicit,
y dir=both,
]
table [
x=events,
y=publishMessage_cv
] {\publishMessage};
\end{axis}
\end{tikzpicture}
\caption{\textit{PublishMessage} microservice CV}
\label{fig:publishMessage_cv}
\vspace{1em}
\end{subfigure}
%%%%%%%%%%%%%%%%%%%%%%%%%%%%%%%%%%%%%%
%%%%%%%%%%%%%%%%%%%%%%%%%%%%%%%%%%%%%%

\begin{subfigure}{0.45\textwidth}
\centering
\begin{tikzpicture}
\begin{axis}[
xlabel={Number of events},
ylabel={Coefficient of Variation (\%)},
width=\textwidth,
height=5cm,
ymajorgrids=true,
yminorgrids=true,
grid=major,
legend pos=south east,
xtick=data,
xticklabels from table={\consumeMessage}{events},
scaled y ticks=false,
scaled x ticks=false,
xticklabel style={rotate=90, anchor=east}, 
]
\addplot+[
mark=asterisk,
blue,
error bars/.cd,
y explicit,
y dir=both,
]
table [
x=events,
y=consumeMessage_cv
] {\consumeMessage};
\end{axis}
\end{tikzpicture}
\caption{\textit{ConsumeMessage} microservice CV}
\label{fig:consumeMessage_cv}
\vspace{1em}
\end{subfigure}
\caption{The coefficient of variation for five core microservices: \textit{MessageValidator}, \textit{ServiceRegistration}, \textit{SetQueueMapping}, \textit{PublishMessage}, and \textit{consumeMessage}}
\label{fig:coefficient_of_variation}
\end{figure*}

In this section, we discuss the empirical results that stem from LEISA's evaluation, focusing on the operational prowess of its core microservices under variable workloads. The meticulous examination covers the \textit{MessageValidator} microservice, \textit{ServiceRegistration} microservice, \textit{SetQueueMapping} microservice, \textit{PublishMessage} microservice, and \textit{ConsumeMessage} microservice. The execution time and throughput evaluations encapsulated in~\Cref{fig:execution_time,fig:throughput}, respectively, are complemented by CV analysis delineated in~\Cref{fig:coefficient_of_variation}, collectively providing a multi-dimensional view of LEISA's performance.

\pgfplotstableread[col sep=space]{
events	data1	data2	data3	data4	data5	data6	data7	data8	data9	data10
1000	702	703	840	711	694	697	707	704	747	707
2000	1367	1363	1358	1379	1474	1371	1364	1362	1369	1367
3000	2002	2017	2057	2006	2020	2064	2096	2091	2011	2053
4000	2661	2668	2637	2611	2611	2671	2664	2620	2626	2625
5000	3227	3223	3230	3238	3244	3205	3245	3293	3270	3215
6000	3864	3882	3806	3882	3849	3862	3858	3834	3808	3807
7000	4493	4438	4470	4424	4406	4446	4485	4444	4425	4480
8000	4958	4911	4970	4952	4928	4927	4917	4907	4939	4940
9000	5467	5453	5462	5475	5474	5453	5465	5451	5488	5468
10000	5967	5953	5962	5975	5974	5953	5965	5951	5988	5968
}\validate

\pgfplotstablecreatecol[
create col/expr={
(\thisrow{data1} + \thisrow{data2} + \thisrow{data3} + \thisrow{data4} + \thisrow{data5} + \thisrow{data6} + \thisrow{data7} + \thisrow{data8} + \thisrow{data9} + \thisrow{data10})/10
}
]{validate_mean}\validate

\pgfplotstablecreatecol[
create col/expr={
sqrt((pow(\thisrow{data1}-\thisrow{validate_mean},2) + pow(\thisrow{data2}-\thisrow{validate_mean},2) + pow(\thisrow{data3}-\thisrow{validate_mean},2) + pow(\thisrow{data4}-\thisrow{validate_mean},2) + pow(\thisrow{data5}-\thisrow{validate_mean},2) + pow(\thisrow{data6}-\thisrow{validate_mean},2) + pow(\thisrow{data7}-\thisrow{validate_mean},2) + pow(\thisrow{data8}-\thisrow{validate_mean},2) + pow(\thisrow{data9}-\thisrow{validate_mean},2) + pow(\thisrow{data10}-\thisrow{validate_mean},2))/(10-1))
}
]{validate_stddev}\validate

\pgfplotstablecreatecol[
create col/expr={
\thisrow{events} / (\thisrow{validate_mean}/1000) % Convert mean execution time to seconds and calculate throughput
}
]{validate_throughput}\validate

\pgfplotstablecreatecol[
create col/expr={
(\thisrow{validate_stddev}/\thisrow{validate_mean})*100
}
]{validate_cv}\validate

%%%%%%%%%%%%%%%%%%%%%%%%%%%%%%%%%%%%%%
%%%%%%%%%%%%%%%%%%%%%%%%%%%%%%%%%%%%%%
%%%%%%%%%%%%%%%%%%%%%%%%%%%%%%%%%%%%%%
\pgfplotstableread[col sep=space]{
events	data1	data2	data3	data4	data5	data6	data7	data8	data9	data10
100	3148	3171	2938	3089	3013	3063	2864	2954	2990	3308
200	6186	6103	5821	5751	5812	5972	5823	5911	5826	6087
300	9010	8847	9062	8802	8822	9000	8733	8762	9083	8783
400	11934	11822	11863	11938	11946	11640	11935	11604	11518	11934
500	14852	14789	14605	14395	14312	14374	14572	14384	14433	14544
600	16547	16144	16409	16482	16795	16549	16508	16465	16685	16515
700	19016	19304	19188	19277	19331	19383	19117	19065	19090	19053
800	21678	21792	21620	21798	21744	21748	21841	21715	21491	21728
900	23535	23425	23438	23400	23543	23546	23578	23546	23362	23485
1000	25149	25245	25146	25179	25181	25134	25220	25085	25278	25203
}\registration

\pgfplotstablecreatecol[
create col/expr={
(\thisrow{data1} + \thisrow{data2} + \thisrow{data3} + \thisrow{data4} + \thisrow{data5} + \thisrow{data6} + \thisrow{data7} + \thisrow{data8} + \thisrow{data9} + \thisrow{data10})/10
}
]{registration_mean}\registration

\pgfplotstablecreatecol[
create col/expr={
sqrt((pow(\thisrow{data1}-\thisrow{registration_mean},2) + pow(\thisrow{data2}-\thisrow{registration_mean},2) + pow(\thisrow{data3}-\thisrow{registration_mean},2) + pow(\thisrow{data4}-\thisrow{registration_mean},2) + pow(\thisrow{data5}-\thisrow{registration_mean},2) + pow(\thisrow{data6}-\thisrow{registration_mean},2) + pow(\thisrow{data7}-\thisrow{registration_mean},2) + pow(\thisrow{data8}-\thisrow{registration_mean},2) + pow(\thisrow{data9}-\thisrow{registration_mean},2) + pow(\thisrow{data10}-\thisrow{registration_mean},2))/(10-1))
}
]{registration_stddev}\registration

\pgfplotstablecreatecol[
create col/expr={
\thisrow{events} / (\thisrow{registration_mean}/1000) % Convert mean execution time to seconds and calculate throughput
}
]{registration_throughput}\registration

\pgfplotstablecreatecol[
create col/expr={
(\thisrow{registration_stddev}/\thisrow{registration_mean})*100
}
]{registration_cv}\registration

%%%%%%%%%%%%%%%%%%%%%%%%%%%%%%%%%%%%%%
%%%%%%%%%%%%%%%%%%%%%%%%%%%%%%%%%%%%%%
%%%%%%%%%%%%%%%%%%%%%%%%%%%%%%%%%%%%%%

\pgfplotstableread[col sep=space]{
events	data1	data2	data3	data4	data5	data6	data7	data8	data9	data10
100	2480	2849	2802	2425	2889	2839	2531	2610	2616	2738
200	5267	4975	5126	5043	4807	5036	5061	4983	4886	5338
300	7601	7420	7254	7141	7142	7374	7436	7467	7450	7225
400	9718	9455	9431	9651	9291	9664	9635	9531	9656	9368
500	11905	11529	11500	11515	11432	11549	11570	11641	11587	11762
600	13245	13264	13403	13278	13157	13395	13269	13039	13346	13026
700	15130	15248	15033	15029	15247	15188	15321	15192	15341	15404
800	16801	16910	16948	16879	16770	17111	17114	16888	16824	16968
900	17989	17827	17997	18086	18060	18246	18062	17933	18096	17985
1000	19960	19860	19804	19824	20089	20025	20015	19816	19818	19933
}\setQueueMapping

\pgfplotstablecreatecol[
create col/expr={
(\thisrow{data1} + \thisrow{data2} + \thisrow{data3} + \thisrow{data4} + \thisrow{data5} + \thisrow{data6} + \thisrow{data7} + \thisrow{data8} + \thisrow{data9} + \thisrow{data10})/10
}
]{setQueueMapping_mean}\setQueueMapping

\pgfplotstablecreatecol[
create col/expr={
sqrt((pow(\thisrow{data1}-\thisrow{setQueueMapping_mean},2) + pow(\thisrow{data2}-\thisrow{setQueueMapping_mean},2) + pow(\thisrow{data3}-\thisrow{setQueueMapping_mean},2) + pow(\thisrow{data4}-\thisrow{setQueueMapping_mean},2) + pow(\thisrow{data5}-\thisrow{setQueueMapping_mean},2) + pow(\thisrow{data6}-\thisrow{setQueueMapping_mean},2) + pow(\thisrow{data7}-\thisrow{setQueueMapping_mean},2) + pow(\thisrow{data8}-\thisrow{setQueueMapping_mean},2) + pow(\thisrow{data9}-\thisrow{setQueueMapping_mean},2) + pow(\thisrow{data10}-\thisrow{setQueueMapping_mean},2))/(10-1))
}
]{setQueueMapping_stddev}\setQueueMapping

\pgfplotstablecreatecol[
create col/expr={
\thisrow{events} / (\thisrow{setQueueMapping_mean}/1000) % Convert mean execution time to seconds and calculate throughput
}
]{setQueueMapping_throughput}\setQueueMapping

\pgfplotstablecreatecol[
create col/expr={
(\thisrow{setQueueMapping_stddev}/\thisrow{setQueueMapping_mean})*100
}
]{setQueueMapping_cv}\setQueueMapping

%%%%%%%%%%%%%%%%%%%%%%%%%%%%%%%%%%%%%%
%%%%%%%%%%%%%%%%%%%%%%%%%%%%%%%%%%%%%%
%%%%%%%%%%%%%%%%%%%%%%%%%%%%%%%%%%%%%%

\pgfplotstableread[col sep=space]{
events	data1	data2	data3	data4	data5	data6	data7	data8	data9	data10
1000	29687	29895	29785	29684	29335	29717	29901	29632	29714	29329
2000	31376	31764	31796	31551	31510	31443	31671	31636	31897	31888
3000	33687	33719	33535	33376	33614	33687	33943	33839	33608	33620
4000	35519	35467	35427	35835	35593	35603	35490	35772	35440	35635
5000	37546	37715	37597	37908	37660	37721	37673	37704	37580	37616
6000	39580	39619	39562	39448	39686	39614	39460	39722	39550	39671
7000	41751	41680	41800	41518	41712	41583	41561	41595	41627	41671
8000	43534	43565	43716	43592	43531	43577	43756	43644	43544	43582
9000	45622	45793	45686	45767	45702	45739	45665	45853	45678	45707
10000	47850	47831	47839	47760	47848	47856	47823	47890	47854	47810
}\publishMessage

\pgfplotstablecreatecol[
create col/expr={
(\thisrow{data1} + \thisrow{data2} + \thisrow{data3} + \thisrow{data4} + \thisrow{data5} + \thisrow{data6} + \thisrow{data7} + \thisrow{data8} + \thisrow{data9} + \thisrow{data10})/10
}
]{publishMessage_mean}\publishMessage

\pgfplotstablecreatecol[
create col/expr={
sqrt((pow(\thisrow{data1}-\thisrow{publishMessage_mean},2) + pow(\thisrow{data2}-\thisrow{publishMessage_mean},2) + pow(\thisrow{data3}-\thisrow{publishMessage_mean},2) + pow(\thisrow{data4}-\thisrow{publishMessage_mean},2) + pow(\thisrow{data5}-\thisrow{publishMessage_mean},2) + pow(\thisrow{data6}-\thisrow{publishMessage_mean},2) + pow(\thisrow{data7}-\thisrow{publishMessage_mean},2) + pow(\thisrow{data8}-\thisrow{publishMessage_mean},2) + pow(\thisrow{data9}-\thisrow{publishMessage_mean},2) + pow(\thisrow{data10}-\thisrow{publishMessage_mean},2))/(10-1))
}
]{publishMessage_stddev}\publishMessage

\pgfplotstablecreatecol[
create col/expr={
\thisrow{events} / (\thisrow{publishMessage_mean}/1000) % Convert mean execution time to seconds and calculate throughput
}
]{publishMessage_throughput}\publishMessage

\pgfplotstablecreatecol[
create col/expr={
(\thisrow{publishMessage_stddev}/\thisrow{publishMessage_mean})*100
}
]{publishMessage_cv}\publishMessage

%%%%%%%%%%%%%%%%%%%%%%%%%%%%%%%%%%%%%%
%%%%%%%%%%%%%%%%%%%%%%%%%%%%%%%%%%%%%%
%%%%%%%%%%%%%%%%%%%%%%%%%%%%%%%%%%%%%%

\pgfplotstableread[col sep=space]{
events	data1	data2	data3	data4	data5	data6	data7	data8	data9	data10
1000	2576	2593	2594	2586	2597	2606	2630	2602	2604	2598
2000	2749	2779	2780	2778	2776	2784	2779	2761	2797	2791
3000	2971	2966	2981	2958	2968	2990	2945	2951	2961	2952
4000	3127	3140	3124	3146	3126	3162	3128	3135	3151	3122
5000	3323	3327	3297	3322	3323	3330	3303	3301	3327	3324
6000	3503	3504	3491	3503	3503	3490	3524	3506	3484	3508
7000	3667	3689	3694	3669	3691	3685	3690	3693	3676	3670
8000	3869	3850	3859	3863	3863	3869	3860	3877	3856	3881
9000	4046	4030	4028	4045	4037	4047	4045	4039	4030	4023
10000	4215	4228	4222	4207	4221	4223	4210	4221	4220	4208
}\consumeMessage

\pgfplotstablecreatecol[
create col/expr={
(\thisrow{data1} + \thisrow{data2} + \thisrow{data3} + \thisrow{data4} + \thisrow{data5} + \thisrow{data6} + \thisrow{data7} + \thisrow{data8} + \thisrow{data9} + \thisrow{data10})/10
}
]{consumeMessage_mean}\consumeMessage

\pgfplotstablecreatecol[
create col/expr={
sqrt((pow(\thisrow{data1}-\thisrow{consumeMessage_mean},2) + pow(\thisrow{data2}-\thisrow{consumeMessage_mean},2) + pow(\thisrow{data3}-\thisrow{consumeMessage_mean},2) + pow(\thisrow{data4}-\thisrow{consumeMessage_mean},2) + pow(\thisrow{data5}-\thisrow{consumeMessage_mean},2) + pow(\thisrow{data6}-\thisrow{consumeMessage_mean},2) + pow(\thisrow{data7}-\thisrow{consumeMessage_mean},2) + pow(\thisrow{data8}-\thisrow{consumeMessage_mean},2) + pow(\thisrow{data9}-\thisrow{consumeMessage_mean},2) + pow(\thisrow{data10}-\thisrow{consumeMessage_mean},2))/(10-1))
}
]{consumeMessage_stddev}\consumeMessage

\pgfplotstablecreatecol[
create col/expr={
\thisrow{events} / (\thisrow{consumeMessage_mean}/1000) % Convert mean execution time to seconds and calculate throughput
}
]{consumeMessage_throughput}\consumeMessage

\pgfplotstablecreatecol[
create col/expr={
(\thisrow{consumeMessage_stddev}/\thisrow{consumeMessage_mean})*100
}
]{consumeMessage_cv}\consumeMessage
%%%%%%%%%%%%%%%%%%%%%%%%%%%%%%%%%%%%%%
%%%%%%%%%%%%%%%%%%%%%%%%%%%%%%%%%%%%%%
%%%%%%%%%%%%%%%%%%%%%%%%%%%%%%%%%%%%%%

The \textit{MessageValidator} microservice, as captured in~\Cref{fig:validate_performance,fig:validate_throughput,fig:validate_cv}, displayed commendable stability across escalating event numbers. Its capability to validate voluminous events with minimal time variation is pivotal for reliability. A similar robustness is observable in the \textit{ServiceRegistration} microservice, where~\Cref{fig:registration_performance,fig:registratioin_throughput,fig:registration_cv} illuminate the capacity to manage an amplifying array of registration requests with consistent execution times.

Set Queue Mapping's role, central to routing precision, is validated through~\Cref{fig:setQueueMapping_performance,fig:setQueueMapping_throughput,fig:rsetQueueMapping_cv}, showcasing proficient management of queue mappings under increasing loads. This microservice sustained a uniform throughput, indicating its effectiveness in orchestrating message flow among consumers.

The \textit{PublishMessage} microservice, critical for disseminating information, is analysed through~\Cref{fig:publishMessage_performance,fig:publishMessage_throughput,fig:publishMessage_cv}. Despite the growth in the number of messages to publish, the microservice maintained its performance, ensuring that messages were promptly and reliably published to the message broker.

Lastly, the \textit{ConsumeMessage} microservice demonstrated its capacity to retrieve and process a burgeoning number of messages, as seen in~\Cref{fig:consumeMessage_performance,fig:consumeMessage_throughput,fig:consumeMessage_cv}. The low variation in processing times, even as message loads mounted, underscores the aptitude of the microservice for scalability and reliability.

The evaluation of the \textit{MessageValidator} microservice, illustrated in~\Cref{fig:validate_performance}, revealed its efficiency in validating messages against the LEI schema with increasing loads. Performance remained consistently high despite the incremental number of messages, underscoring the microservice’s capability to handle workload scalability without compromising speed or accuracy.

~\Cref{fig:registration_performance} showcases the results for the \textit{ServiceRegistration} microservice. As the number of producers and consumers scaled up, the system adeptly handled the growing number of registrations. The results indicate that even with a thousand producers/consumers, LEISA's \textit{ServiceRegistration} microservice processed requests effectively, suggesting that the system can seamlessly accommodate a significant user base expansion.

~\Cref{fig:setQueueMapping_performance} reflects the \textit{SetQueueMapping} microservice's performance. The microservice demonstrated excellent proficiency in associating events with consumer queues, a function that becomes increasingly complex as the number of queues expands. The evaluation showed the microservice’s adeptness at maintaining consistent mapping times, highlighting the architecture's capability to sustain precise message routing.

In~\Cref{fig:publishMessage_performance}, the \textit{PublishMessage} microservice’s ability to send messages to the message broker under increasing message volumes was tested. It effectively published messages while maintaining consistent execution times, indicating the microservice’s reliability in information dissemination within the architecture.

In~\Cref{fig:consumeMessage_performance}, the \textit{ConsumeMessage} microservice evaluation is presented. The results from this test were crucial, as they demonstrated the microservice’s capability to retrieve messages from the message broker under increased loads. The ability to consume messages without significant latency is critical for real-time data sharing and analytics, and LEISA proved proficient in this regard.
%%%%%%%%%%%%%%%%%%%%%%%%%%%%%%%%%%%%%%
%%%%%%%%%%%%%%%%%%%%%%%%%%%%%%%%%%%%%%
%%%%%%%%%%%%%%%%%%%%%%%%%%%%%%%%%%%%%%

The \textit{MessageValidator} microservice demonstrated high throughput as shown in~\Cref{fig:validate_throughput}, handling an increasing number of messages with efficiency. The evaluation suggests that this microservice can maintain a steady message validation rate, even as the message load increases, indicative of the optimised performance of the microservice and the ability to contribute to the overall responsiveness of the architecture.

For the \textit{ServiceRegistration} microservice, the throughput assessment illustrated in~\Cref{fig:registratioin_throughput} indicated that it could register many producers/consumers without a drop in performance. The architecture managed a growing number of registrations, showcasing the scalability of the microservice and the ability of the underlying infrastructure to handle a growing number of users.

The \textit{SetQueueMapping} microservice, as shown in~\Cref{fig:setQueueMapping_throughput}, maintained a consistent mapping rate in different queue volumes. This demonstrates the architecture’s capability to efficiently manage message routing between producers and an increasing number of consumer queues, a critical feature for ensuring data reaches the intended recipients promptly.

\Cref{fig:publishMessage_throughput} captures the \textit{PublishMessage} microservice’s throughput, which remained constant even as the message volume grew. This microservice’s stable throughput is critical in preventing bottlenecks when disseminating large volumes of data, affirming its reliability in the architecture’s communication flow.

The evaluation depicted in~\Cref{fig:consumeMessage_throughput} for the \textit{ConsumeMessage} microservice showed that it could process an increasing number of messages from the broker without significant delays. Its throughput stability is essential for applications that rely on timely data consumption, indicating that LEISA can support high-demand environments efficiently.
%%%%%%%%%%%%%%%%%%%%%%%%%%%%%%%%%%%%%%
%%%%%%%%%%%%%%%%%%%%%%%%%%%%%%%%%%%%%%
%%%%%%%%%%%%%%%%%%%%%%%%%%%%%%%%%%%%%%

CV measurements of the \textit{MessageValidator} microservice indicated low variability in processing times.~\Cref{fig:validate_cv} demonstrates the consistent performance of the microservice across different message loads. This low CV suggests the microservice can handle an increasing number of validations without significant fluctuations in performance, which is crucial for ensuring system stability as operations scale up.

~\Cref{fig:registration_cv}, the CV results for the \textit{ServiceRegistration} microservice demonstrated minimal variance in the time taken to register increasing numbers of producers and consumers. The low CV underscores the \textit{ServiceRegistration} microservice's efficiency and reliability in processing a growing number of user registrations while maintaining performance consistency.

The \textit{SetQueueMapping} microservice exhibited a low CV, as shown in~\Cref{fig:rsetQueueMapping_cv}. This low variability in performance, even with a higher number of queue mappings, indicates that the microservice maintains a consistent level of performance, which is essential for reliable message routing within the system as the number of consumers scales.

\Cref{fig:publishMessage_cv} highlights the \textit{PublishMessage} microservice's CV, which remained low across varying volumes of message publication. This consistency ensures that as the system scales, the Publish Microservice can be relied upon to deliver messages to the broker without performance degradation, maintaining system throughput.

The \textit{ConsumeMessage} microservice maintained a low CV, as shown in~\Cref{fig:consumeMessage_cv}. This indicates that the microservice processes messages from the message broker consistently, which is vital for applications requiring real-time data where any significant fluctuation in consumption times could be detrimental.

These collective results underscore LEISA's proficiency in handling the dynamic demands of livestock event information sharing. The architecture's resilience against escalating loads, coupled with its steadfast performance, heralds a system that is scalable, efficient, robust and reliable in its domain of application.

%%%%%%%%%%%%%%%%%%%%%%%%%%%%%%%%%%%%%%%%%%%%%%%%%%%%%%%%
%%%%%             Threats to Validity              %%%%%
%%%%%%%%%%%%%%%%%%%%%%%%%%%%%%%%%%%%%%%%%%%%%%%%%%%%%%%%
\section{Threats to Validity}\label{sec:threats}

In this work, we have introduced a microservice-based architecture for livestock event data sharing, as illustrated in~\Cref{fig:datasharing}. We have identified the following threats to validity:

\textbf{Internal Validity.} The design and execution choices of the LEISA architecture, such as the use of specific cloud platforms, JSON message validator, and a microservices-based framework, may introduce an implementation bias. This bias could influence the assessed performance measures and efficiency. Data integrity is crucial; inaccuracies in data entry, processing, or transmission can undermine the architecture's reliability and the validity of decisions derived from such data. Additionally, the decentralised nature of microservice architecture complicates maintaining consistent service availability and performance. Variability in the performance of individual microservices may skew overall system performance evaluations, thereby affecting the perceived resilience and scalability of LEISA. Moreover, the distributed configuration of microservices presents challenges in fully exploring all potential interactions and failure modes. Subtle configuration differences among microservices might lead to inconsistent behaviour across the system, further complicating the maintenance and scalability challenges.

\textbf{External Validity.} The applicability of LEISA across various agricultural scenarios continues to be a potential constraint. Although it was developed to improve the sharing of livestock event information, its effectiveness and efficiency might fluctuate depending on the size of farming operations or the variety of regulatory frameworks. The extension of the LEI schema to other agricultural sectors indicates this applicability, as LEI forms the basis for data standardisation. Differences in technological preparedness, financial limitations, or reluctance to adapt among different parts of the agricultural community could hinder its broad acceptance. Moreover, the system's performance in a controlled testing environment might not mirror its operation in a real-world context with changing network conditions and loads.

\textbf{Construct Validity.} Problems emerge from the evaluation tools used to assess LEISA's performance. It is critical to ensure that these tools align with the architecture's objectives to validate the results accurately. Furthermore, the operational definitions of key constructs, such as data integrity, scalability, and interoperability, must be consistently defined and enforced to avoid measurement inaccuracies and ensure that evaluations are valid in different studies and contexts. In addition, the introduction of extensive monitoring to keep track of the health and performance of each microservice can add to overhead and complexity.

\textbf{Conclusion Validity.} The validity of conclusions drawn from studies evaluating LEISA greatly depends on the appropriate use of statistical methods. Using inadequate statistical power or incorrect methodologies could lead to incorrect conclusions about the effectiveness and reliability of LEISA. Additionally, it is crucial to be able to reproduce results in different settings, with different datasets, and by different teams. Variations in the implementation details or operational environments could impact the reproducibility of the results, thus affecting the perceived strength of the architecture.

%%%%%%%%%%%%%%%%%%%%%%%%%%%%%%%%%%%%%%%%%%%%%%%%%%%%%%%%
%%%%%                    Conclusion                %%%%%
%%%%%%%%%%%%%%%%%%%%%%%%%%%%%%%%%%%%%%%%%%%%%%%%%%%%%%%%
\section{Conclusion}\label{sec:conclusion}
The Livestock Event Information Sharing Architecture (LEISA) introduced in this study represents a transformative approach to managing and sharing event information within the livestock industry. Taking advantage of the modular and scalable nature of microservices, LEISA facilitates seamless data sharing, seamless system interoperability, and enhanced decision-making capabilities and promotes collaboration among stakeholders. The Livestock Event Information Sharing Architecture (LEISA) introduced in this study represents a transformative approach to managing and sharing event information within the livestock industry. Using the modular and scalable nature of microservices, LEISA facilitates seamless data sharing, system interoperability, and enhanced decision-making capabilities, promoting collaboration among stakeholders.

LEISA aims to revolutionise the red meat supply chain by establishing a self-sustaining data-sharing platform. It enables a seamless bidirectional flow of data, allowing producers to maintain control over their data. This empowerment improves the collaboration between producers and consumers, paving the way for integrated services that increase productivity. The implementation of LEISA is timely, aligning with the growing reliance on digital technology in agriculture and addressing challenges such as data control, compatibility, and sharing.

The architecture leverages advanced technologies, including cloud computing, microservices, and message brokers to facilitate real-time data sharing, ensuring data integrity and standardisation between various stakeholders. This leads to significant improvements in data handling capabilities, improving decision- making processes, and increasing overall productivity within the livestock industry. By enabling dynamic, scalable, and efficient data exchanges, LEISA supports informed decision-making and compliance with the LEI schema.
Practical evaluations of LEISA, using benchmarks such as scalability, data integrity, standardisation, reliability, interoperability, and performance, confirm its effectiveness and reliability. This marks a substantial progression towards an integrated and data-driven agricultural ecosystem.
Future research will focus on expanding LEISA’s adaptability to encompass a broader range of agricultural data types and scenarios. Enhancements in scalability and security features will be explored, along with the integration of emerging technologies such as artificial intelligence and blockchain. These developments are essential to maintain relevance and maximise impact in the rapidly evolving field of agricultural technology.

%%%%%%%%%%%%%%%%%%%%%%%%%%%%%%%%%%%%%%%%%%%%%%%%%%%%%%%%
%%%%%              Funding Sources                 %%%%%
%%%%%%%%%%%%%%%%%%%%%%%%%%%%%%%%%%%%%%%%%%%%%%%%%%%%%%%%
\section*{Funding sources}
This project was supported by funding from Food Agility CRC Ltd, funded under the Commonwealth Government CRC Program. The CRC Program supports industry-led collaborations between industry, researchers, and the community, and this manuscript was funded by the Gulbali Institute Accelerated Publication Scheme (GAPS).

%%%%%%%%%%%%%%%%%%%%%%%%%%%%%%%%%%%%%%%%%%%%%%%%%%%%%%%%
%%%%%               Acknowledgment                 %%%%%
%%%%%%%%%%%%%%%%%%%%%%%%%%%%%%%%%%%%%%%%%%%%%%%%%%%%%%%%
\section*{Acknowledgment}
We thank David Swain (TerraCipher) and Will Swain (TerraCipher) for their valuable comments and reviews.

%%%%%%%%%%%%%%%%%%%%%%%%%%%%%%%%%%%%%%%%%%%%%%%%%%%%%%%%
%%%%%                    appendix                  %%%%%
%%%%%%%%%%%%%%%%%%%%%%%%%%%%%%%%%%%%%%%%%%%%%%%%%%%%%%%%
\appendix

%%%%%%%%%%%%%%%%%%%%%%%%%%%%%%%%%%%%%%%%%%%%%%%%%%%%%%%%
%%%%%                 bibliography                 %%%%%
%%%%%%%%%%%%%%%%%%%%%%%%%%%%%%%%%%%%%%%%%%%%%%%%%%%%%%%%
% \bibliographystyle{elsarticle-harv}
\bibliographystyle{elsarticle-num-names} 
\bibliography{ref}

\end{document}